%
\catcode`@=11 
%
%
%

\font\fourteenrm=cmr10 scaled\magstep2
\font\twelverm=cmr12
\font\elvrm=cmr10 scaled\magstephalf
\font\ninerm=cmr9            \font\sixrm=cmr6
\font\egtrm=cmr8

\font\fourteenbf=cmbx10 scaled\magstep2
\font\twelvebf=cmbx12
\font\elvbf=cmbx10 scaled\magstephalf
\font\ninebf=cmbx9            \font\sixbf=cmbx6
\font\egtbf=cmbx8
\font\seventeeni=cmmi10 scaled\magstep3     \skewchar\seventeeni='177
\font\fourteeni=cmmi10 scaled\magstep2      \skewchar\fourteeni='177
\font\twelvei=cmmi12                        \skewchar\twelvei='177
\font\elvi=cmmi10 scaled\magstephalf     \skewchar\elvi='177
\font\ninei=cmmi9                           \skewchar\ninei='177
\font\egti=cmmi8                           \skewchar\egti='177
\font\sixi=cmmi6                            \skewchar\sixi='177
\font\seventeensy=cmsy10 scaled\magstep3    \skewchar\seventeensy='60
\font\fourteensy=cmsy10 scaled\magstep2     \skewchar\fourteensy='60
\font\twelvesy=cmsy10 scaled\magstep1       \skewchar\twelvesy='60
\font\elvsy=cmsy10 scaled\magstephalf    \skewchar\elvsy='60
\font\ninesy=cmsy9                          \skewchar\ninesy='60
\font\egtsy=cmsy8                          \skewchar\egtsy='60
\font\sixsy=cmsy6                           \skewchar\sixsy='60

\font\fourteenex=cmex10 scaled\magstep2
\font\twelveex=cmex10 scaled\magstep1

\font\elvex=cmex10 scaled\magstephalf
\font\ninex=cmex10

\font\fourteensl=cmsl10 scaled\magstep2
\font\twelvesl=cmsl12
\font\elvsl=cmsl10 scaled\magstephalf
\font\ninesl=cmsl9

\font\fourteenit=cmti10 scaled\magstep2
\font\twelveit=cmti12
\font\elvit=cmti10 scaled \magstephalf
\font\nineit=cmti9

\font\twelvett=cmtt12
\font\elvtt=cmtt10 scaled \magstephalf
\font\ninett=cmtt9

 \font\twelvecp=cmcsc10 scaled\magstep1
 \font\elvcp=cmcsc10 scaled\magstephalf
 \font\tencp=cmcsc10
 \font\ninecp=cmcsc10
 
 \newfam\cpfam
%
%
\newcount\f@ntkey            \f@ntkey=0
\def\samef@nt{\relax \ifcase\f@ntkey \rm \or\oldstyle \or\or
         \or\it \or\sl \or\bf \or\tt \or\caps \fi }
\def\fourteenpoint{\relax
    \textfont0=\fourteenrm          \scriptfont0=\tenrm
    \scriptscriptfont0=\sevenrm
     \def\rm{\fam0 \fourteenrm \f@ntkey=0 }\relax
    \textfont1=\fourteeni           \scriptfont1=\teni
    \scriptscriptfont1=\seveni
     \def\oldstyle{\fam1 \fourteeni\f@ntkey=1 }\relax
    \textfont2=\fourteensy          \scriptfont2=\tensy
    \scriptscriptfont2=\sevensy
    \textfont3=\fourteenex     \scriptfont3=\fourteenex
    \scriptscriptfont3=\fourteenex
    \def\it{\fam\itfam \fourteenit\f@ntkey=4 }\textfont\itfam=\fourteenit
    \def\sl{\fam\slfam \fourteensl\f@ntkey=5 }\textfont\slfam=\fourteensl
    \scriptfont\slfam=\tensl
    \def\bf{\fam\bffam \fourteenbf\f@ntkey=6 }\textfont\bffam=\fourteenbf
    \scriptfont\bffam=\tenbf     \scriptscriptfont\bffam=\sevenbf
    \def\tt{\fam\ttfam \twelvett \f@ntkey=7 }\textfont\ttfam=\twelvett
    \h@big=11.9\p@{} \h@Big=16.1\p@{} \h@bigg=20.3\p@{} \h@Bigg=24.5\p@{}
    \def\caps{\fam\cpfam \twelvecp \f@ntkey=8 }\textfont\cpfam=\twelvecp
    \setbox\strutbox=\hbox{\vrule height 12pt depth 5pt width\z@}
    \samef@nt}
\def\twelvepoint{\relax
    \textfont0=\twelverm          \scriptfont0=\ninerm
    \scriptscriptfont0=\sixrm
     \def\rm{\fam0 \twelverm \f@ntkey=0 }\relax
    \textfont1=\twelvei           \scriptfont1=\ninei
    \scriptscriptfont1=\sixi
     \def\oldstyle{\fam1 \twelvei\f@ntkey=1 }\relax
    \textfont2=\twelvesy          \scriptfont2=\ninesy
    \scriptscriptfont2=\sixsy
    \textfont3=\twelveex          \scriptfont3=\twelveex
    \scriptscriptfont3=\twelveex
    \def\it{\fam\itfam \twelveit \f@ntkey=4 }\textfont\itfam=\twelveit
    \def\sl{\fam\slfam \twelvesl \f@ntkey=5 }\textfont\slfam=\twelvesl
    \scriptfont\slfam=\ninesl
    \def\bf{\fam\bffam \twelvebf \f@ntkey=6 }\textfont\bffam=\twelvebf
    \scriptfont\bffam=\ninebf     \scriptscriptfont\bffam=\sixbf
    \def\tt{\fam\ttfam \twelvett \f@ntkey=7 }\textfont\ttfam=\twelvett
    \h@big=10.2\p@{}
    \h@Big=13.8\p@{}
    \h@bigg=17.4\p@{}
    \h@Bigg=21.0\p@{}
    \def\caps{\fam\cpfam \twelvecp \f@ntkey=8 }\textfont\cpfam=\twelvecp
    \setbox\strutbox=\hbox{\vrule height 10pt depth 4pt width\z@}
    \samef@nt}
\def\elvpoint{\relax
    \textfont0=\elvrm          \scriptfont0=\egtrm
    \scriptscriptfont0=\sixrm
    \def\rm{\fam0 \elvrm \f@ntkey=0 }\relax
    \textfont1=\elvi           \scriptfont1=\egti
    \scriptscriptfont1=\sixi
    \def\oldstyle{\fam1 \elvi \f@ntkey=1 }\relax
    \textfont2=\elvsy          \scriptfont2=\egtsy
    \scriptscriptfont2=\sixsy
    \textfont3=\elvex          \scriptfont3=\elvex
    \scriptscriptfont3=\elvex
    \def\it{\fam\itfam \elvit \f@ntkey=4 }\textfont\itfam=\elvit
    \def\sl{\fam\slfam \elvsl \f@ntkey=5 }\textfont\slfam=\elvsl
    \def\bf{\fam\bffam \elvbf \f@ntkey=6 }\textfont\bffam=\elvbf
    \scriptfont\bffam=\egtbf     \scriptscriptfont\bffam=\sixbf
    \def\tt{\fam\ttfam \elvtt \f@ntkey=7 }\textfont\ttfam=\elvtt
    \def\caps{\fam\cpfam \elvcp \f@ntkey=8 }\textfont\cpfam=\elvcp
    \setbox\strutbox=\hbox{\vrule height 9.2pt depth 3.7pt width\z@}
    \samef@nt}
\def\tenpoint{\relax
    \textfont0=\tenrm          \scriptfont0=\sevenrm
    \scriptscriptfont0=\fiverm
    \def\rm{\fam0 \tenrm \f@ntkey=0 }\relax
    \textfont1=\teni           \scriptfont1=\seveni
    \scriptscriptfont1=\fivei
    \def\oldstyle{\fam1 \teni \f@ntkey=1 }\relax
    \textfont2=\tensy          \scriptfont2=\sevensy
    \scriptscriptfont2=\fivesy
    \textfont3=\tenex          \scriptfont3=\tenex
    \scriptscriptfont3=\tenex
    \def\it{\fam\itfam \tenit \f@ntkey=4 }\textfont\itfam=\tenit
    \def\sl{\fam\slfam \tensl \f@ntkey=5 }\textfont\slfam=\tensl
    \def\bf{\fam\bffam \tenbf \f@ntkey=6 }\textfont\bffam=\tenbf
    \scriptfont\bffam=\sevenbf     \scriptscriptfont\bffam=\fivebf
    \def\tt{\fam\ttfam \tentt \f@ntkey=7 }\textfont\ttfam=\tentt
    \def\caps{\fam\cpfam \tencp \f@ntkey=8 }\textfont\cpfam=\tencp
    \setbox\strutbox=\hbox{\vrule height 8.5pt depth 3.5pt width\z@}
    \samef@nt}
\def\ninpoint{\relax
    \textfont0=\ninerm          \scriptfont0=\sevenrm
    \scriptscriptfont0=\fiverm
    \def\rm{\fam0 \ninerm \f@ntkey=0 }\relax
    \textfont1=\ninei           \scriptfont1=\seveni
    \scriptscriptfont1=\fivei
    \def\oldstyle{\fam1 \ninei \f@ntkey=1 }\relax
    \textfont2=\ninesy          \scriptfont2=\sevensy
    \scriptscriptfont2=\fivesy
    \textfont3=\ninex          \scriptfont3=\ninex
    \scriptscriptfont3=\ninex
    \def\it{\fam\itfam \nineit \f@ntkey=4 }\textfont\itfam=\nineit
    \def\sl{\fam\slfam \ninesl \f@ntkey=5 }\textfont\slfam=\ninesl
    \def\bf{\fam\bffam \ninebf \f@ntkey=6 }\textfont\bffam=\ninebf
    \scriptfont\bffam=\sevenbf     \scriptscriptfont\bffam=\fivebf
    \def\tt{\fam\ttfam \ninett \f@ntkey=7 }\textfont\ttfam=\ninett
    \def\caps{\fam\cpfam \ninecp \f@ntkey=8 }\textfont\cpfam=\ninecp
    \setbox\strutbox=\hbox{\vrule height 8. pt depth 3.3pt width\z@}
    \samef@nt}
%
%
%
%
\newdimen\h@big  \h@big=8.5\p@
\newdimen\h@Big  \h@Big=11.5\p@
\newdimen\h@bigg  \h@bigg=14.5\p@
\newdimen\h@Bigg  \h@Bigg=17.5\p@
\def\big#1{{\hbox{$\left#1\vbox to\h@big{}\right.\n@space$}}}
\def\Big#1{{\hbox{$\left#1\vbox to\h@Big{}\right.\n@space$}}}
\def\bigg#1{{\hbox{$\left#1\vbox to\h@bigg{}\right.\n@space$}}}
\def\Bigg#1{{\hbox{$\left#1\vbox to\h@Bigg{}\right.\n@space$}}}
%
%
%
\normalbaselineskip = 20pt plus 0.2pt minus 0.1pt
\normallineskip = 1.5pt plus 0.1pt minus 0.1pt
\normallineskiplimit = 1.5pt
\newskip\normaldisplayskip
\normaldisplayskip = 8pt plus 4pt minus 3pt
\newskip\normaldispshortskip
\normaldispshortskip = 6pt plus 2pt
\newskip\normalparskip
\normalparskip = 3pt plus 1pt minus 1pt
\newskip\skipregister
\skipregister = 4pt plus 1pt minus .5pt
\newif\ifsingl@    \newif\ifdoubl@
\newif\iftwelv@    \twelv@true
\def\singlespace{\singl@true\doubl@false\spaces@t}
\def\doublespace{\singl@false\doubl@true\spaces@t}
\def\normalspace{\singl@false\doubl@false\spaces@t}
\def\Elvpoint{\elvpoint\twelv@false\spaces@t}
\def\Tenpoint{\tenpoint\twelv@false\spaces@t}
\def\Ninpoint{\ninpoint\twelv@false\spaces@t}
\def\Twelvepoint{\twelvepoint\twelv@true\spaces@t}
\def\spaces@t{\relax%
 \iftwelv@ \ifsingl@\subspaces@t3:4;\else\subspaces@t1:1;\fi%
 \else \ifsingl@\subspaces@t3:5;\else\subspaces@t4:5;\fi \fi%
 \ifdoubl@ \multiply\baselineskip by 5%
 \divide\baselineskip by 4 \fi \unskip}
\def\subspaces@t#1:#2;{
      \baselineskip = \normalbaselineskip
      \multiply\baselineskip by #1 \divide\baselineskip by #2
      \lineskip = \normallineskip
      \multiply\lineskip by #1 \divide\lineskip by #2
      \lineskiplimit = \normallineskiplimit
      \multiply\lineskiplimit by #1 \divide\lineskiplimit by #2
      \parskip = \normalparskip
      \multiply\parskip by #1 \divide\parskip by #2
      \abovedisplayskip = \normaldisplayskip
      \multiply\abovedisplayskip by #1 \divide\abovedisplayskip by #2
      \belowdisplayskip = \abovedisplayskip
      \abovedisplayshortskip = \normaldispshortskip
      \multiply\abovedisplayshortskip by #1
        \divide\abovedisplayshortskip by #2
      \belowdisplayshortskip = \abovedisplayshortskip
      \advance\belowdisplayshortskip by \belowdisplayskip
      \divide\belowdisplayshortskip by 2
      \smallskipamount = \skipregister
      \multiply\smallskipamount by #1 \divide\smallskipamount by #2
      \medskipamount = \smallskipamount \multiply\medskipamount by 2
      \bigskipamount = \smallskipamount \multiply\bigskipamount by 4 }
\def\normalbaselines{ \baselineskip=\normalbaselineskip
   \lineskip=\normallineskip \lineskiplimit=\normallineskip
   \iftwelv@\else \multiply\baselineskip by 4 \divide\baselineskip by 5
     \multiply\lineskiplimit by 4 \divide\lineskiplimit by 5
     \multiply\lineskip by 4 \divide\lineskip by 5 \fi }
\Twelvepoint  
\interlinepenalty=50
\interfootnotelinepenalty=5000
\predisplaypenalty=9000
\postdisplaypenalty=500
\hfuzz=1pt
\vfuzz=0.2pt
%
%
%
\def\pagecontents{
   \ifvoid\topins\else\unvbox\topins\vskip\skip\topins\fi
   \dimen@ = \dp255 \unvbox255
   \ifvoid\footins\else\vskip\skip\footins\footrule\unvbox\footins\fi
   \ifr@ggedbottom \kern-\dimen@ \vfil \fi }
\def\makeheadline{\vbox to 0pt{ \skip@=\topskip
      \advance\skip@ by -12pt \advance\skip@ by -2\normalbaselineskip
      \vskip\skip@ \line{\vbox to 12pt{}\the\headline} \vss
      }\nointerlineskip}
\def\makefootline{\baselineskip = 1.5\normalbaselineskip
                 \line{\the\footline}}
\newif\iffrontpage
\newif\ifletterstyle
\newif\ifp@genum
\def\nopagenumbers{\p@genumfalse}
\def\pagenumbers{\p@genumtrue}
\pagenumbers
\newtoks\paperheadline
\newtoks\letterheadline
\newtoks\letterfrontheadline
\newtoks\lettermainheadline
\newtoks\paperfootline
\newtoks\letterfootline
\newtoks\date
\footline={\ifletterstyle\the\letterfootline\else\the\paperfootline\fi}
\paperfootline={\hss\iffrontpage\else\ifp@genum\tenrm\folio\hss\fi\fi}
\letterfootline={\hfil}
\headline={\ifletterstyle\the\letterheadline\else\the\paperheadline\fi}
\paperheadline={\hfil}
\letterheadline{\iffrontpage\the\letterfrontheadline
     \else\the\lettermainheadline\fi}
\lettermainheadline={\rm\ifp@genum page \ \folio\fi\hfil\the\date}
\def\monthname{\relax\ifcase\month 0/\or January\or February\or
   March\or April\or May\or June\or July\or August\or September\or
   October\or November\or December\else\number\month/\fi}
\date={\monthname\ \number\day, \number\year}
\countdef\pagenumber=1  \pagenumber=1
\def\advancepageno{\global\advance\pageno by 1
   \ifnum\pagenumber<0 \global\advance\pagenumber by -1
    \else\global\advance\pagenumber by 1 \fi \global\frontpagefalse }
\def\folio{\ifnum\pagenumber<0 \romannumeral-\pagenumber
           \else \number\pagenumber \fi }
\def\footrule{\dimen@=\prevdepth\nointerlineskip
   \vbox to 0pt{\vskip -0.25\baselineskip \hrule width 0.35\hsize \vss}
   \prevdepth=\dimen@ }
\newtoks\foottokens
\foottokens={\Tenpoint\singlespace}
\newdimen\footindent
\footindent=24pt
\def\vfootnote#1{\insert\footins\bgroup  \the\foottokens
   \interlinepenalty=\interfootnotelinepenalty \floatingpenalty=20000
   \splittopskip=\ht\strutbox \boxmaxdepth=\dp\strutbox
   \leftskip=\footindent \rightskip=\z@skip
   \parindent=0.5\footindent \parfillskip=0pt plus 1fil
   \spaceskip=\z@skip \xspaceskip=\z@skip
   \Textindent{$ #1 $}\footstrut\futurelet\next\fo@t}
\def\Textindent#1{\noindent\llap{#1\enspace}\ignorespaces}
\def\footnote#1{\attach{#1}\vfootnote{#1}}

\def\foot{\attach\footsymbolgen\vfootnote{\footsymbol}}
\let\footsymbol=\star
\newcount\lastf@@t           \lastf@@t=-1
\newcount\footsymbolcount    \footsymbolcount=0
\newif\ifPhysRev
\def\footsymbolgen{\relax \ifPhysRev \iffrontpage \NPsymbolgen\else
      \PRsymbolgen\fi \else \NPsymbolgen\fi
   \global\lastf@@t=\pageno \footsymbol }
\def\NPsymbolgen{\ifnum\footsymbolcount<0 \global\footsymbolcount=0\fi
   {\iffrontpage \else \advance\lastf@@t by 1 \fi
    \ifnum\lastf@@t<\pageno \global\footsymbolcount=0
     \else \global\advance\footsymbolcount by 1 \fi }
   \ifcase\footsymbolcount \fd@f\star\or \fd@f\dagger\or \fd@f\ast\or
    \fd@f\ddagger\or \fd@f\natural\or \fd@f\diamond\or \fd@f\bullet\or
    \fd@f\nabla\else \fd@f\dagger\global\footsymbolcount=0 \fi }
\def\fd@f#1{\xdef\footsymbol{#1}}
\def\PRsymbolgen{\ifnum\footsymbolcount>0 \global\footsymbolcount=0\fi
      \global\advance\footsymbolcount by -1
      \xdef\footsymbol{\sharp\number-\footsymbolcount} }
\def\space@ver#1{\let\@sf=\empty \ifmmode #1\else \ifhmode
   \edef\@sf{\spacefactor=\the\spacefactor}\unskip${}#1$\relax\fi\fi}
\def\attach#1{\space@ver{\strut^{\mkern 2mu #1} }\@sf\ }
%
%
\def\smallsize{\relax
\font\eightrm=cmr8
\font\eightbf=cmbx8
\font\eighti=cmmi8
\font\eightsy=cmsy8
\font\eightsl=cmsl8
\font\eightit=cmti8
\font\eightt=cmtt8
\def\eightpoint{\relax
\textfont0=\eightrm  \scriptfont0=\sixrm
\scriptscriptfont0=\sixrm
\def\rm{\fam0 \eightrm \f@ntkey=0}\relax
\textfont1=\eighti  \scriptfont1=\sixi
\scriptscriptfont1=\sixi
\def\oldstyle{\fam1 \eighti \f@ntkey=1}\relax
\textfont2=\eightsy  \scriptfont2=\sixsy
\scriptscriptfont2=\sixsy
\textfont3=\tenex  \scriptfont3=\tenex
\scriptscriptfont3=\tenex
\def\it{\fam\itfam \eightit \f@ntkey=4 }\textfont\itfam=\eightit
\def\sl{\fam\slfam \eightsl \f@ntkey=5 }\textfont\slfam=\eightsl
\def\bf{\fam\bffam \eightbf \f@ntkey=6 }\textfont\bffam=\eightbf
\scriptfont\bffam=\sixbf   \scriptscriptfont\bffam=\sixbf
\def\tt{\fam\ttfam \eightt \f@ntkey=7 }
\def\caps{\fam\cpfam \tencp \f@ntkey=8 }\textfont\cpfam=\tencp
\setbox\strutbox=\hbox{\vrule height 7.35pt depth 3.02pt width\z@}
\samef@nt}
\def\Eightpoint{\eightpoint \relax
  \ifsingl@\subspaces@t2:5;\else\subspaces@t3:5;\fi
  \ifdoubl@ \multiply\baselineskip by 5
            \divide\baselineskip by 4\fi }
\parindent=16.67pt
\itemsize=25pt
\thinmuskip=2.5mu
\medmuskip=3.33mu plus 1.67mu minus 3.33mu
\thickmuskip=4.17mu plus 4.17mu
\def\thinspace{\kern .13889em }
\def\negthinspace{\kern-.13889em }
\def\enspace{\kern.416667em }
 
\def\enskip{\hskip.416667em\relax}
\def\quad{\hskip.83333em\relax}
\def\qquad{\hskip1.66667em\relax}
\def\crr{\cropen{8.3333pt}}
\foottokens={\Eightpoint\singlespace}
\def\papersize{\vsize=38.67pc\hsize=29.17pc\hoffset=3.44pc\voffset=3.7pc
               \skip\footins=\bigskipamount}
\def\lettersize{\hsize=5.417in\vsize=7.08in\hoffset=0in\voffset=.834in
   \skip\footins=\smallskipamount \multiply\skip\footins by 3 }
\def\attach##1{\space@ver{\strut^{\mkern 1.6667mu ##1} }\@sf\ }
\def\PH@SR@V{\doubl@true\baselineskip=20.08pt plus .1667pt minus .0833pt
             \parskip = 2.5pt plus 1.6667pt minus .8333pt }
\def\author##1{\vskip\frontpageskip\titlestyle{\tencp ##1}\nobreak}
\def\address##1{\par\kern 4.16667pt\titlestyle{\tenpoint\it ##1}}
\def\andaddress{\par\kern 4.16667pt \centerline{\sl and} \address}
\def\SLAC{\address{Stanford Linear Accelerator Center\break
      Stanford University, Stanford, California, 94305}}
\def\abstract{\vskip\frontpageskip\centerline{\twelverm ABSTRACT}
              \vskip\headskip }
\def\submit##1{\par\nobreak\vfil\nobreak\medskip
   \centerline{Submitted to \sl ##1}}
\def\doeack{\foot{Work supported by the Department of Energy,
      contract $\caps DE-AC03-76SF00515$.}}
\def\cases##1{\left\{\,\vcenter{\Tenpoint\m@th
    \ialign{$####\hfil$&\quad####\hfil\crcr##1\crcr}}\right.}
\def\matrix##1{\,\vcenter{\Tenpoint\m@th
    \ialign{\hfil$####$\hfil&&\quad\hfil$####$\hfil\crcr
      \mathstrut\crcr\noalign{\kern-\baselineskip}
     ##1\crcr\mathstrut\crcr\noalign{\kern-\baselineskip}}}\,}
\Tenpoint \paperstyle
}
%
%
%
\newcount\chapternumber      \chapternumber=0
\newcount\sectionnumber      \sectionnumber=0
\newcount\equanumber         \equanumber=0
\let\chapterlabel=0
\newtoks\chapterstyle        \chapterstyle={\Number}
\newskip\chapterskip         \chapterskip=\bigskipamount
\newskip\sectionskip         \sectionskip=\medskipamount
\newskip\headskip            \headskip=6pt plus 1pt minus 3pt
\newdimen\chapterminspace    \chapterminspace=5pc
\newdimen\sectionminspace    \sectionminspace=3pc
\newdimen\referenceminspace  \referenceminspace=5pc
\def\chapterreset{\global\advance\chapternumber by 1
   \ifnum\the\equanumber<0 \else\global\equanumber=0\fi
   \figurecount=0
   \tablecount=0
   \sectionnumber=0 \makel@bel}
\def\makel@bel{\xdef\chapterlabel{%
\the\chapterstyle{\the\chapternumber}.}}
\def\sectionlabel{\number\sectionnumber \quad }
\def\alphabetic#1{\count255='140 \advance\count255 by #1\char\count255}
\def\Alphabetic#1{\count255='100 \advance\count255 by #1\char\count255}
\def\Roman#1{\uppercase\expandafter{\romannumeral #1}}
\def\roman#1{\romannumeral #1}
\def\Number#1{\number #1}
\def\unnumberedchapters{\let\makel@bel=\relax \let\chapterlabel=\relax
\let\sectionlabel=\relax \equanumber=-1 }
\def\titlestyle#1{\par\begingroup \interlinepenalty=9999
     \leftskip=0.02\hsize plus 0.23\hsize minus 0.02\hsize
     \rightskip=\leftskip \parfillskip=0pt
     \hyphenpenalty=9000 \exhyphenpenalty=9000
     \tolerance=9999 \pretolerance=9000
     \spaceskip=0.333em \xspaceskip=0.5em
     \iftwelv@\fourteenpoint\else\twelvepoint\fi
   \noindent #1\par\endgroup }
\def\spacecheck#1{\dimen@=\pagegoal\advance\dimen@ by -\pagetotal
   \ifdim\dimen@<#1 \ifdim\dimen@>0pt \vfil\break \fi\fi}
\def\chapter#1{\par \penalty-300 \vskip\chapterskip
   \spacecheck\chapterminspace
   \chapterreset \titlestyle{\chapterlabel \ #1}
   \nobreak\vskip\headskip \penalty 30000
   \wlog{\string\chapter\ \chapterlabel} }

\def\section#1{\par \ifnum\the\lastpenalty=30000\else
   \penalty-200\vskip\sectionskip \spacecheck\sectionminspace\fi
   \wlog{\string\section\ \chapterlabel \the\sectionnumber}
   \global\advance\sectionnumber by 1  \noindent
   {\caps\enspace\chapterlabel \sectionlabel #1}\par
   \nobreak\vskip\headskip \penalty 30000 }
\def\subsection#1{\par
   \ifnum\the\lastpenalty=30000\else \penalty-100\smallskip \fi
   \noindent\undertext{#1}\enspace \vadjust{\penalty5000}}
\let\subsec=\subsection
\def\undertext#1{\vtop{\hbox{#1}\kern 1pt \hrule}}
\def\APPENDIX#1#2{\par\penalty-300\vskip\chapterskip
   \spacecheck\chapterminspace \chapterreset \xdef\chapterlabel{#1}
   \titlestyle{APPENDIX #2} \nobreak\vskip\headskip \penalty 30000
   \wlog{\string\Appendix\ \chapterlabel} }
\def\Appendix#1{\APPENDIX{#1}{#1}}
\def\appendix{\APPENDIX{A}{}}
%
%
%
\def\eqname#1{\relax \ifnum\the\equanumber<0%
     \xdef#1{{\noexpand\rm(\number-\equanumber)}}%
     \global\advance\equanumber by -1%
    \else \global\advance\equanumber by 1%
      \xdef#1{{\noexpand\rm(\chapterlabel \number\equanumber)}} \fi}

\def\eqn#1{\eqno\eqname{#1}#1}

\def\eqinsert#1{\noalign{\dimen@=\prevdepth \nointerlineskip
   \setbox0=\hbox to\displaywidth{\hfil #1}
   \vbox to 0pt{\vss\hbox{$\!\box0\!$}\kern-0.5\baselineskip}
   \prevdepth=\dimen@}}
\def\sequentialequations{\globaleqnumbers}
%
%
\def\GENITEM#1;#2{\par \hangafter=0 \hangindent=#1
    \Textindent{$ #2 $}\ignorespaces}
\outer\def\newitem#1=#2;{\gdef#1{\GENITEM #2;}}
\newdimen\itemsize                \itemsize=30pt
\newitem\item=1\itemsize;
\newitem\sitem=1.75\itemsize;     
\newitem\ssitem=2.5\itemsize;     
\outer\def\newlist#1=#2&#3&#4;{\toks0={#2}\toks1={#3}%
   \count255=\escapechar \escapechar=-1
   \alloc@0\list\countdef\insc@unt\listcount     \listcount=0
   \edef#1{\par
      \countdef\listcount=\the\allocationnumber
      \advance\listcount by 1
      \hangafter=0 \hangindent=#4
      \Textindent{\the\toks0{\listcount}\the\toks1}}
   \expandafter\expandafter\expandafter
    \edef\c@t#1{begin}{\par
      \countdef\listcount=\the\allocationnumber \listcount=1
      \hangafter=0 \hangindent=#4
      \Textindent{\the\toks0{\listcount}\the\toks1}}
   \expandafter\expandafter\expandafter
    \edef\c@t#1{con}{\par \hangafter=0 \hangindent=#4 \noindent}
   \escapechar=\count255}
\def\c@t#1#2{\csname\string#1#2\endcsname}
\newlist\point=\Number&.&1.0\itemsize;
\newlist\subpoint=(\alphabetic&)&1.75\itemsize;
\newlist\subsubpoint=(\roman&)&2.5\itemsize;
%

%
%
%
\newcount\referencecount     \referencecount=0
\newif\ifreferenceopen       \newwrite\referencewrite
\newtoks\rw@toks
\def\NPrefmark#1{\attach{\scriptstyle [ #1 ] }}
\let\PRrefmark=\attach
\def\refmark#1{\relax\ifPhysRev\PRrefmark{#1}\else\NPrefmark{#1}\fi}
\def\refend{\refmark{\number\referencecount}}
\newcount\lastrefsbegincount \lastrefsbegincount=0
\def\refsend{\refmark{\count255=\referencecount
   \advance\count255 by-\lastrefsbegincount
   \ifcase\count255 \number\referencecount
   \or \number\lastrefsbegincount,\number\referencecount
   \else \number\lastrefsbegincount-\number\referencecount \fi}}
\def\refch@ck{\chardef\rw@write=\referencewrite
   \ifreferenceopen \else \referenceopentrue
   \immediate\openout\referencewrite=referenc.texauxil \fi}
%
{\catcode`\^^M=\active 
  \gdef\obeyendofline{\catcode`\^^M\active \let^^M\ }}%
%
{\catcode`\^^M=\active 
  \gdef\ignoreendofline{\catcode`\^^M=5}}
{\obeyendofline\gdef\rw@start#1{\def\t@st{#1} \ifx\t@st\blankend%
\endgroup \@sf \relax \else \ifx\t@st\bl@nkend \endgroup \@sf \relax%
\else \rw@begin#1
\backtotext
\fi \fi } }
{\obeyendofline\gdef\rw@begin#1
{\def\n@xt{#1}\rw@toks={#1}\relax%
\rw@next}}
\def\blankend{}
{\obeylines\gdef\bl@nkend{
}}
\newif\iffirstrefline  \firstreflinetrue
\def\rwr@teswitch{\ifx\n@xt\blankend \let\n@xt=\rw@begin %
 \else\iffirstrefline \global\firstreflinefalse%
\immediate\write\rw@write{\noexpand\obeyendofline \the\rw@toks}%
\let\n@xt=\rw@begin%
      \else\ifx\n@xt\rw@@d \def\n@xt{\immediate\write\rw@write{%
        \noexpand\ignoreendofline}\endgroup \@sf}%
             \else \immediate\write\rw@write{\the\rw@toks}%
             \let\n@xt=\rw@begin\fi\fi \fi}
\def\rw@next{\rwr@teswitch\n@xt}
\def\rw@@d{\backtotext} \let\rw@end=\relax
\let\backtotext=\relax

\newdimen\refindent     \refindent=30pt
\def\refitem#1{\par \hangafter=0 \hangindent=\refindent \Textindent{#1}}
\def\REFNUM#1{\space@ver{}\refch@ck \firstreflinetrue%
 \global\advance\referencecount by 1 \xdef#1{\the\referencecount}}
\def\refnum#1{\space@ver{}\refch@ck \firstreflinetrue%
 \global\advance\referencecount by 1 \xdef#1{\the\referencecount}\refend}

\def\REF#1{\REFNUM#1%
 \immediate\write\referencewrite{%
 \noexpand\refitem{#1.}}%
\begingroup\obeyendofline\rw@start}
\def\ref{\refnum\?%
 \immediate\write\referencewrite{\noexpand\refitem{\?.}}%
\begingroup\obeyendofline\rw@start}
\def\Ref#1{\refnum#1%
 \immediate\write\referencewrite{\noexpand\refitem{#1.}}%
\begingroup\obeyendofline\rw@start}
\def\REFS#1{\REFNUM#1\global\lastrefsbegincount=\referencecount
\immediate\write\referencewrite{\noexpand\refitem{#1.}}%
\begingroup\obeyendofline\rw@start}

\def\REFSCON#1{\REF#1}
\newcount\figurecount     \figurecount=0
\newif\iffigureopen       \newwrite\figurewrite
\def\figch@ck{\chardef\rw@write=\figurewrite \iffigureopen\else
   \immediate\openout\figurewrite=figures.texauxil
   \figureopentrue\fi}
\def\FIGNUM#1{\space@ver{}\figch@ck \firstreflinetrue%
 \global\advance\figurecount by 1 \xdef#1{\the\figurecount}}
\def\FIG#1{\FIGNUM#1
   \immediate\write\figurewrite{\noexpand\refitem{#1.}}%
   \begingroup\obeyendofline\rw@start}
\def\fig{\FIGNUM\? fig.~\?%
\immediate\write\figurewrite{\noexpand\refitem{\?.}}%
\begingroup\obeyendofline\rw@start}
\def\figure{\FIGNUM\? figure~\?
   \immediate\write\figurewrite{\noexpand\refitem{\?.}}%
   \begingroup\obeyendofline\rw@start}
\def\Fig{\FIGNUM\? Fig.~\?%
\immediate\write\figurewrite{\noexpand\refitem{\?.}}%
\begingroup\obeyendofline\rw@start}
\def\Figure{\FIGNUM\? Figure~\?%
\immediate\write\figurewrite{\noexpand\refitem{\?.}}%
\begingroup\obeyendofline\rw@start}
\newcount\tablecount     \tablecount=0
\newif\iftableopen       \newwrite\tablewrite
\def\tabch@ck{\chardef\rw@write=\tablewrite \iftableopen\else
   \immediate\openout\tablewrite=tables.texauxil
   \tableopentrue\fi}
\def\TABNUM#1{\space@ver{}\tabch@ck \firstreflinetrue%
 \global\advance\tablecount by 1 \xdef#1{\the\tablecount}}
\def\TABLE#1{\TABNUM#1
   \immediate\write\tablewrite{\noexpand\refitem{#1.}}%
   \begingroup\obeyendofline\rw@start}
\def\Table{\TABNUM\? Table~\?%
\immediate\write\tablewrite{\noexpand\refitem{\?.}}%
\begingroup\obeyendofline\rw@start}
%

%
%
%
\def\masterreset{\global\pagenumber=1 \global\chapternumber=0
   \ifnum\the\equanumber<0\else \global\equanumber=0\fi
   \global\sectionnumber=0
   \global\referencecount=0 \global\figurecount=0 \global\tablecount=0 }
\def\FRONTPAGE{\ifvoid255\else\vfill\penalty-2000\fi
      \masterreset\global\frontpagetrue
      \global\lastf@@t=0 \global\footsymbolcount=0}
\let\Frontpage=\FRONTPAGE
\def\paperstyle{\letterstylefalse\normalspace\papersize}
\def\letterstyle{\letterstyletrue\singlespace\lettersize}
\def\papersize{\hsize=161mm\vsize=240mm\hoffset=0mm\voffset=0mm
               \skip\footins=\bigskipamount}
\def\lettersize{\hsize=161mm\vsize=230mm\hoffset=0mm\voffset=10mm
   \skip\footins=\smallskipamount \multiply\skip\footins by 3 }
\paperstyle   
%
%
\def\MEMO{\letterstyle\FRONTPAGE \letterfrontheadline={\hfil}
    \line{\quad\fourteenrm KLOE MEMORANDUM\hfil\twelverm\the\date\quad}
    \medskip \memod@f}

\def\memit@m#1{\smallskip \hangafter=0 \hangindent=1in
      \Textindent{\caps #1}}
\def\memod@f{\xdef\to{\memit@m{To:}}\xdef\from{\memit@m{From:}}%
     \xdef\topic{\memit@m{Topic:}}\xdef\subject{\memit@m{Subject:}}%
     \xdef\rule{\bigskip\hrule height 1pt\bigskip}}
\memod@f

\font\tenrm=cmr10


%
\newskip\lettertopfil
\lettertopfil = 0pt plus .5in minus 0pt
\newskip\letterbottomfil
\letterbottomfil = 0pt plus 2.3in minus 0pt
\newskip\spskip \setbox0\hbox{\ } \spskip=-1\wd0
\def\addressee#1{\medskip\rightline{\the\date\hskip 30pt} \smallskip
   \vskip\lettertopfil
   \ialign to\hsize{\strut ##\hfil\tabskip 0pt plus \hsize \cr #1\crcr}
   \medskip\noindent\hskip\spskip}
\newskip\signatureskip       \signatureskip=40pt
\def\signed#1{\par \penalty 9000 \bigskip \dt@pfalse
  \everycr={\noalign{\ifdt@p\vskip\signatureskip\global\dt@pfalse\fi}}
  \setbox0=\vbox{\singlespace \halign{\tabskip 0pt \strut ##\hfil\cr
   \noalign{\global\dt@ptrue}#1\crcr}}
  \line{\hskip 0.5\hsize minus 0.5\hsize \box0\hfil} \medskip }

\def\endletter{\ifnum\pagenumber=1 \vskip\letterbottomfil\supereject
\else \vfil\supereject \fi}
\newbox\letterb@x
\def\lettertext{\par\unvcopy\letterb@x\par}
\def\multiletter{\setbox\letterb@x=\vbox\bgroup
      \everypar{\vrule height 1\baselineskip depth 0pt width 0pt }
      \singlespace \topskip=\baselineskip }
\def\letterend{\par\egroup}
%
%
%
\newskip\frontpageskip
\newtoks\pubtype
\newtoks\Pubnum
\newtoks\pubnum
\newif\ifp@bblock  \p@bblocktrue
\def\PH@SR@V{\doubl@true \baselineskip=24.1pt plus 0.2pt minus 0.1pt
             \parskip= 3pt plus 2pt minus 1pt }
\def\PHYSREV{\paperstyle\PhysRevtrue\PH@SR@V}
\def\titlepage{\FRONTPAGE\paperstyle\ifPhysRev\PH@SR@V\fi
   \ifp@bblock\p@bblock\fi}
\def\nopubblock{\p@bblockfalse}
\def\endpage{\vfil\break}
\frontpageskip=1\medskipamount plus .5fil
\pubtype={\tensl Preliminary Version}
\Pubnum={$\caps SLAC - PUB - \the\pubnum $}
\pubnum={0000}
\def\p@bblock{\begingroup \tabskip=\hsize minus \hsize
   \baselineskip=1.5\ht\strutbox \topspace-2\baselineskip
   \halign to\hsize{\strut ##\hfil\tabskip=0pt\crcr
   \the\Pubnum\cr \the\date\cr \the\pubtype\cr}\endgroup}
\def\title#1{\vskip\frontpageskip \titlestyle{#1} \vskip\headskip }
\def\author#1{\vskip\frontpageskip\titlestyle{\twelvecp #1}\nobreak}

\def\address#1{\par\kern 5pt\titlestyle{\twelvepoint\it #1}}
\def\andaddress{\par\kern 5pt \centerline{\sl and} \address}
\def\SLAC{\address{Stanford Linear Accelerator Center\break
      Stanford University, Stanford, California, 94305}}
\def\abstract{\vskip\frontpageskip\centerline{\fourteenrm ABSTRACT}
              \vskip\headskip }
\def\submit#1{\par\nobreak\vfil\nobreak\medskip
   \centerline{Submitted to \sl #1}}
\def\doeack{\foot{Work supported by the Department of Energy,
      contract $\caps DE-AC03-76SF00515$.}}
%
%
%
\def\ie{\hbox{\it i.e.}}

\def\\{\relax\ifmmode\backslash\else$\backslash$\fi}
\def\globaleqnumbers{\relax\ifnum\the\equanumber<0%
\else\global\equanumber=-1\fi}
\def\nextline{\unskip\nobreak\hskip\parfillskip\break}

\def\journal#1&#2(#3){\unskip, \sl #1~\bf #2 \rm (19#3) }
\def\cropen#1{\crcr\noalign{\vskip #1}}
\def\crr{\cropen{10pt}}
\def\topspace{\hrule height 0pt depth 0pt \vskip}

\let\int=\intop         
\def\prop{\mathrel{{\mathchoice{\pr@p\scriptstyle}{\pr@p\scriptstyle}{
                \pr@p\scriptscriptstyle}{\pr@p\scriptscriptstyle} }}}
\def\pr@p#1{\setbox0=\hbox{$\cal #1 \char'103$}
   \hbox{$\cal #1 \char'117$\kern-.4\wd0\box0}}
\def\lsim{\mathrel{\mathpalette\@versim<}}
\def\gsim{\mathrel{\mathpalette\@versim>}}
\def\@versim#1#2{\lower0.2ex\vbox{\baselineskip\z@skip\lineskip\z@skip
  \lineskiplimit\z@\ialign{$\m@th#1\hfil##\hfil$\crcr#2\crcr\sim\crcr}}}
%
%
%
\let\sec@nt=\sec
\def\sec{\relax\ifmmode\let\n@xt=\sec@nt\else\let\n@xt\section\fi\n@xt}
\def\obsolete#1{\message{Macro \string #1 is obsolete.}}
\def\firstsec#1{\obsolete\firstsec \section{#1}}
\def\firstsubsec#1{\obsolete\firstsubsec \subsection{#1}}
\def\thispage#1{\obsolete\thispage \global\pagenumber=#1\frontpagefalse}
\def\thischapter#1{\obsolete\thischapter \global\chapternumber=#1}
\def\nextequation#1{\obsolete\nextequation \global\equanumber=#1
   \ifnum\the\equanumber>0 \global\advance\equanumber by 1 \fi}
\def\BOXITEM{\afterassigment\B@XITEM\setbox0=}
\def\B@XITEM{\par\hangindent\wd0 \noindent\box0 }
%

%
\newdimen\xshift      \xshift=0in
\newdimen\xpos        \xpos=0.5in
\newdimen\yshift      \yshift=0in
\newdimen\ypos        \ypos=0.75in

\def\translate#1#2#3{
   \vbox to 0pt{\offinterlineskip
      \kern-#2\hbox to 0pt{\kern#1{#3}\hss}\vss } }
\def\insertimPRESS#1#2#3{%
    \advance\xshift by -\xpos
    \advance\yshift by -\ypos
      \hbox{%
      \translate{\xshift}{\yshift}{\special{mergeug(#3)}}%
     \blankbox{#1}{#2} }}
\def\blankbox#1#2{\vrule width 0pt depth 0pt height #2
                   \vrule height 0pt depth 0pt width #1}

\def\binsertimPRESS#1{%
    \advance\xshift by -\xpos
    \advance\yshift by -\ypos
      \hbox{%
      \translate{\xshift}{\yshift}{\special{mergeug(#1)}}%
       }}
\def\blankbox#1#2{\vrule width 0pt depth 0pt height #2
                   \vrule height 0pt depth 0pt width #1}

%
\newbox\figbox
\newdimen\zero  \zero=0pt
\newdimen\figmove
\newdimen\figwidth
\newdimen\figheight
\newdimen\figrefheight
\newdimen\textwidth
\newtoks\figtoks
\newcount\figcounta
\newcount\figlines
\def\figreset{\global\figmove=\baselineskip \global\figcounta=0
\global\figlines=1 \global\figtoks={ } }
%
%
%

\def\picture#1by#2:#3{\global\setbox\figbox=\vbox{\vskip #1
\hbox{\vbox{\hsize=#2 \noindent #3}}}
\global\setbox\figbox=\vbox{\kern 5pt
\hbox{\kern 5pt \box\figbox \kern 5pt }\kern 7.5pt}
\global\figwidth=1\wd\figbox
\global\figheight=1\ht\figbox
\global\figrefheight=\figheight
\global\textwidth=\hsize
\global\advance\textwidth by - \figwidth }

\def\figtoksappend{\edef\temp##1{\global\figtoks=%
{\the\figtoks ##1}}\temp}
\def\figparmsa#1{\loop \global\advance\figcounta by 1%
\ifnum \figcounta < #1 \figtoksappend{0pt \the\hsize}%
\global\advance\figlines by 1 \repeat }
\newdimen\figstep
\def\figst@p{\global\figstep = \baselineskip}
\def\figparmsb{\loop \ifdim\figrefheight > 0pt%
\figtoksappend{ \the\figwidth \the\textwidth}%
\global\advance\figrefheight by -\figstep%
\global\advance\figlines by 1%
\repeat }
%
%
%
%
%
\def\figtext#1:#2{\figreset \figst@p%
\figparmsa{#1}%
\figparmsb%
\multiply\figmove by #1%
\global\setbox\figbox=\vbox to 0pt{\vskip\figmove\hbox{\box\figbox}\vss}%
\parshape=\the\figlines\the\figtoks\the\zero\the\hsize%
\noindent\rlap{\box\figbox} #2}
%
%
\catcode`@=12 
\message{ by V.K.}
\everyjob{\input myphyx }

\hsize=160mm\vsize=240mm
\hoffset=0mm\voffset=0mm
\def\ifm#1{\relax\ifmmode#1\else$#1$\fi}
\def\eps{\ifm{\epsilon}} \def\epm{\ifm{e^+e^-}}
\def\rep{\ifm{\Re(\eps'/\eps)}}  \def\imp{\ifm{\Im(\eps'/\eps)}}  
\def\DAF{DA$\Phi$NE}  \def\sig{\ifm{\sigma}}
\def\gam{\ifm{\gamma}} \def\to{\ifm{\rightarrow}}
\def\pip{\ifm{\pi^+}} \def\pim{\ifm{\pi^-}}
\def\po{\ifm{\pi^0}} 
\def\pic{\ifm{\pi^+\pi^-}} \def\pio{\ifm{\pi^0\pi^0}} 
\def\ks{\ifm{K_S}} \def\kl{\ifm{K_L}} \def\kls{\ifm{K_{L,\,S}}} 
\def\ksl{\ifm{K_S,\ K_L}} \def\ko{\ifm{K^0}}
\def\K{\ifm{K}} 
\def\Kb{\ifm{\rlap{\kern.3em\raise1.9ex\hbox to.6em{\hrulefill}} K}}
\def\ab{\ifm{\sim}}  \def\x{\ifm{\times}}
\def\ff{$\phi$--factory}
\def\sta#1{\ifm{|\,#1\,\rangle}} 
\def\amp#1,#2,{\ifm{\langle#1|#2\rangle}}
\def\kob{\ifm{\Kb\vphantom{K}^0}}
\def\f{\ifm{\phi}}   \def\pb{{\bf p}}
\def\L{\ifm{{\cal L}}}  
\def\up#1{$^{#1}$}  \def\dn#1{$_{#1}$}
\def\etal{{\it et al.}}

\def\deg{\ifm{^\circ}}

\def\dt{ \ifm{{\rm d}t} }  
 
\def\kkb{\ifm{\ko\kob}} 

\def\ppc{\ifm{\pi^+\pi^-}}
\def\ppo{\ifm{\pi^0\pi^0}}
\def\pppco{\ifm{\pi^+\pi^-\pi^0}}
\def\pppo{\ifm{\pi^0\pi^0\pi^0}}

\def\figure#1]#2]#3]{\par\vbox to #1cm{\vfill\centerline{
\tenpoint{\bf Fig. #2.} #3} }}

\def\pt#1,#2,{#1\x10\up{#2}}

\input tables
\newcount\subsecno           \subsecno=0
\def\section#1{\par \ifnum\the\lastpenalty=30000\else
   \penalty-200\vskip\sectionskip  \spacecheck\sectionminspace\fi
   \wlog{\string\section\ \chapterlabel \the\sectionnumber}
   \global\advance\sectionnumber by 1  \noindent
   \subsecno=0  
   {\caps\enspace\chapterlabel \sectionlabel #1}\par
   \nobreak\vskip\headskip \penalty 30000 }
\def\subsec#1{\par 
   \global\advance\subsecno by 1  \noindent 
   {\it\enspace\chapterlabel 
    \number\sectionnumber.\number\subsecno \quad #1}\par 
   \nobreak\smallskip } 

\input epsfig.tex

\let\cl=\centerline

 \def\plm{\ifm{\pm}}

\def\undertext#1{$\underline{\hbox{#1}}$}

\def\chapter#1{\par \penalty-300 \vskip12pt
   \spacecheck\chapterminspace
   \chapterreset
\noindent 
   {\bf\chapterlabel \ #1}\par
   \nobreak\vskip8pt \penalty 30000
   \wlog{\string\chapter\ \chapterlabel} }
\def\sectionlabel{\number\sectionnumber.}
\def\section#1{\par \ifnum\the\lastpenalty=30000\else
   \penalty-200\vskip\sectionskip \spacecheck\sectionminspace\fi
   \wlog{\string\section\ \chapterlabel \the\sectionnumber}
   \global\advance\sectionnumber by 1  \noindent
   {\it\chapterlabel \sectionlabel \ #1}\par
   \nobreak\vskip\headskip \penalty 30000 }

\hsize=16.3 cm 
\vsize=22.25 cm

\Frontpage

\vbox to 3.5cm{
\vglue1cm
\line{\hfill\undertext{LNF-97/032 (P)}}
\line{\hfill 2 September 1997}
\vfill}
\font\fortnmitb=cmbxti10 at 14pt

\vglue 4mm
\vbox to 8cm{
{\fourteenpoint\cl{\bf {\fortnmitb CP} VIOLATION in the 
{\fortnmitb K}-SYSTEM}}
\vglue4mm
\baselineskip 15pt
\cl{JULIET LEE-FRANZINI}
\vglue 4mm
\cl{\it Laboratori di Frascati dell'INFN}
\cl{\it CP-13, Via Enrico Fermi 40, I-00044, Frascati Roma}
\cl{\it and}
\cl{\it Physics Department, SUNY at Stony Brook, N.Y. 11794, 
U.S.A.}
\cl{\it and}
\cl{PAOLO FRANZINI}
\cl{Universit\`a di Roma 1, {\it La Sapienza}}
\cl{Roma, Italy}
\cl{\it E-mail:JULIET@LNF.INFN.IT}
\cl{\it E-mail:PAOLO@LNF.INFN.IT}

\vfill}

\baselineskip 13pt
\vglue 6mm
\cl{\bf Abstract}
\parshape 1 12mm 128mm
\noindent
$CP$ violation in the \K\ system is pedagogically reviewed.
We discuss its manifestations in
the neutral \K\ meson systems, in rare \K\ meson decays and in decays of 
charged \K\ mesons. Results from classical experiments, and 
perspectives for 
upcoming experiments are included. We also briefly discuss the 
possibility of $CPT$ tests.
\vglue2cm
\noindent
PACS: 11.30.Er; 13.20.Jf; 29.30.Gx; 29.40.Vj

\vfill

{\it Invited Lectures at the XXVth ITEP Winter School, February 
18-27th, 1997. To be Published in the Proceedings.}

\endpage
\def\C{\ifm{C}}  \def\P{\ifm{P}}  \def\T{\ifm{T}}
\def\noc{\hglue.1pt\rlap{\raise .3mm
\hbox{\kern.5mm\ifm{\backslash}\kern.7mm}}\C}
\def\nop{\hglue.1pt\rlap{\raise .3mm
\hbox{\kern.5mm\ifm{\backslash}\kern.7mm}}\P}
\def\noT{\hglue.1pt\rlap{\raise .3mm
\hbox{\kern.5mm\ifm{\backslash}\kern.7mm}}\T}
\def\nocp{\noc\nop} \def\nocpt{\noc\nop\noT}
\def\kpm{\ifm{K^\pm}}

\twelvepoint
\singlespace
\def\C{$C$}
\def\P{$P$}
\def\S{$S$}
\def\noc{\rlap{$C$}\raise.3mm\hbox{\kern.5mm\\\kern.7mm}}
\def\nop{\rlap{$P$}\raise.3mm\hbox{\kern.5mm\\\kern.7mm}}
\def\noT{\rlap{$T$}\raise.3mm\hbox{\kern.5mm\\\kern.7mm}}
\def\kon{\ifm{K^0_1}} \def\ktw{\ifm{K^0_2}} 
\def\minus{$-$}  \def\dif{\hbox{d}}   \def\Gam{\ifm{\Gamma}}

\def\noT{\rlap{$T$}\raise.3mm\hbox{\kern.5mm\\\kern.7mm}}
\def\nocpt{\noc\nop\noT}
\def\B{\ifm{BR}}
\def\epss{\ifm{\eps_S}}  \def\epsl{\ifm{\eps_L}}
\def\nocp{\noc\nop}
\def\nocpt{\noc\nop\noT} \def\CPT{$CPT$}
\def\B{\ifm{BR}}
\def\prl{Phys. Rev. Lett}
\def\qq{$q\bar q$} \def\qqq{$qqq$}
\FIG\figoscil
\FIG\figregen
\FIG\figfitch
\FIG\eetofi
\FIG\foneftwo
\FIG\fpicpio
\FIG\fellell
\FIG\fellpi
\FIG\boxdia
\FIG\penguin
\FIG\kkdec
\FIG\decdif
\FIG\dmphi
\FIG\epseps
\FIG\epseta
\FIG\etarho
\FIG\nafe
\FIG\ktev
\FIG\klosec
\sequentialequations

\chapter{Introduction}
\section{\K mesons and strangeness}

\K\ mesons were discovered in 1944 in the cosmic radiation\Ref\cosmic{L. 
LePrince-Ringuet and M. Lheritier,``Existence Probable
d'une Particule de Masse 990 {\it me} dans le Rayonnement Cosmique",
Compt. Rend. {\bf 219} (1944) 618.}
and have been responsible for many new ideas in particle physics. 
$K$
mesons led to the the concept of strangeness and flavor mixing in 
the
weak interactions. Parity violation was first observed through the
$\theta$--$\tau$ decay modes of \K\ mesons. All these concepts are 
fundamental parts of the standard model, SM.

The strange properties, of \K\ mesons and certain other particles,
the hyperons, led to the introduction of a new quantum number, 
the strangeness, 
$S$\rlap.\Ref\gmann{M. Gell-Mann,``Isotopic Spin and New Unstable
Particles", Phys. Rev. {\bf 92} (1953) 833.} 
Strangeness is conserved in strong interactions while 
first order weak interaction can induce transitions in which 
strangeness is changed by one unit. Today we describe these 
properties in terms of
quarks with different ``flavors'', first 
suggested in 1964 independently by
Gell-Mann and Zweig\rlap.\Ref\gmzw{M. Gell-Man, 
``A Schematic Model of Baryons and Mesons", Phys. Lett. {\bf 8} (1964) 214; 
G. Zweig,``An $SU_3$ Model for Strong Interaction Symmetry and Its Breaking",
CERN Report 8182/th (1964) 401.}

The ``normal particles'' are bound states of \qq, the mesons, or of \qqq, 
baryons,
where
$$q=\pmatrix{u\cr d\cr}=\pmatrix{\rm up\cr \rm down}.$$
\K's, hyperons and hypernuclei contain a strange quark, $s$:
$$
\eqalign{
&\ko=d\bar s  \cr &\K^+=u\bar s \cr &S=+1\cr}
\quad\eqalign{
&\kob=\bar ds \cr &\K^-=\bar us \cr &S=-1\cr}$$
Assigning  negative strangeness to the strange quark $s$ is 
totally arbitrary but  maintains today the original assignment of 
positive strangeness for \ko, \K\up+ and negative for the 
$\Lambda$ and $\Sigma$ hyperons and for \kob\ and \K\up-. An 
important consequence of the fact that \K\ mesons carry strangeness, a 
new additive quantum number, is that the neutral \K\ and anti 
neutral \K\ meson are distinct particles:
$$C\sta{\ko}=\sta{\kob},\quad S\sta{\ko}=\sta{\ko},\quad 
S\sta{\kob}=-\sta{\kob}$$

An apocryphal story says that upon hearing of this hypothesis, 
Fermi
challenged Gell-Mann to devise an experiment which shows an 
observable
difference between the \ko\ and the \kob. Today we know that it is 
trivial
to do so. For example,  the process
$p\bar p\to\pim K^+\kob$,   produces \kob 's which interact with 
matter
in a totally different manner than the \ko 's produced in  $p\bar 
p\to\pip K^-\ko$.

Since the fifties \K\ mesons have been produced at accelerators, 
first amongst them was the Cosmotron. 

\section{Mass and \C\P\ eigenstates}
While the strong interactions conserve strangeness, the weak 
interactions do
not. In fact, not only do they violate \S\ with $\Delta S=1$, they 
also violate
charge conjugation, \C, and parity, \P, though except for a very 
peculiar
case as we shall see, not the combined \C\P\ symmetry. We assume 
for now that \C\P\ is a symmetry of the world. We define an 
arbitrary, unmeasurable  phase by:
$$CP\sta{\ko}=\sta{\kob}$$
Then the physical mass eigenstates are\rlap:\Ref\PGM{M. Gell-Mann and 
A. Pais,``Behavior of Neutral Particles Under Charge Conjugation",
Phys. Rev. {\bf 97} (1955) 1387. The argument of these authors
were based on C invariance, but remains unchanged assuming CP invariance.}

$$\eqalign{
\sta{K_1}&\equiv{ \sta{ \ko}+\sta{ \kob}\over\sqrt2}\cr
\sta{K_2}&\equiv{ \sta{ \ko}-\sta{ \kob}\over\sqrt2},\cr
}\eqn\eqkokt$$
where \kon\ has \C\P=+1 and \ktw\ has \C\P=\minus1. While \ko\ and 
\kob\ are degenerate states, as required by $CPT$ invariance, the 
weak interactions, which induces to second order 
\ko$\leftrightarrow$\kob\ transitions, induces a small mass 
difference between \kon\ and \ktw. We expect $\Delta 
m\ab\Gamma$, since the decay is a first order process, but we must 
take the square of the appropriate matrix element while the mass 
difference is just the second order matrix element.

\section{\kon\ and \ktw\ lifetimes and mass difference}
\K-mesons have numerous decay modes. For neutral \K's one of the 
principal decay modes is into two or three pions. 
The relevant properties of the 
neutral two and three pion systems with zero total angular 
momentum are given below.
\pointbegin \ppc, \ppo: $P=+1,\ C=+1,\  CP=+1.$
\point  \pppco: $P=-1,\ C=(-1)^l,\ CP=\pm1$, where $l$ is the 
angular momentum of the charged pions in their center of mass. 
States with $l>0$ are suppressed by the angular momentum barrier.
\point \pppo: $P=-1,\ C=+1,\ CP=-1.$ Bose statistics requires 
that $l$ for any pion pair be even in this case.

If the total Hamiltonian conserves $CP$, \ie\ $[H,CP]=0$,
the decays of the $K_1$'s and $K_2$'s must conserve $CP$. Thus
the $K_1$'s with $CP=1$, must decay into two pions (and three 
pions
in an $l=1$ state, surmounting an angular momentum barrier), 
while the $K_2$'s with $CP=-1$, must decay into three pion final 
states. 
Since the energy available in the two pion decay mode is 
approximately 220 MeV, while that for the three pion decay mode is 
only 
about 90 MeV, the lifetime of the $K_1$ is much much shorter than 
that of 
the $K_2$. 

The first verification of the above consideration was obtained as 
early as 1956 by Ledermann \etal\Ref\longk{K. Lande \etal, 
``Observation of a Long Lived Neutral V Particle", Phys. Rev. 
{\bf 103} (1956) 1901.} who 
observed that, in fact, neutral \K\ mesons were still present at times 
much larger than the then accepted value of the neutral \K\ 
lifetime, which was in fact the \kon\ lifetime.
Today we know, ignoring for the moment \C\P\ violation, 

$$\eqalign{
\Gamma_1&=(0.892 \pm 0.002)\x 10^{-10} {\rm\ s}^{-1}\cr
\Gamma_2&=(1.72 \pm 0.02\x)10^{-3}\x\Gamma_1\cr
\Delta m&=m(K_2)- m(K_1)= (0.477\pm 0.003)\x\Gamma_1.\cr
}\eqn\kpara$$

\section{Strangeness oscillations}
The mass eigenstates \kon\ and \ktw\ evolve in vacuum and in their 
rest frame according to
$$i{\dif\over\dif t}\Psi=H\Psi={\cal M}\psi,\eqn\timeev$$
where the complex mass ${\cal M}_{1,2}=m_{1,2}-i\Gamma_{1,2}/2$, 
with $\Gamma=1/\tau$. The state evolution is therefore given by:
$$\sta{K_{1,2},\ t}=\sta{K_{1,2},\ t=0}e^{-i(m_{1,2}-
\Gamma_{1,2}/2)t}$$

If the initial state has definite strangeness, say it is a \ko\ as 
from the production process $\pim p\to \ko\Lambda^0$, it must 
first be rewritten in terms of the mass eigenstates \kon\ and 
\ktw\ which then evolve in time as above. Since the \kon\ and 
\ktw\ amplitudes change differently in time, the pure \S=1 state 
at $t$=0 acquires an \S=\minus1 component at $t>0$.
Using equations \eqkokt\ the wave function at time $t$ is:
$$\eqalign{
\Psi(t)&=\sqrt{1/2}\big[e^{(im_1-\Gamma_1/2)t}\sta{K_1}  
+e^{(im_2-\Gamma_1/2)t}\sta{K_2}\big]\cr
     &=1/2\big[\big(e^{i(m_1-\Gamma_1/2)t}+e^{i(m_2-
\Gamma_2/2)t}\big)\sta{\ko}
     +\big(e^{i(m_1-\Gamma_1/2)t}-e^{i(m_2-
\Gamma_2/2)t}\big)\sta{\kob}\big].\cr
}$$
The intensity of \ko\ (\kob) at time t is given by:
$$I(\ko\ (\kob), t)=|\langle\ko\ 
(\kob)\sta{\Psi(t)}|^2={1\over4}\big[e^{-t\Gam_1}+e^{-t\Gam_2}+(-
)2e^{-t(\Gam_1+\Gam_2)/2}\big]\cos\Delta m\,t$$
which exhibits  oscillations whose frequency depends on the mass 
difference, see fig. \figoscil.
\vglue3mm 
\vbox{\centerline{\epsfig{file=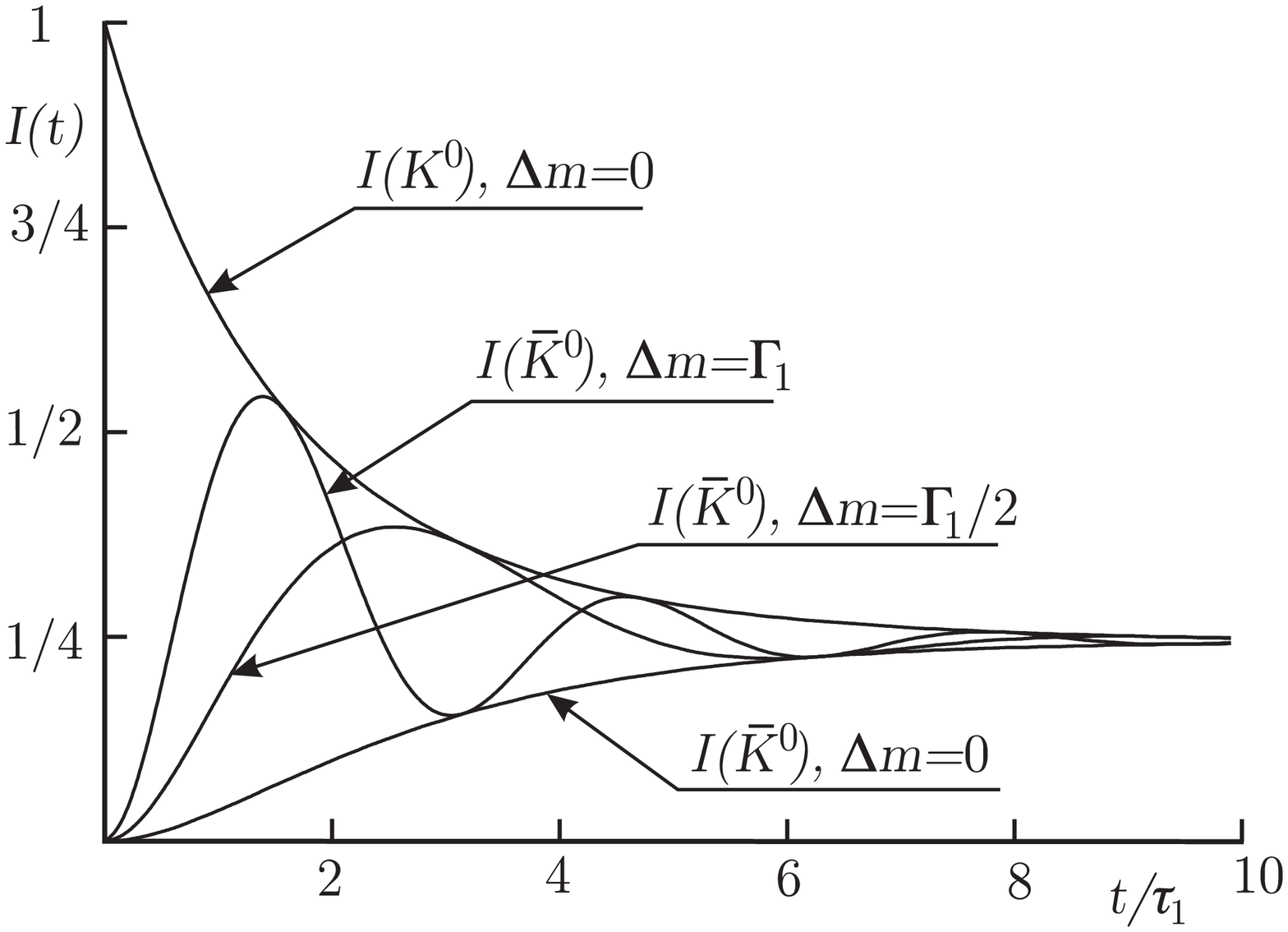,width=13truecm}}
\centerline{\tenpoint{\bf Figure \figoscil.} Evolution in time of  
a pure $S=1$ state at time $t=0$.}}
\endpage
\noindent
The appearance of \kob's from an initially pure
\ko\ beam can be detected by the production of hyperons,
according to the reactions
$$\eqalign{ \kob p &\to \pip\Lambda^0,\quad \to \pip\Sigma^+,\quad 
\to\po\Sigma^+ ,\cr
\kob n &\to \po\Lambda^0,\quad \to \po\Sigma^0,\quad 
\to\pim\Sigma^- .\cr}$$
and the \kl-\ks\ mass difference can be obtained from the 
oscillation frequency.
\section{Regeneration}
Another interesting, and extremely useful phenomenon, is that it 
is 
possible to regenerate $\K_1$'s by placing a piece of material in
the path of a $\K_2$ beam. Let's take our standard reaction,
$$\pim p \to \ko \Lambda^0,$$
the initial state wave function of the \ko's is
$$\Psi(t=0)\equiv\sta{\ko}={\sta{K_1}+\sta{K_2}\over\sqrt 2}.$$
Note that it is composed equally of $\K_1$'s and $\K_2$'s. The
$K_1$ component decays away quickly via the two pion decay modes, 
leaving
a virtually pure $\K_2$ beam. This $\K_2$ beam has equal \ko\ and 
\kob\
components, which interact differently in matter, for example,
the \ko's undergo elastic scattering, charge exchange etc. whereas 
the
\kob's can in addition produce hyperons via strangeness conserving 
transitions. Thus we have 
emerging, from a target material placed in front of the $\K_2$ 
beam,
see fig. \figregen, an apparent rebirth of $\K_1$'s!
\vglue3mm 
\vbox{\centerline{\epsfig{file=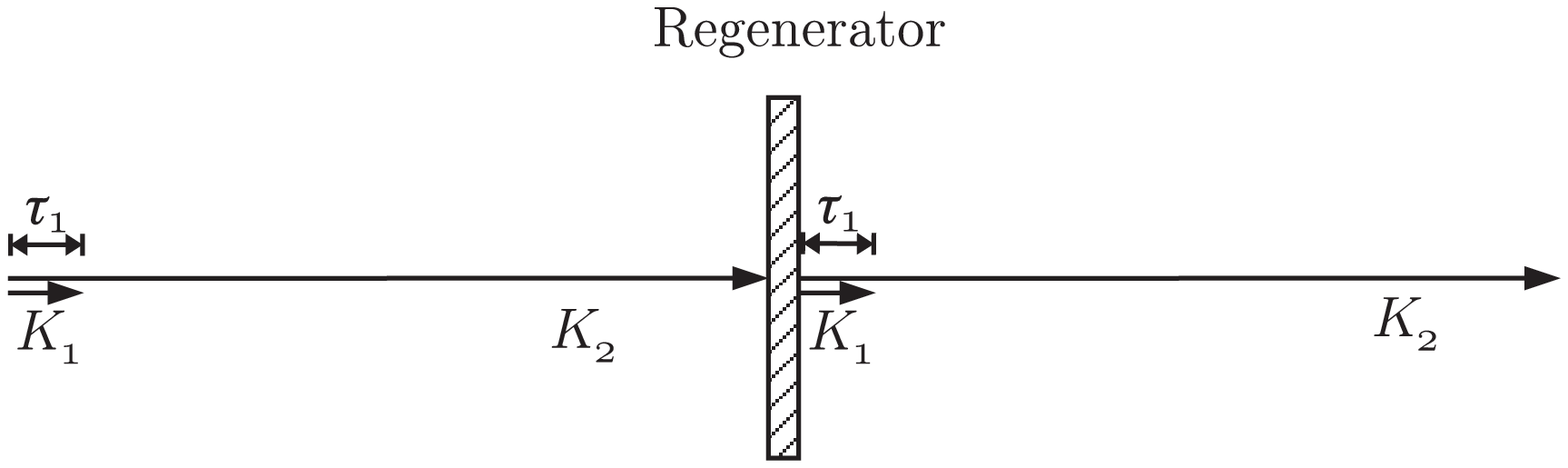,width=14truecm}}  
\centerline{\tenpoint{\bf Figure \figregen.} \kon\ regeneration.}  
}
\vskip3mm
Virtually all past and present
experiments, with the exception of a couple which will be 
mentioned
explicitly, use this method to obtain a source of \kon's (or \ks's, as 
we shall see later). 
Denoting the amplitudes for \ko\ and  \kob\ scattering on nuclei 
by $f$ and $\bar f$ respectively, the scattered amplitude for an 
initial \ktw\ state is given by:
$$\eqalign{
\sqrt{1/2}(f\sta{\ko}-\bar f\sta{\kob})&={f+\bar 
f\over2\sqrt2}(\sta{\ko}-
       \sta{\kob})+{f-\bar f\over2\sqrt2}(\sta{\ko}+\sta{\kob})\cr
       &=1/2(f+\bar f)\sta{K_2}+1/2(f-\bar f)\sta{K_1}.\cr
}$$
The so called regeneration amplitude for \ktw\to\kon, $f_{21}$ is 
given by $1/2(f-\bar f)$ which of course would be 0 if $f=\bar f$.

Another important property of regeneration is that when the \kon\ 
is produced at non-zero angle to the incident \ktw\ beam, 
regeneration on different nuclei in a regenerator is incoherent, 
while at zero degree the amplitudes from different nuclei add up 
coherently.
The intensity for coherent regeneration depends on the \kon, \ktw\ 
mass difference. Precision mass measurements have been performed 
by measuring the ratio of coherent to diffraction regeneration. 
The interference of \kon\ waves from two or more generators has 
also allowed us to determine that the \ktw\ meson is heavier than the 
\kon\ meson. This perhaps could be expected but is nice to have it 
measured.

\kon\ and \ktw\ amplitudes after regeneration are coherent and can 
interfere if $CP$ is violated.

\chapter{CP Violation in Two Pion Decay Modes}
\section{Discovery}
For some years after the discovery that \C\ and \P\ are violated
in the weak interactions, it was thought that \C\P\ might still be
conserved.
$CP$ violation was discovered in '64\rlap,\Ref\fitch{J. H. 
Christenson \etal,''Evidence for the 2$\pi$ Decay of the \ktw\ Meson",
Phys. Rev. Lett. {\bf13} (1964) 138.}
through the observation of the unexpected decay \ktw\to\pic. 
This beautiful experiment is conceptually very simple, see fig. 
\figfitch. 
\vskip3mm
\vbox{\centerline{\epsfig{file=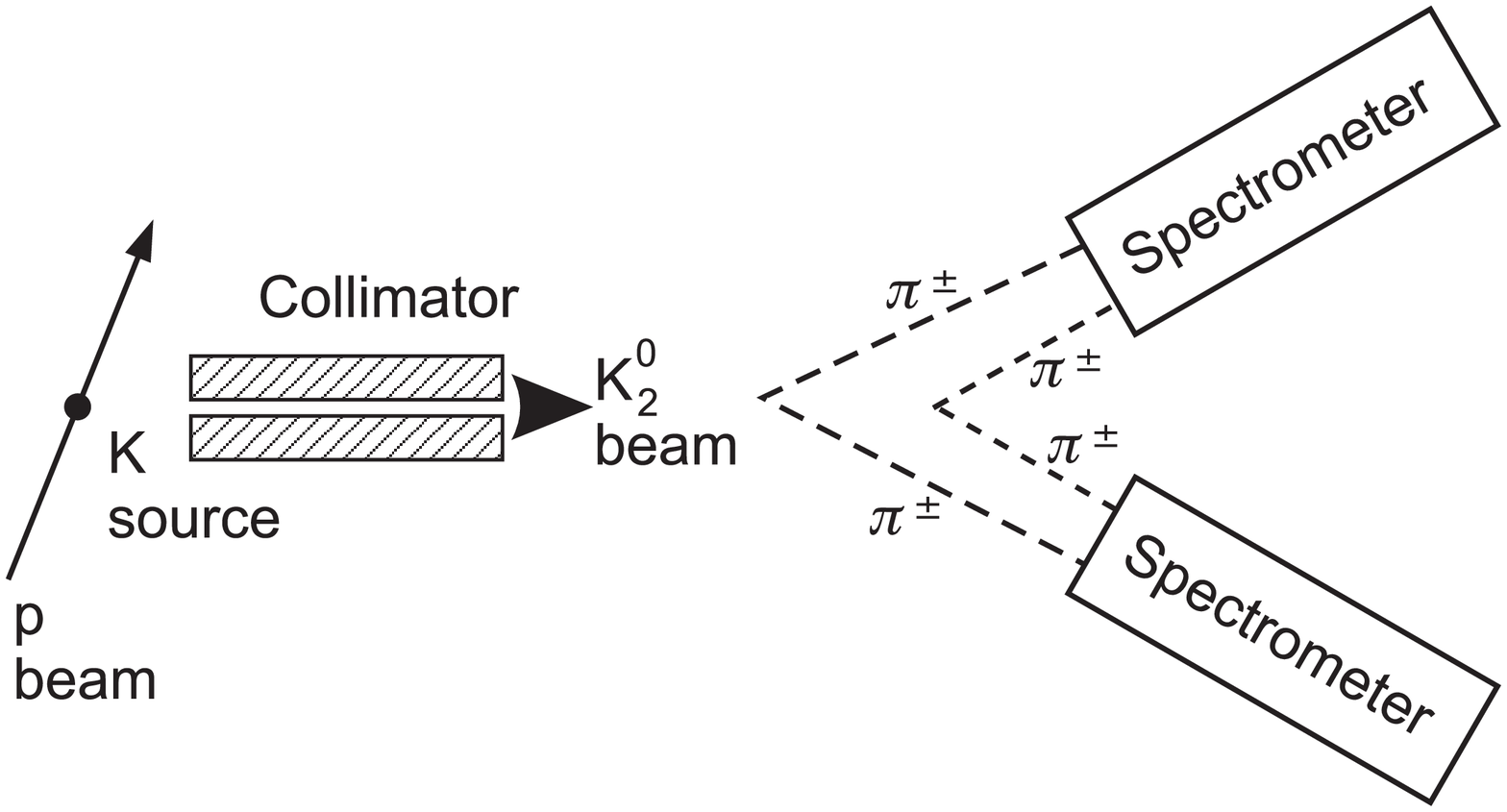,width=16truecm}}
\centerline{\tenpoint{\bf Figure \figfitch.} The setup of the experiment of 
Christenson \etal.} }
\noindent
Let a \ktw\
beam pass through a long collimator and decay in an empty space
(actually a big helium bag) in front of two spectrometers. The 
decay products
are viewed by spark chambers and scintillator hodoscopes
in the spectrometers placed on either side of the beam. Two 
pion decay 
modes are distinguished
from three pion and leptonics decay modes by the reconstructed 
invariant mass
$M_{\pi\pi}$, and the direction $\theta$ of their resultant 
momentum
vector relative to the beam.  In the mass interval between 490 MeV
and 510 MeV, 50 events were found which were exactly collinear 
with
the beam ($\cos\theta > 0.999$), which demonstrated for the first 
time
that \ktw's decayed into two pions, with a branching ratio of the 
order
1/$10^{-3}$, thus $CP$ is shown to be violated!
The $CP$ violating decay \kl\to\ppo\ has also been observed.

\section{Neutral \K\ Decays with \C\P\ Violation}
Since \C\P\ is  violated in \K\ decays,
the mass eigenstates are no more \C\P\ eigenstates and can be 
written, assuming \C\P\T\ invariance, as:
$$\eqalign{
\sta{\ks}=&{\sta{\kon}+\eps\sta{\ktw}\over\sqrt{1+|\eps|^2}}\cr
\sta{\kl}=&{\sta{\ktw}+\eps\sta{\kon}\over\sqrt{1+|\eps|^2}}\cr 
}\eqn\eqkskl$$
with $|\eps|=(2.259\pm0.018)\x10^{-3}$ from experiment. Note that 
the \ks\ and \kl\ states are not orthogonal states, contrary to the case of 
\kon\ and \ktw.
Equation \timeev\ can be rewritten, to lowest order, as:
$${\rm d\over\dt}\sta{K_{S,L}}=-i{\cal M}_{S,L}\sta{K_{S,L}},\quad{\cal 
M}_{S,L}=M_{S,L}-
i\Gamma_{S,L}/2 $$
and the values of masses and decay widths given in eq. \kpara\ belong to 
\ks\ and \kl\ rather than to \kon\ and \ktw.

Since 1964 we have been left with an
unresolved problem: is \C\P\ violated directly in \ko\ decays,
\ie\ is the $|\Delta S|$=1 amplitude $\langle\pi\pi\sta{K_2}\ne0$ 
or
the only manifestation of \noc\nop\ is to introduce a small
impurity of $K_1$ in the \kl\ state, via \ko$\leftrightarrow$\kob,
$|\Delta S|$=2 transitions?
With the standard definitions, using the phase choice of Wu and 
Yang\rlap,\Ref\etaref{T. T. Wu and C. N. Yang,
``Phenominological Analysis of Violation of $CP$ Invariance on
Decay of \ko\ and \kob", \prl. {\bf 13} (1964) 380.}
the two pion decay amplitude ratios $\eta$'s can be written as
$$\eqalign{
{\langle\pi^+\pi^-\sta{\kl}\over\langle\pi^+\pi^-\sta{\ks}}=&
\eta_{+-}=\eps-2\eps'\cr
{\langle\pi^0\pi^0\sta{\kl}\over\langle\pi^0\pi^0\sta{\ks}}=&
\eta_{00}=\eps+\eps'\cr},$$
where $\eps$ is defined above and $\eps'$ is essentially
$${A(\ktw\to\pi\pi)\over A(\kon\to\pi\pi)}.$$
The question above is then the same as: is $\eps'\ne0$?

Since 1964, experiments searching for a difference in $\eta_{+-}$ and
$\eta_{00}$ have been going on. If $\eta_{+-} \ne\eta_{00}$ the
ratios of branching ratios for \kls\to\pic\ and \pio\ are different. 
Most experiments
measure the quantity ${\cal R}$, the so called double ratio of the four
rates for \kls\to\pio,\pic, which is related, to lowest order in
\eps\ and $\eps'$, to \eps\ and $\eps'$ by 
$${\cal R}\equiv{\Gamma(\kl\to\pio)/\Gamma(\ks\to\pio)\over
\Gamma(\kl\to\pic)/\Gamma(\ks\to\pic)}\equiv
\Big|{\eta_{00}\over\eta_{+-}}\Big|^2=1-6\rep.$$ 
Observation of ${\cal R}\ne0$ is proof that \rep$\ne$0 and therefore of 
``direct'' $CP$ violation, \ie\
that the amplitude for $|\Delta S|$=1, $CP$ violating transitions
$A(K_2\to2\pi)\ne0$. 

All present observations of \C\P\ violation, \nocp\ for short, 
\ie\ the decays
\kl\to2$\pi$, \pic\gam\ and the charge asymmetries in $K_{\ell 3}$ 
decays are
examples of so called ``indirect'' violation, due to $|\Delta 
S|$=2
\ko$\leftrightarrow$\kob\ transitions introducing a small $CP$ 
impurity in the
mass eigenstates \ks\ and \kl.
Because of the smallness of $\eps'$, most formula given above for 
\kon\ and \ktw\ remain valid substituting \kon\to\ks\ and 
\ktw\to\kl.

There is no new information on direct \nocp\
from the last round of precision experiments. 
One of the two, NA31, was 
performed at CERN and reported a tantalizing positive 
result.\Ref\epscern{G. D. Barr \etal, 
``A New Measurement of Direct $CP$ Violation in the neutral Kaon System",
Phys. Lett. {\bf 317} (1993) 233.} NA31 alternated \ks\ and \kl\
data taking by the insertion of a \ks\ target every other run,
while the experimental apparatus collected both
charged and neutral two pion decay modes simultaneously.
The other experiment, E731, was done at Fermilab and reported
an  essentially null result.\Ref\epsfnal{L. K. Gibbons \etal, 
``Measurement of the $CP$ Violation Parameter \rep",
\prl. {\bf 70} (1993) 1203.} E731 had a fixed
thick \ks\ target in front of one of the two parallel \kl\
beams which entered the detector which, however, collected
alternately the neutral and charged two pion decay modes.

Therefore, at present we are confronted with the following experimental 
situation:$$\eqalign{
\rep=&\hbox{\pt(23$\pm$6.5),-4,}\cr
\rep=&\hbox{\pt(7.4$\pm$5.9),-4,}\cr
}$$
Taking the Particle Data Group's\Ref\pdg{Particle Data Group,
``Review of Particle Properties", 
Phys. Rev. D {\bf 50} (1994) 1173.} average at face value, 
we could say that the confidence level that $0<\rep<\hbox{\pt3,-
3,}$ is 94\%.
We will come back to what is being done to overcome this problem.

\chapter{$CP$ violation in other modes}

\section{Semileptonic decays}

$K$-mesons also decay semileptonically, into a hadron, with charge $Q$
and strangeness zero, and a pair of lepton-neutrino. These decays at
quark levels  are due to the elementary processes
$$\eqalign{
s&\to W^- u\to\ell^-\bar\nu u\cr
\bar s&\to W^+\bar u\to\ell^+ \nu \bar u\cr
}$$
which for the physical \K-mesons correspond to the decays
$$\eqalign{
\ko\to &\pim \ell^+ \nu,\ \Delta S=-1,\ \Delta Q=-1\cr
\kob\to &\pip \ell^- \bar\nu,\ \Delta S=+1,\ \Delta Q=+1\cr
\kob\to &\pim \ell^+ \nu,\ \Delta S=+1,\ \Delta Q=-1\cr
\ko\to &\pip \ell^- \bar\nu,\ \Delta S=-1,\ \Delta Q=+1\cr
}$$
Therefore \ko\ should decay only to $\ell^+$ and \kob\ to $\ell^-$.
This is commonly referred to as the $\Delta S=\Delta Q$ rule, 
experimentally established in the very early days of strange particle 
studies.

The leptonic asymmetry
$${\cal A}_\ell={\ell^- -\ell^+\over\ell^- +\ell^+}$$
in \kl\ decays should therefore be 2$\Re\eps$\ab$\sqrt2\eps$. 
The measured value of ${\cal A}_\ell$ for \kl\ decays
is (0.327$\pm$0.012)\%, in good agreement with the above
expectation, a proof that \C\P\ violation is in the mass term.

In a situation where the neutral $K$-mesons are
produced in a strangeness tagged state as in
$$p+\bar p\to K+K^\pm +\pi^\mp$$
the charge of the charged kaon (pion) defines the strangeness
of the neutral $K$.
In the leptonic kaon decay, assuming the $\Delta S=\Delta Q$ rule, the
lepton charge defines the strangeness of the neutral $K$.

\section{$CP$ violation in \ks\ decays}

So far $CP$ violation has only been seen in $K_L$ decays
($K_L \rightarrow \pi\pi$ and semileptonic decays).
At a \ff\ such as \DAF, where ${\cal O}$(10\up{10}) \ks/y are produced, one
can look for $K_S \rightarrow \pi^0\pi^0\pi^0$, the
counterpart to $K_L\rightarrow \pi\pi$.  The branching ratio 
for this process is proportional to $\epsilon + \epsilon'_{000}$
where $\epsilon'_{000}$ is a quantity similar to $\epsilon'$,
signalling direct $CP$ violation.  It is not as suppressed as the
normal $\epsilon'$, perhaps a factor of twenty less.
Nonetheless, as the expected \B\ is $2\times 10^{-9}$, the whole
signal will be at the 30 event level, and therefore there is here
only the possibility to see the $CP$ impurity of \ks, never observed 
before, not direct $CP$ violation.  The current limit on this 
\B\ is $3.7\times 10^{-5}$. Finally the leptonic asymmetry ${\cal A}_
\ell$(\ks) in \ks\ decays has never been measured. The expected value
is \pt3.2,-4, and at \DAF\ it can be measured to an accuracy of 
\ab\pt2.5,-4,.
Again this would be only a measurement of $\epsilon$, not $\epsilon'$,
but the observation for the first time of $CP$ violation in two
new channels of \ks\ decay would be nonetheless of considerable interest.

\section{$CP$ violation in charged $K$ decays}

Evidence for direct $CP$ violation can be also be obtained from the
decays of charged \K\ mesons. 
CP invariance requires equality of the partial rates for
$K^{\pm}\to\pi^{\pm}\pi^+\pi^-$ ($\tau^{\pm}$) and for
$K^{\pm}\to\pi^{\pm}\pi^0\pi^0$ ($\tau'^{\pm}$). With the
luminosities obtainable at \DAF\
one can improve the present rate asymmetry measurements by two orders of 
magnitude. 
There, one could also observe differences in the
Dalitz plot distributions for $K^+$ and $K^-$ decays in both the $\tau$
and $\tau'$ modes and 
reach sensitivities of \ab10$^{-4}$. 
Finally, differences in rates in the radiative two pion decays of
$K^\pm$, \kpm\to $\pi^\pm\po\gam$, are also proof of direct $CP$
violation. 

\chapter{$CP$ violation at a \ff}

\section{\epm\to$\phi$, $\phi\to\kkb$}
\noindent
The cross section for production of a bound $q\bar q$ pair of mass 
$M$ and total width $\Gamma$ with $J^{PC}=1^{--}$ \ie\ a so vector 
meson $V$ in \epm\ 
annihilation (see fig. \eetofi) is given by:
$$\sig_{q\bar q, {\rm res}}={12\pi\over 
s}{\Gamma_{ee}\Gamma{M^2}\over(M^2-s)^2+M^2\Gamma^2}
={12\pi\over s}B_{ee}B_{q\bar q} {M^2\Gamma^2\over(M^2-
s)^2+M^2\Gamma^2}$$
\vglue3mm 
\vbox{\centerline{\epsfig{file=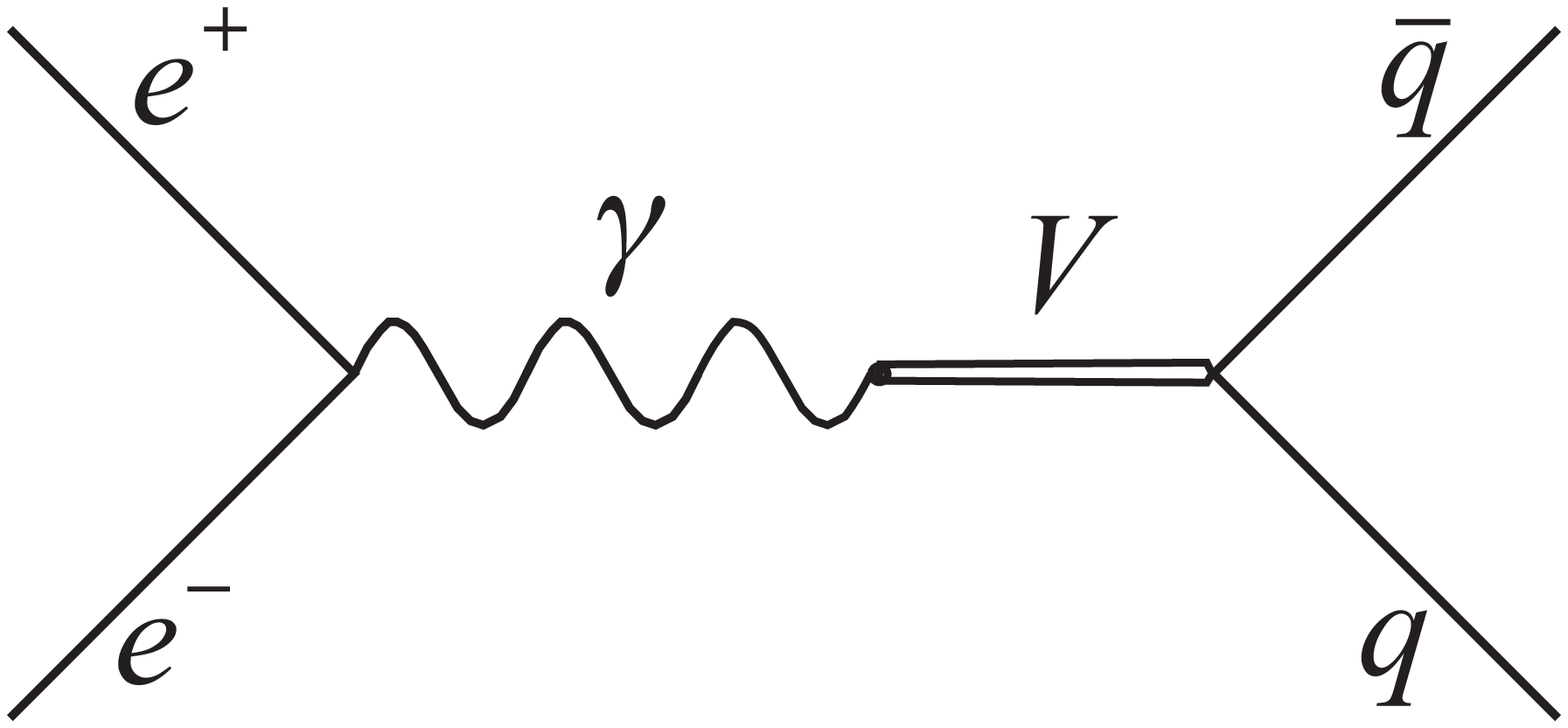,width=5truecm}}
\vglue2mm
\centerline{\tenpoint{\bf Figure \eetofi.} Amplitude for 
production of a bound $q\bar q$ pair.}}
\vskip3mm
\noindent
The $\phi$ meson is a $s\bar s$ \up3S\dn1 bound state with 
$J^{PC}$=1\up{--} same as a photon. The cross section for its 
production in \epm\ annihilations at 1020 MeV is
$$\sig_{s\bar s}(s=(1.02)^2\hbox{ GeV\up2})\ab{12\pi\over 
s}B_{ee}= 36.2\x(1.37/4430)=0.011\hbox{ GeV\up{-2}}\ab4000\hbox{ 
nb}$$
The Frascati \ff, \DAF, will have a luminosity ${\cal  
L}=10^{33}\hbox{ cm\up{-2} s\up{-1}}=1\hbox{ nb\up{-1}s\up{-1}}$. 
Integrating over one year, taken as 10\up7 s or one third of a 
calendar year, we find 
$$ \int_{\rm 1\ y}{\cal  L}\dt=10^7\hbox{ nb\up{-1}},$$
corresponding to the production at \DAF\ of
\ab$4000\x10^7=4\x10^{10}$ $\phi$ mesons per year or approximately 
\pt1.3,10, \ko, \kob\ pairs, a large number indeed.
One of the advantages of studying $K$ mesons at a \ff\ is 
that
they are produced in a well defined quantum and kinematical state. 
Neutral \K\
mesons are produced in collinear pairs, with a momentum of about 
110
MeV/c, thus detection of one $K$ gives the direction of the other.
In addition in the reaction 
$\epm\to\hbox{``\gam''}\to\phi\to\kkb$,
$C(\kkb)=C(\phi)=C(\gam)=-1.$
Let \sta{i}=\sta{K\Kb,\ t=0,\ C=-1}, then:
$$\sta{i}={\sta{\ko,\pb}\sta{\kob,-\pb}-\sta{\kob,\pb}\sta{\ko,-
\pb}
\over\sqrt2}$$
From eq. \eqkskl\ the relations between \ks, \kl\ and \ko, \kob, 
to lowest order in \eps, are:
$$\sta{\ks\ (\kl)}={(1+\eps)\sta{\ko}+(-)(1-
\eps)\sta{\kob}\over\sqrt2},\qquad
\sta{\ko\ (\kob)}={\sta{\ks}+(-)\sta{\kl}\over(1+(-)\eps)\sqrt2}$$
from which
$$\sta{i}={1\over\sqrt2}\,
\big(\sta{\ks,-\pb}\sta{\kl,\pb}-\sta{\ks,\pb}\sta{\kl,-
\pb}\big)$$
so that the neutral kaon pair produced in \epm\ annihilations is a 
pure \ko, \kob\ as well as a pure \ks, \kl\ for {\it all times,} 
in vacuum. This is valid to all orders in \eps\ and also for 
\nocpt.
\section{Correlations in \ks, \kl\ decays}
\noindent
To obtain the amplitude for decay of $K(\pb)$ into a final state 
$f_1$
at time $t_1$ and of $K(-\pb)$ to $f_2$ at time $t_2$, see the 
diagram in fig. \foneftwo, we time evolve the initial state in 
our usual way:
\setbox2=\vbox to1.5cm{\vfil\hsize=15truecm
\noindent\strut
\hbox{\rlap{\kern2.1truecm\rlap{$\bullet$}\kern
4truecm\rlap{$\bullet$}\kern 7truecm\rlap{$\bullet$}}\rlap{\raise
1truemm\hbox{\kern2.17truecm\vrule width11truecm
height.7truept depth.7truept}}\rlap{\raise 3truemm\hbox{\kern
4truecm\rlap{$t_1$}\kern5.5truecm\rlap{$t_2$}}}\rlap{\lower
6truemm\hbox{\kern3truecm\rlap{\ks, \kl}\kern5.5truecm\rlap{\kl,
\ks}}}\rlap{\kern1.2truecm\rlap{$f_1$}\kern
12.5truecm$ f_2$}\rlap{\raise 5truemm\hbox{\kern
6truecm$\f$}}}
\strut\vskip.4truecm\vfil
}
$$\eqalign{&
\sta{t_1,\ \pb;\ t_2,\ -\pb}
={1+|\eps^2|\over(1-\eps^2)\sqrt2}\times\cr
&\qquad\Big(\sta{\ks(-\pb)}\sta{\kl(\pb)}
e^{-i({\cal M}_St_2+{\cal M}_Lt_1)}\ -
\sta{\ks(\pb)}\sta{\kl(-\pb)}
e^{-i({\cal M}_St_1+{\cal M}_Lt_2)}\Big)\cr
}$$
$$\box2$$
\centerline{\tenpoint{\bf Fig. \foneftwo.} \f\to\kl, \ks\to$f_1$ ,$f_2$}
\vglue3mm
\noindent
where ${\cal M}_{S,L}=M_{S,L}-i\Gamma_{S,L}/2$ are the complex 
\ks,
\kl\ masses.
In terms of the previously mentioned ratios
$\eta_i=\langle\,f_i\sta{\kl}/\langle\,f_i\sta{\ks}$ and defining
$\Delta t=t_2-t_1$, $t=t_1+t_2$,
$\Delta{\cal M=\cal M}_L-{\cal M}_S$ and ${\cal M=M}_L+{\cal
M}_S$
we get the amplitude for decay to states 1 and 2:
$$
A(f_1, f_2, t_1, t_2)=
\langle\,f_1\sta{\ks}\langle\,f_2\sta{\ks}e^{-i{\cal M}t/2}
\Big(\eta_1e^{i\Delta {\cal M}\Delta t/2}-
\eta_2e^{-i\Delta {\cal M}\Delta t/2}\Big)/\sqrt2.\eqn\eqfive$$
This implies
$A(\epm\to\phi\to\kkb\to f_1f_2)=0$
for $t_1=t_2$ and $f_1=f_2$ (Bose statistics).
For $t_1=t_2$, $f_1=\pi^+\pi^-$ and $f_2=\pi^0\pi^0$ instead,
$A\propto\eta_{+-}-\eta_{00}=3\times\epsilon'$ which suggests a
(unrealistic) way to measure $\eps'$.
The intensity for decay to final states  $f_1$ and $f_2$ at times 
$t_1$ and $t_2$ obtained taking the modulus squared of eq. 
\eqfive\ depends on magnitudes and arguments of  $\eta_1$ and 
$\eta_2$ as well as on $\Gam_{L,S}$ and $\Delta M$. The intensity is 
given by
$$\eqalign{I(f_1, f_2, t_1, 
t_2)&=|\langle\,f_1\sta{\ks}|^2|\langle\,f_2\sta{\ks}|^2e^{-
\Gam_S\,t/2}\x\cr
&\big(|\eta_1|^2e^{\Gam_S\Delta t/2}+|\eta_2|^2e^{-\Gam_S\Delta 
t/2}-
2|\eta_1||\eta_2|\cos(\Delta m\, t+\phi_1-\phi_2)\big)\cr}$$
where we have everywhere neglected \Gam\dn L with respect to 
\Gam\dn S.
Thus the study of the decay of \K\ pairs at a \ff\ offers the 
unique possibility of observing interference patterns in time, or 
space, in the intensity observed at two different points in space. 
This fact is the source of endless excitement and frustration to some 
people.
Rather than studying the intensity above, which is a function of 
two times or distances, it is more convenient to consider the once 
integrated distribution. In particular one can integrate the 
intensity over all times $t_1$ and $t_2$ for fixed time difference 
$\Delta t=t_1-t_2$, to obtain the intensity as a function of 
$\Delta t$.
Performing the integrations yields, for $\Delta t>0$,
$$\eqalign{
I(f_1,\ f_2;\ \Delta t)=&
{1\over2\Gamma}|\langle f_1\sta{\ks}\langle f_2\sta{\ks}|^2\x\cr
&\Big(|\eta_1|^2e^{-\Gamma_L\Delta t}+
|\eta_2|^2e^{-\Gamma_S\Delta t}-
2|\eta_1||\eta_2|e^{-\Gamma\Delta t/2}\cos(\Delta m\Delta
t+\f_1-\f_2)\Big)\cr
}$$
and a similar expression is obtained for $\Delta t<0$.

The interference pattern is quite different according to the choice of $f_1$ 
and $f_2$ as illustrated in figs. \fpicpio-\fellpi.
\vglue3mm 
\vbox{
\centerline{\epsfig{file=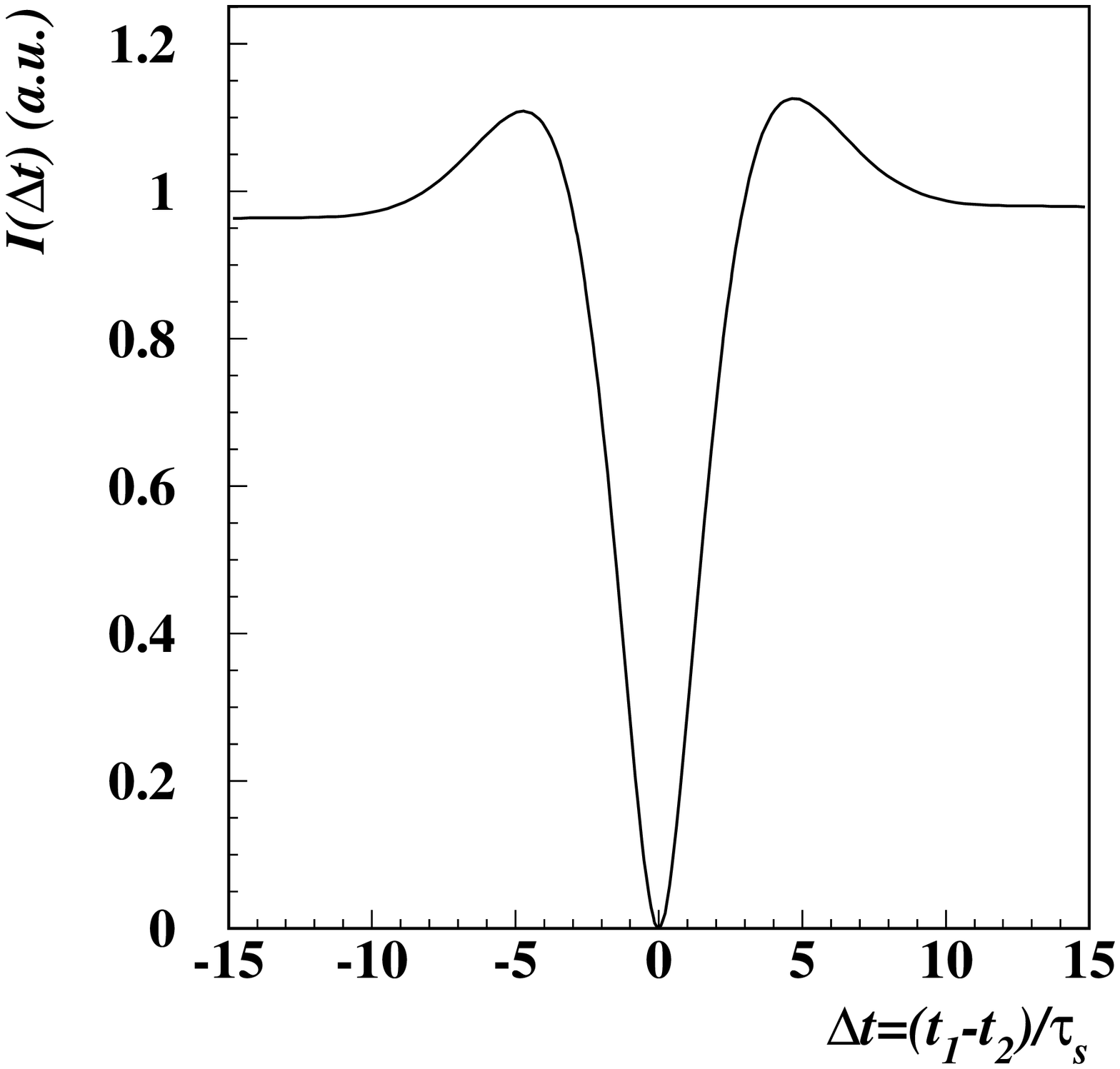,width=8.1cm} 
\hglue1mm\epsfig{file=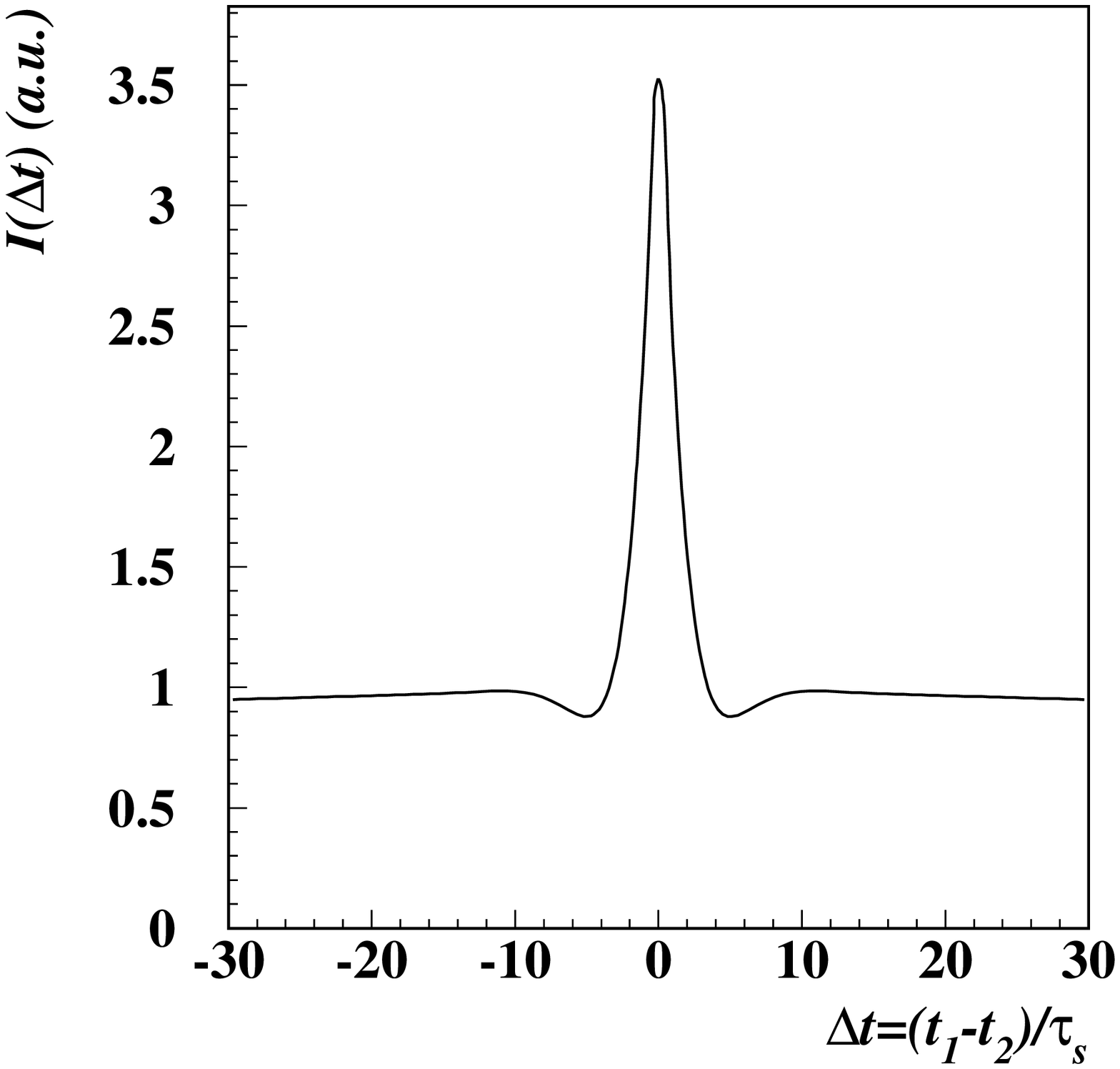,width=8.1cm}}
\centerline{\tenpoint\hbox to7cm{{\bf Fig. \fpicpio.} Interference 
for $f_1$=\pic,$f_2$=\pio.}
\hglue1cm \hbox to7cm{{\bf Fig. \fellell.} Interference for 
$f_1=\ell^-,f_2=\ell^+$.}}
}
\vglue2mm
One can thus perform a whole spectrum of
precision ``kaon-interferometry'' experiments at \DAF\ by 
measuring the above decay intensity distributions 
for appropriate choices of the final states $f_1,\ f_2$.
Four examples are listed below.
\pointbegin With $f_1$=$f_2$ one measures $\Gamma_S$, $\Gamma_L$ 
and 
$\Delta m$, since all phases cancel. Rates can be measured with a 
\x10
improvement in accuracy and $\Delta m$ to \ab\x 2.
\point With $f_1$=\pic,\ $f_2$=\pio, one measures \rep\
at large time differences, and \imp\
for $|\Delta t|\le 5 \tau_s$. Fig. \fpicpio\ shows the
interference pattern for this case. The strong destructive 
interference at zero time difference is due to the antisymmetry of 
the initial \K\K\ state, decay amplitude phases being identical. 
\point With $f_1=\pi^+\ell^-\nu$ and $f_2=\pi^- \ell^+\nu$, one
can measure the $CPT$--violation parameter $\delta$, see our 
discussion later concerning tests of \C\P\T. Again the real part 
of $\delta$ is measured 
at large time differences and the
imaginary part for $|\Delta t|\le10\tau_s$. Fig. \fellell\
shows the 
interference pattern. The destructive interference at zero time 
difference becomes positive since the amplitude for \ko\to$\ell^+$ 
has opposite sign to that for \kob\to$\ell^-$ thus making the 
overall amplitude symmetric.
\point For $f_1=2\pi$, $f_2=\pi^+\ell^-\nu$ or $\pi^- \ell^+\nu$,
small time differences yield $\Delta m$, $|\eta_{\pi\pi}|$ 
and $\phi_{\pi\pi}$, while
at large time differences, the asymmetry
in \kl\ semileptonic decays provides tests
of $T$ and $CPT$. The {\it vacuum regeneration}
interference is shown in fig. \fellpi.

\centerline{\epsfig{file=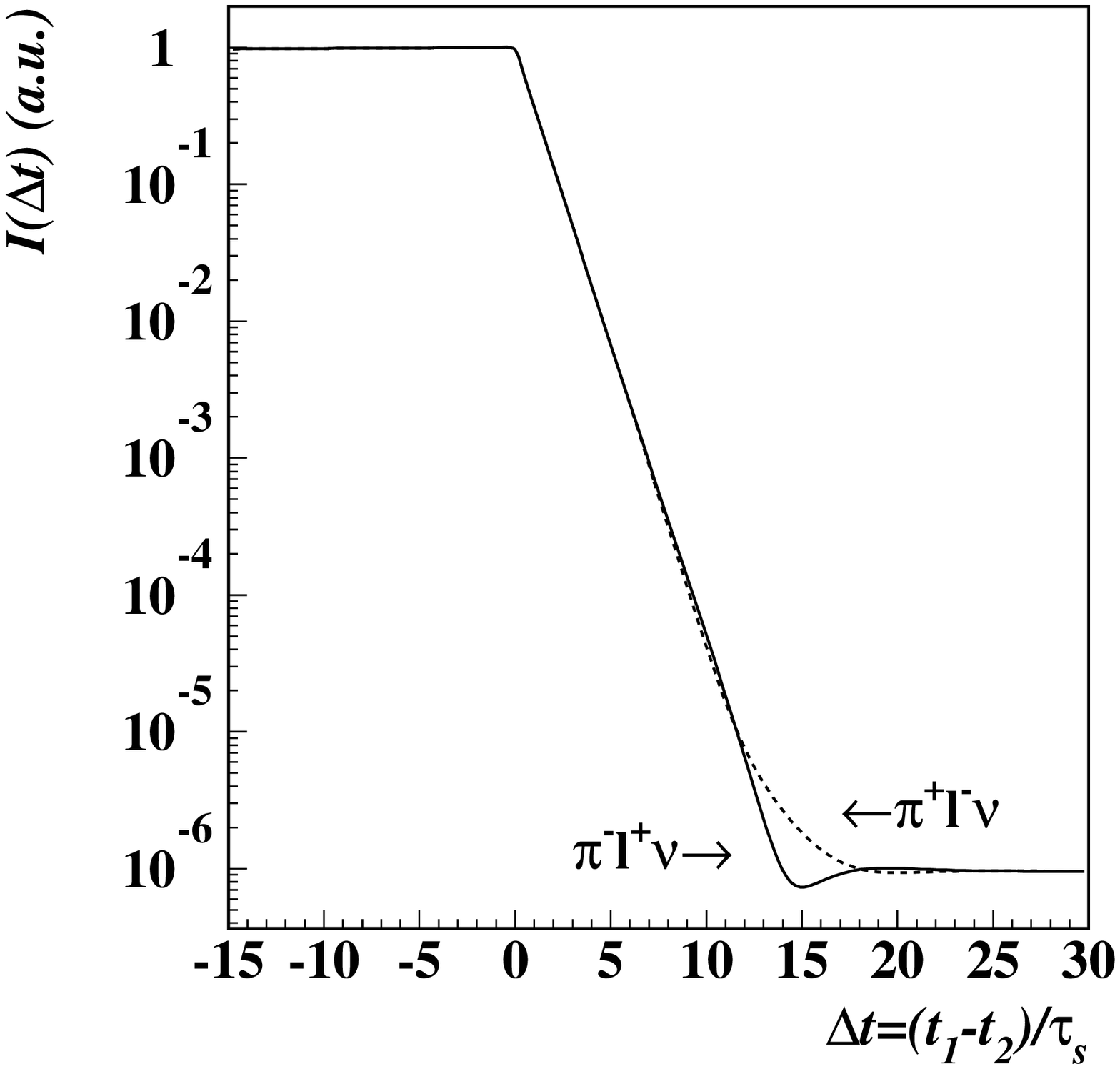,width=9cm}}
\centerline{{\bf Fig. \fellpi.}
Interference pattern for $f_1=2\pi,f_2=\ell^{\pm}$.}
\vskip2mm

\chapter{\C\P\ violation in the standard model}

\noindent
The Standard Model has a {\it natural} place for \C\P\ violation
(Cabibbo, Kobayashi-Maskawa, Maiani).
A phase can be introduced in the unitary matrix {\bf V} which 
mixes the quarks 
$$\pmatrix{d'\cr s'\cr b'\cr}=
  \pmatrix{V_{ud}&V_{us}&V_{ub}\cr
           V_{cd}&V_{cs}&V_{cb}\cr
           V_{td}&V_{ts}&V_{tb}\cr}
  \pmatrix{d\cr s\cr b\cr}$$
but the theory does not predict the magnitude of the effect. The 
constraint that the mixing matrix be unitary corresponds to the 
desire of having a universal weak interaction.
Our present knowledge of the magnitude of the $V_{ij}$ elements is 
given below.
$$\pmatrix{0.9745-0.9757&0.219-0.224    &0.002-0.005\cr
                   0.218-0.224    &0.9736-0.9750&0.036-0.047\cr
                   0.004-0.014    &0.034-0.046    &0.9989-
.9993\cr}$$
The diagonal elements are close 
but definitely not equal to unity. If 
such were the case there could be no \C\P\ violation.

However, if the violation of
\C\P\ which results in $\eps\ne0$ is explained in this way then, 
in 
general, we expect
$\eps'\ne0$. For {\it technical} reasons, it is difficult to 
compute
the value of $\eps'$. Predictions are $\eps'/\eps\le10^{-3}$, but
cancellations can occur, depending on the value of the top mass 
and the 
values of appropriate matrix elements, mostly connected with 
understanding the light hadron structure. 

A fundamental task of experimental physics today is the 
determination of the four parameters of the CKM mixing matrix, 
including the phase which results in \nocp. A knowledge of all 
parameters is required to confront experiments. Rather, many 
experiments 
are necessary to complete our knowledge of the parameters and 
prove 
the uniqueness of the model or maybe finally break beyond it.
As it happens rare \K\ decays can be crucial to this task.
We will therefore discuss the following topics:
recent measurements of \ks, \kl\ parameters and searches for 
symmetry violations; new rare \K\ decay results;
other searches for \nocp\ and \noT as well as present limits on 
\nocpt. 
We will also briefly describe
perspectives for developments in the near future.

\def\lam{\ifm{\lambda}}
To this end it is convenient to parameterize the mixing matrix 
above in a way which reflects more immediately our present 
knowledge of the value of some of the elements and has the \C\P\ 
violating phase appearing in only two off-diagonal elements. The 
Wolfenstein\Ref\wolf{L. Wolfenstein, 
``Present Status of $CP$ Violation",
Ann. Rev. Nucl. Part. Sci., {\bf 36
} (1986) 137.} {\it approximate} parameterization of 
the mixing matrix 
expanded up to \lam\up3 is
$$ \pmatrix{V_{ud}&V_{us}&V_{ub}\cr
                    V_{cd}&V_{cs}&V_{cb}\cr
                    V_{td}&V_{ts}&V_{tb}\cr}
=\pmatrix{ 1-{\lam^2\over2}    &\lam                     
&A\lam^3(\rho-i\eta)\cr
                 -\lam                          &1-
{\lam^2\over2}&A\lam^2\cr
                 A\lam^3(1-\rho-i\eta) &-A\lam^2             &   1 
\cr}.$$
\lam=0.2215$\pm$0.0015 is the Cabibbo angle, a real number 
describing mixing of $s$ and $d$ quarks. $A$, also real, is close 
to one $A$\ab0.84$\pm$0.06 and $|\rho-i\eta|$\ab0.3. 
\C\P\ violation occurs only if $\eta$ does not vanish. $\eta$ and $\rho$ 
are not really known.
Several constraints on $\eta$ and $\rho$ can however be obtained from 
the values of measured parameters. The value of \eps\ can be calculated 
from the $\Delta S$=2 amplitude of fig. \boxdia, the so called box 
diagram. At the quark level the calculations is straightforward, but 
complications arise in estimating the correct matrix element between 
\ko\ and \kob\ states. Apart from the uncertainties in these estimates 
\eps\ depends on $\eta$ and $\rho$ as:
$$|\eps|=a\eta+b\eta\rho$$ 
which is a hyperbola in the $\eta,\rho$ plane 
whose central value is shown in figure \etarho.
The calculation of $\eps'$ is more complicated. There are three 
$\Delta S$=1 amplitudes that contribute to \K\to$\pi\pi$ decays:
$$A(s\to u\bar u d)\propto U_{us}U^*_{ud}\ab\lambda\eqn\equud$$
$$A(s\to c\bar c d)\propto U_{cs}U^*_{cd}\ab-\lambda+i\eta A^2\lambda^5
\eqn\eqccd$$
$$A(s\to t\bar t d)\propto U_{ts}U^*_{td}\ab-A^2\lambda^5(1-\rho+i\eta)
\eqn\eqttd$$
where the amplitude \equud\ correspond to the natural way for computing \K\to
$\pi\pi$ in the standard model and the amplitudes \eqccd, \eqttd\ account 
for direct 
\nocp. If the latter amplitudes were zero there would be no direct 
\C\P\ violation in the standard model. The flavor changing neutral 
current (FCNC) diagram of fig. \penguin\ 
called, for no good reason in the world, the penguin diagram, 
contributes to the amplitudes \eqccd, \eqttd. The calculation of the 
hadronic matrix elements is even more difficult because there is a 
cancellation between the electroweak (\gam, $Z$) and the gluonic 
penguins, for $m_t$ around 200 GeV, close to the now known top mass. 
Estimates of \rep\ range from few\x10\up{-3} to 10\up{-4}.
\vglue3mm 
\vbox{
\centerline{\epsfig{file=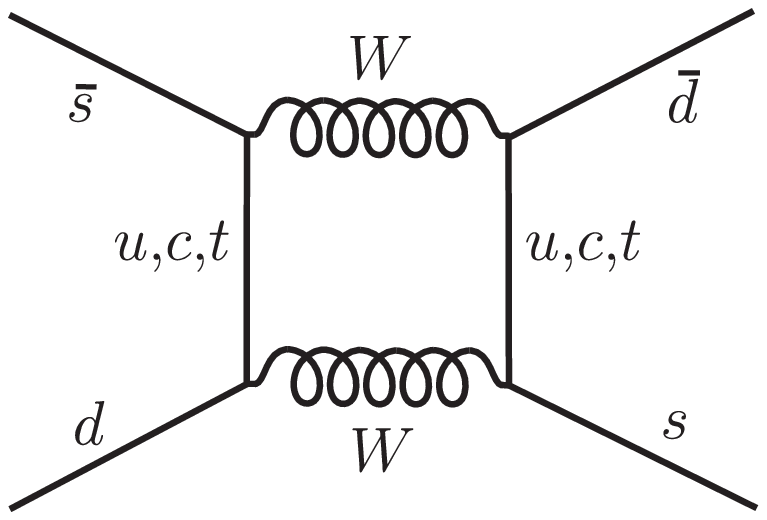,width=5cm} 
\hglue3cm\epsfig{file=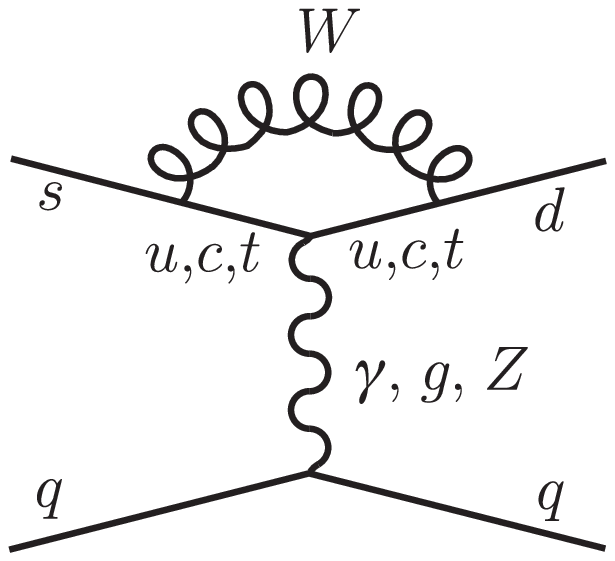,width=5cm}\hglue5mm}
\centerline{\hglue1cm\tenpoint\hbox to7cm{{\bf Fig. \boxdia.} 
Box diagram for \ko
\to\kob.\hfil}
\hglue1cm \hbox to7cm{{\bf Fig. \penguin.} ``Penguin'' FCNC diagram.\hfil}}
}
\vglue3mm
\chapter{New Measurements of the Neutral Kaon Properties}

\section{CPLEAR}
The CPLEAR experiment\Ref\CPlear{R. Adler \etal, 
``CPLEAR Detector at CERN",
NIM {\bf A379} (1996) 76.} studies 
neutral \K\ mesons produced in equal numbers in proton-antiproton 
annihilations at rest:
$$\eqalign{p\bar p\to&K^-\pi^+\ko\quad\hbox{\B=\pt2,-3,}\cr
                  \to&K^+\pi^-\kob\quad\hbox{\B=\pt2,-3,}\cr}$$
\REFS\cpleara{R. Adler \etal, 
``$CP$ Violation Parameter $\eta_{+-}$ Using
Tagged \ko\ and \kob", Phys. Lett. {\bf B363} (1995) 237; 
``\kl\-\ks\ Mass Difference using Semileptonic Decays
of Tagged Neutral Kaons", {\it ibid} 243.}
\REFSCON\cplearb{R. Adler \etal, 
``$CP$ Violation in \ks\to\pppco",
Phys. Lett. {\bf B370} (1996) 167.}
The charge of $K^\pm(\pi^\pm)$ tags the strangeness $S$ of the 
neutral 
\K\ at $t$=0. They have recently presented several new 
results\refsend
from studying \pic, \pic\po\ and $\pi^\pm\ell^\mp\bar\nu(\nu)$ 
final states.
Their measurement of the \kl--\ks\ mass difference $\Delta m$
is independent of the value of $\phi_{+-}$, unlike in most other 
experiments. They have improved limits on the
possible violation of the $\Delta S=\Delta Q$ rule, quantified by 
the 
amplitude's ratio
$x=A(\Delta S=-\Delta Q)/A(\Delta S=\Delta Q)$, without assuming 
$CPT$ invariance. A direct test of $CPT$ invariance has also been 
obtained.
The data require small corrections for background asymmetry 
\ab1\%,  
differences in tagging efficiency, $\varepsilon(K^+\pi^-)-
\varepsilon(K^-\pi^+)$\ab10\up{-3} and in detection,
$\varepsilon(\pi^+e^-)-\varepsilon(\pi^-e^+)$\ab\pt3,-3,. They 
also 
correct for some regeneration in the detector.

\subsec{$\ko(\kob)\to e^+ (e^-)$}
Of particular interest are the study of the decays 
$\ko(\kob)\to e^+ (e^-)$. One can define the four decay 
intensities:
$$\eqalign{
I^+(t)\hbox{ for }&\ko\to e^+\cr
\overline I^-(t)\hbox{ for }&\kob\to e^-\cr}
\Big\}\eqalign{\Delta S=0\quad\cr}
\eqalign{
\overline I^+(t)\hbox{ for }&\kob\to e^+\cr
I^-(t)\hbox{ for }&\ko\to e^-\cr}
\Big\}\eqalign{&|\Delta S|=2\cr}$$
where $\Delta S=0,2$ mean that the strangeness of the decaying \K\ 
is 
the same as it was at $t$=0 or has changed by 2, because of 
$\ko\leftrightarrow\kob$ transitions. 
One 
can then define four asymmetries:
$$\eqalign{
A_1(t)&={I^+(t)+\overline I^-(t)-(\overline I^+(t)+I^-(t)) \over 
I^+(t)+\overline I^-(t)+\overline I^+(t)+I^-(t)}\cr
A_2(t)&={\overline I^-(t)+\overline I^+(t)-(I^+(t)+I^-(t)) \over 
\overline I^-(t)+\overline I^+(t)+I^+(t)+I^-(t)}\cr}$$
$$A_T(t)={\overline I^+(t)-I^-(t)\over \overline I^+(t)+I^-
(t)},\quad
A_{CPT}(t)={\overline I^-(t)-I^+(t)\over \overline I^-
(t)+I^+(t)}$$
From the time dependence of $A_1$ they obtain:
$\Delta m=(0.5274\pm0.0029\pm0.0005)\x10^{10}$ s\up{-1}, a result 
which 
is independent of $\phi_{+-}$ and $\Re 
x=(12.4\pm11.9\pm6.9)\x10^{-3}$, 
without assuming $CPT$. From $A_2$ and assuming $CPT$ they obtain 
$\Im x
=(4.8\pm4.3)\x10^{-3}$, a result \ab5 times more stringent than 
the 
PDG94 
world average. $A_T$ gives a direct measurement of $T$ violation. 
Assuming $CPT$, the 
expected value for $A_T$ is \pt6.52,-3,. The CPLEAR result 
is $A_T=(6.3\pm2.1\pm1.8)\x10^{-3}$. From a study of the $CPT$ 
violating 
asymmetry, $A_{CPT}(t)$, they obtain
$\Re\delta_{CPT}=(0.07\pm0.53\pm0.45)\x10^{-3}$. 
We will come 
back later to the definition of $\delta_{CPT}$, which we will simply 
call $\delta$.

\vbox{\cl{\epsfig{file=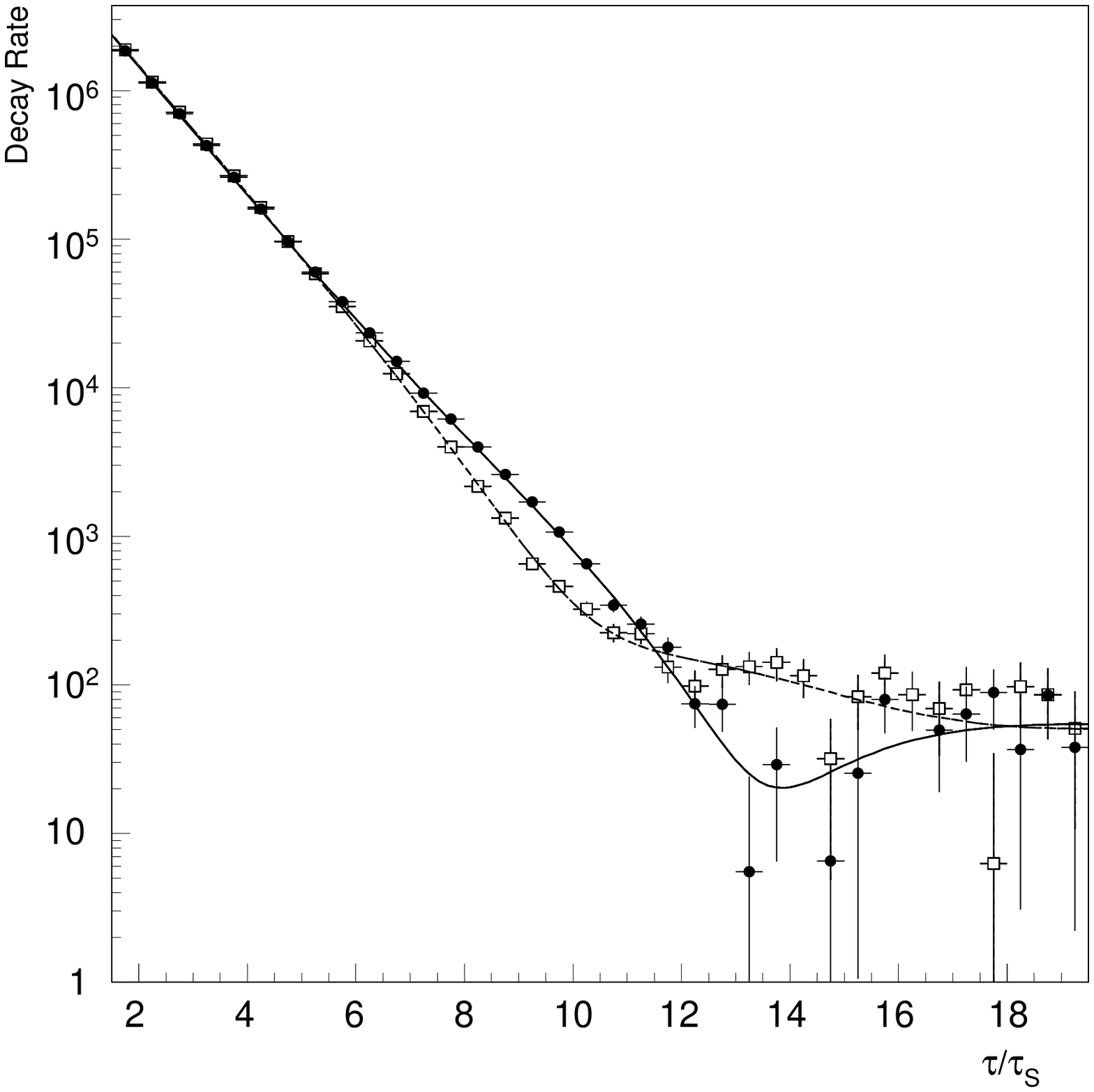,width=12.5cm}}
\cl{\tenpoint{\bf Fig. \kkdec.} \ Decay distributions 
for \ko and \kob.}}
\endpage

\subsec{\pic\ Final State}
From an analysis of \pt1.6,7, \pic\ decays of \ko\ 
and \kob\ they determine $|\eta_{+-
}|=(2.312\pm0.043\pm0.03\pm0.011_{
\tau_S})\x10^{-3}$ and 
$\phi_{+-}=42.6\deg\pm0.9\deg\pm0.6\deg\pm0.9\deg_{\Delta m}$.
The {\it errors} in the values quoted reflect uncertainties in the 
knowledge of the \ks\ lifetime and the \ks--\kl\ mass difference, 
respectively. Fig. \kkdec\ shows the decay intensities of \ko\ and \kob, 
while fig. \decdif\ is a plot of the time dependent asymmetry 
$A_{+-}=
\big(I(\kob\to\pic)-\alpha I(\ko\to\pic)\big)/\big(I(\kob\to\pic)+
\alpha I(\ko\to\pic)\big)$. Most systematics cancel in the ratio 
and the 
residual difference in efficiencies for \ko\ and \kob\ decays is 
determined from a fit to the same data: $\alpha=0.9989\pm0.0006$.

\vglue2mm
\vbox{\cl{\epsfig{file=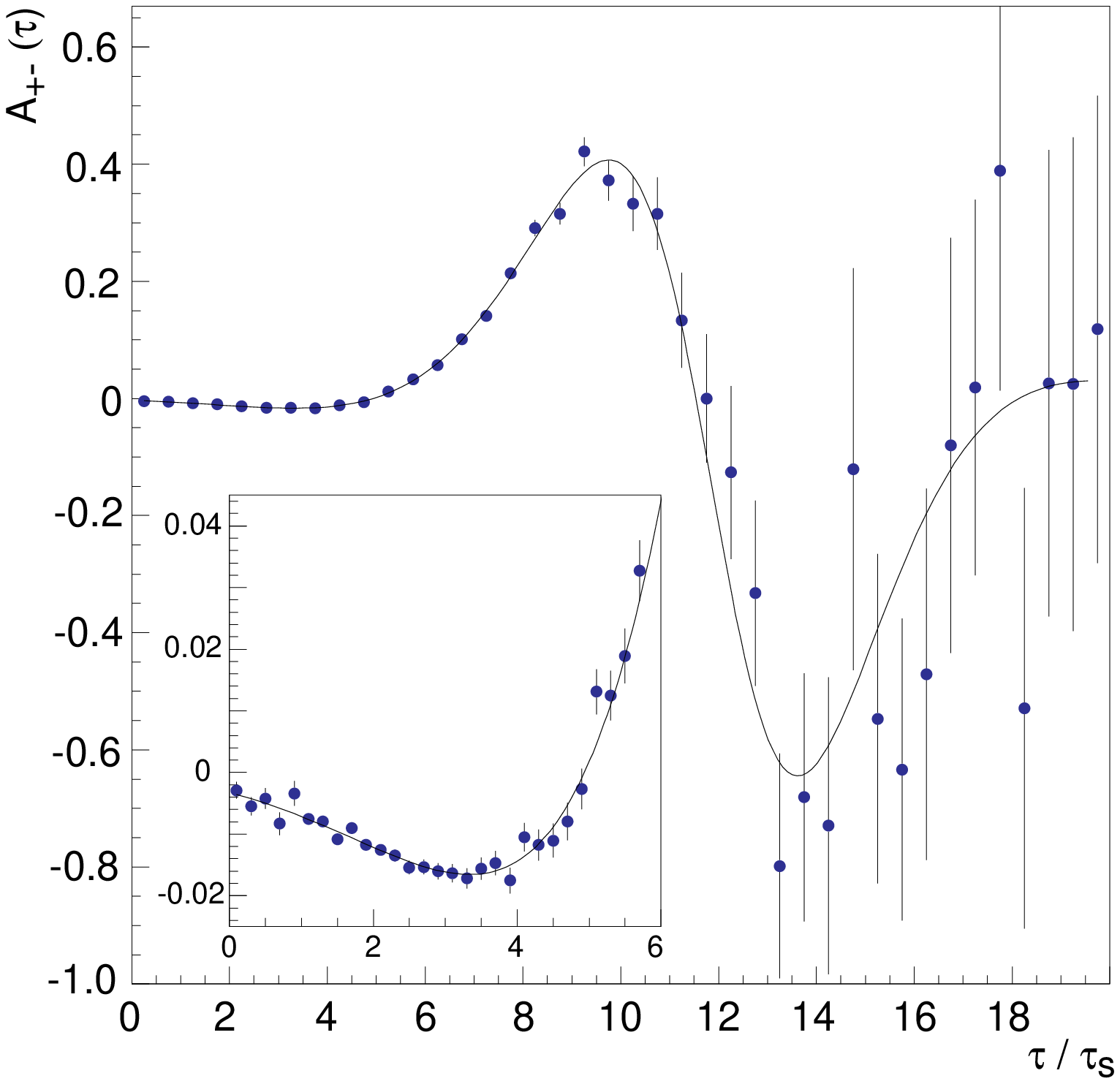,width=13cm}}
\cl{\tenpoint{\bf Fig. \decdif.} \ Difference of 
decay distributions for \ko and \kob.}}
\vglue3mm

\subsec{\pic\po\ Final States}
Studies of \ko-\kob\to\pic\po\ decays give the results 
$\Re\eta_{+-0}=
(-4\pm17\pm3)\x10^{-3}$ and $\Im\eta_{+-0}=(-16\pm20\pm8)\x10^{-
3}$, 
where $\eta_{+-0}=A(\kl\to\pic\po)/A(\ks\to\pic\po)$. 
By setting $\Re\eta_{+-0}=\Re\eta_{+-}$ they obtain $\Im(\eta_{+-
0}=(-11
\pm14\pm8)\x10^{-3}$. These results are significantly more precise 
than any 
previous ones. 

For completeness the results by E621 at FNAL 
for \K\to\pic\po\ must 
be mentioned\rlap.\Ref\pmzero{Y. Zou \etal, 
``$CP$ Violation in the Decay \ks\to\pppco",
Phys. Lett. {\bf B329} (1994) 519; G. B. Thomson \etal, 
``Measurement of the Amplitude of the $CP$ Conserving Decay \ks\to\pppco", 
Phys. Lett. {\bf B337} (1994) 411.}
In this experiment the \C\P\ conserving amplitude 
$A$(\ks\to\pic\po) is measured, obtaining
$$\eqalign{|\rho_{\pic\po}|&=\Big|{A(\ks\to\pic\po,I=2)
\over A(\kl\to\pic\po)}\Big|=0.035^{+0.019}_{-0.011}\pm0.004\cr
\phi_\rho&=-59\deg\pm48\deg\cr
BR(\ks\to\pic\po)&=(3.9^{+0.54+0.8}_{-1.8-0.7})\x 10^{-7}\cr
\Im(\eta_{+-0})&=-0.015\pm0.017\pm0.025,\quad
          {\rm assuming}\ \Re(\eta_{+-0})=\Re(\eps).\cr}$$

\section{E773 at FNAL}
E773 is a modified E731 setup, with a downstream regenerator added.
New results have been obtained on $\Delta m$, $\tau_S$, 
$\phi_{00}-\phi_{+-}$ and $\phi_{+-}$ from a study of \K\to\pic, 
\pio\ 
decays\rlap.\Ref\Schwin{B. Schwingenheuer \etal. 
``$CPT$ Test in the Neutral Kaon System",
\prl. {\bf74} (1995) 4376.}

\subsec{Two Pion Final States}
This study of \K\to$\pi\pi$ is a classic experiment where one 
beats
the amplitude $A(\kl\to\pi\pi\big]_i)$=$\eta_iA(\ks\to\pi\pi)$ 
with the 
coherently
regenerated \ks\to$\pi\pi$ amplitude $\rho A(\ks\to\pi\pi)$, 
resulting in
the decay intensity
$$\eqalign{I(t)=&|\rho|^2e^{-\Gamma_S t}+|\eta|^2e^{-\Gamma_L 
t}+\cr
&\qquad 2|\rho||\eta|e^{-\Gamma t}\cos(\Delta mt+\phi_\rho-
\phi_{+-})\cr
}$$

Measurements of the time dependence of $I$ for the \pic\ final 
state
yields $\Gamma_S$, $\Gamma_L$, $\Delta m$ and $\phi_{+-}$. They 
give the 
following results:
$\tau_S=(0.8941\pm0.0014\pm0.009)$\x10\up{-10} s
setting $\phi_{+-}=\phi_{SW}=\tan^{-1}2\Delta m/\Delta\Gamma$ and
floating $\Delta m$;  
$\Delta m=(0.5297\pm0.0030\pm0.0022)$\x10\up{10} s\up{-1} 
using for $\tau_S$ the PDG94 value, leaving $\phi_{+-}$ free in 
the fit;
$\phi_{+-}=43.53\deg\pm0.58\deg\pm0.40\deg$, using for $\tau_S$ 
the PDG94 value and for the mass difference the combined values of 
E731 
and E773, $\Delta m=(0.5282\pm0.0030)\x10^{10}$ s\up{-1}. 
Including the 
uncertainties on $\Delta m$ and $\tau_S$ and the correlations in 
their 
measurements they finally quote $\phi_{+-}=43.53\deg \pm0.97\deg$

From a simultaneous fit to the \pic\ and \pio\ data they obtain
$\Delta\phi=\phi_{00}-\phi_{+-}=0.62\deg\pm0.71\deg\pm0.75\deg$, 
which
combined with the E731 result gives $\Delta\phi=-
0.3\deg\pm0.88\deg$.

\subsec{\K\to\pic\gam}
From a study of \pic\gam\ final states $|\eta_{+-\gam}|$ and 
$\phi_{+-\gam}$ are obtained.
The time dependence of the this decay, like that for two pion 
case, 
allows extraction of the corresponding parameters $|\eta_{+-\gam}|$ and 
$\phi_{+-\gam}$.
The elegant point of this measurement is that because 
interference is observed (which vanishes between orthogonal 
states) 
one truly measures the ratio
$$\eta_{+-\gam}={A(\kl\to\pic\gam,\hbox{ \nocp\ }) 
\over A(\ks\to\pic\gam, \hbox{ \C\P\ OK })}$$
which is dominated by E1, inner bremsstrahlung transitions. Thus, 
again, one 
is measuring the $CP$ 
impurity of \kl. Direct $CP$ could contribute via E1, direct 
photon
emission \kl\ decays, 
but it is not observed within the sensitivity of the 
measurement.

The results obtained are\rlap:\Ref\pmgam{J. N. Mathews \etal. 
``New Measurements of the $CP$ Violation Parameter $\eta_{+-\gamma}$",
\prl. {\bf75} (1995) 2803.}
$|\eta_{+-\gam}|=(2.362\pm0.064\pm0.04)\x10^{-3}$ and 
$\phi_{+-\gam}=43.6\deg\pm3.4\deg\pm1.9\deg$.
Comparison with $|\eta_{+-}|\ab|\eps|\ab2.3$, $\phi_{+-}\ab43\deg$ 
gives excellent agreement. This implies that the decay is 
dominated by radiative 
contributions and that all one sees is the $CP$ impurity of the \K\ 
states.

\section{Combining Results for $\Delta m$ and $\phi_{+-}$ from 
Different Experiments}

\REF\cplearc{R. Adler \etal, 
``Evaluation of the phase of the CP Violation Parameter $\eta_{+-}$
and the \kl\-\ks\ Mass Difference from a Correlation Analysis of
Different Experiments", Phys. Lett. {\bf B369} (1996) 367.}
\REF\dac{R. Carosi \etal, 
``A Measurement of the Phases of the $CP$ Violating
Amplitudes in \ko\to2$\pi$ Decays and a Test of CPT Invariance",
Phys. Lett. {\bf B237} (1990) 303.}
\REF\dagz{W. C. Carithers \etal, 
``Measurement of the Phase of the $CP$
Nonconservation Parameter $\eta_{+-}$ and the \ks\ Total Decay Rate",
Phys. Rev. Lett. {\bf 34} (1975) 1244.}
\REF\dab{S. Gjesdal \etal, ``A Measurement of the \kl-\ks\ Mass Difference
from the Charge Asymmetry in Semileptonic Kaon Decays",
Phys. Lett. {\bf B52} (1974) 113.}
\REF\dae{C. Geweniger \etal, 
``Measurement of the Kaon Mass Difference $M$(\kl)-$M$(\ks)
by the Two Regenerator Method", Phys. Lett. {\bf B52} (1974) 108.}
\REF\daf{C. Geweniger \etal, 
``A New Determination of the \ko\to\pic\ Decay Parameters",
Phys. Lett. {\bf B48} (1974) 487.}
\REF\dad{M. Cullen \etal, `` A Precision Determination of the
\kl\ - \ks\ Mass Difference", Phys. Lett. {\bf B32} (1970) 523.}
\REF\dbc{L. K. Gibbons \etal, 
``New Measurements of the Neutral Kaon Parameters $\Delta m$,
$\tau_s$, $\phi_{00}-\phi_{+-}$ and $\phi_{+-}$",
Phys. Rev. Lett. {\bf 70}, (1993) 1199.}
The CPLEAR collaboration\refmark{\cplearc} has performed an
analysis for
obtaining the best value for $\Delta m$ and $\phi_{+-}$, taking 
properly 
into account the fact that different 
experiments have different correlations between the two variables. 
The 
data\refmark{\cpleara,\cplearb,\Schwin,\dac-\dbc} with their 
correlations are 
shown in fig. \dmphi. 

\vbox{\cl{\epsfig{file=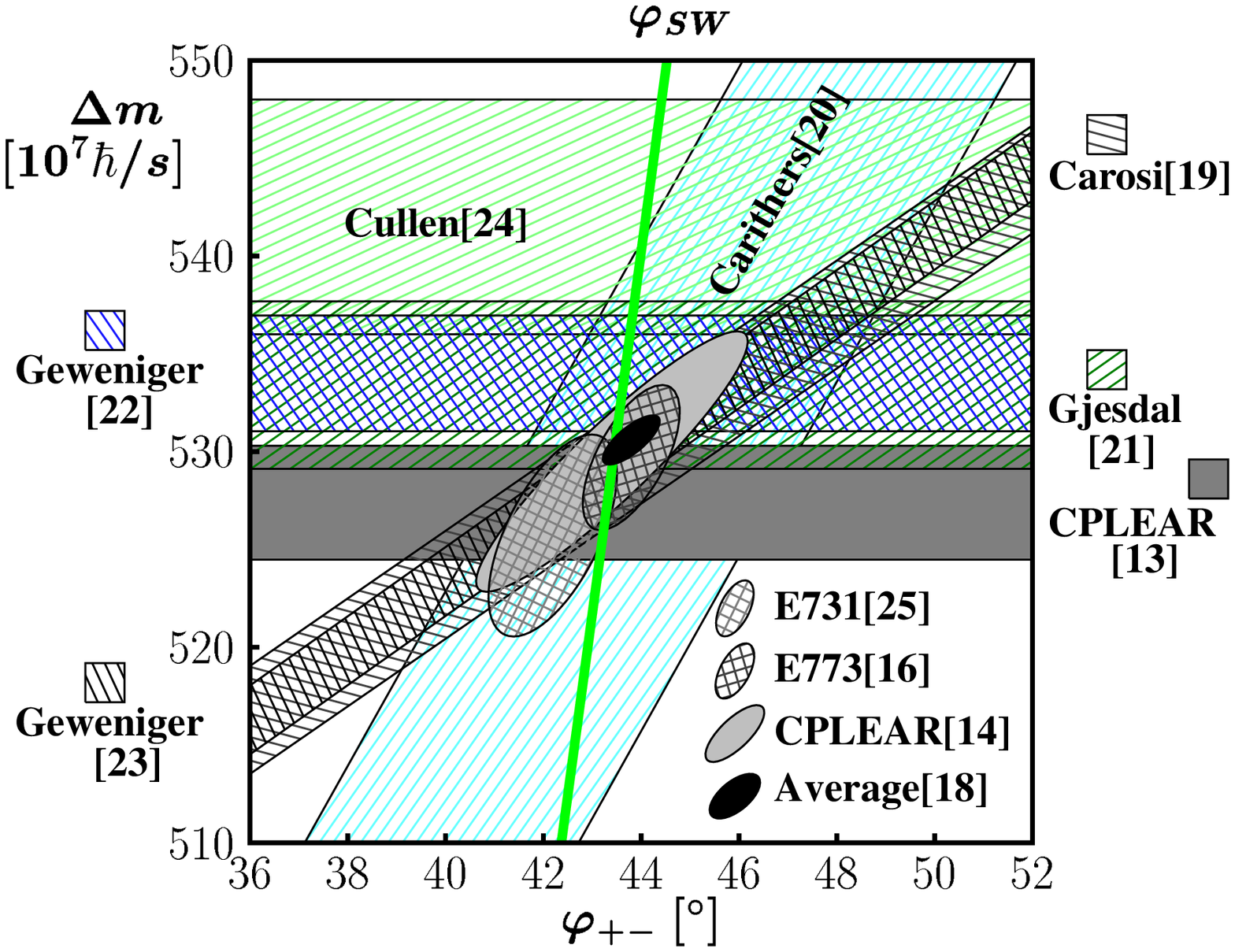,width=16cm}}
\cl{\tenpoint{\bf Fig. \dmphi.} \ A compilation of $\Delta m$ and 
$\phi_{+-}$ results, from ref. \cplearc.}}

\vglue1mm
\noindent
A maximum likelihood analysis of all data gives 
$\Delta m$=(530.6\plm1.3)\x10\up7 s\up{-1} and 
$\phi_{+-}$=\nextline
43.75\deg\plm0.6\deg. $\phi_{+-}$ is very close to the 
{\it superweak phase} 
\f\dn{\rm SW}=43.44\deg\plm0.09\deg.           

\chapter{Tests of $CPT$ Invariance}

\REF\cptnoa{P. Huet, 
``$CPT$ Violation from Planck Scale Physics",
in {\it Proc. Workshop on K Physics}, ed. 
L. Iconomidou-Fayard, (Edition Fronti\`eres, 1997) 413; A. Kostelecky, 
``$CPT$ Strings and Neutral Meson Systems", 
{\it ibid.}, 407.} 
In field theory, $CPT$ invariance is a consequence of quantum 
mechanics and Lorentz invariance. Experimental evidence that $CPT$ 
invariance might be violated would therefore invalidate our belief 
in either or both quantum mechanics and Lorentz invariance. We might 
not be so ready to 
abandon them, although recent ideas\rlap,\refmark{\cptnoa}
such as distortions of the metric 
at the Planck mass scale or the loss of coherence due to the 
properties of black holes might make the acceptance somewhat more 
palatable. Very sensitive tests of $CPT$ invariance, or lack 
thereof, can be carried out investigating the neutral \K\ system 
at a \ff. One should not however forget other 
possibilities\rlap.\Ref\okun{L.B. Okun, ``Tests of $CPT$", in {\it Proc. 
Workshop on K Physics}, ed. 
L. Iconomidou-Fayard, (Edition Fronti\`eres, 1997) 419.}

\section{\C\P\T\ at a \ff}
A \ff\ such as \DAF\ can produce of the order of 10\up{10} neutral \K\ 
pairs per year which allow study of $CP$, $T$ and $CPT$ 
invariance. An advantage of $\phi$-factories, in this respect, is 
that the \K\ pair is produced in a well defined quantum state, 
allowing more refined tests than otherwise.
\K\ mesons are produced via the reaction 
$\epm\to\hbox{``\gam''}\to\phi\to\kkb$ in a $C=-1$ state. 
Therefore the two kaons are in a pure \ko, \kob\ or \ks, \kl\ state 
with a \kl\kl\ or \ks\ks\ impurity of $\ll10^{-5}$. In addition only 
at a \ff\ it is possible to obtain a pure \ks\ beam using the 
observation of a decay at long time as a tag for the presence of a 
\ks.
In general, $CPT$ requires 
$$M_{11}-M_{22}=M(\ko)-M(\kob)=0$$
and in the following we discuss present  
experimental limits on $(M(\ko)-M(\kob))/\langle M\rangle$ and possible 
future improvements.

\section{Neutral \K\ decays without assuming \C\P\T}
One of the problems in dealing with the neutral \K\ system is the 
large number of parameters which are necessary for its 
description\rlap.\Ref\maia{L. Maiani, ``$CP$ and $CPT$ Violation in 
Neutral Kaon Decays", {\it The Second DA${\mit \Phi}$NE Physics 
Handbook},  Eds. L. Maiani \etal, (S.I.S, INFN, LNF, Frascati, 
1995) 3.} Moreover different authors use different notations. For 
consistency we will redefine, following Maiani's 
analysis\refmark{\maia} but with some different symbols, all the 
relevant parameters which will be used below.
To lowest order in ``\eps'' we write the \ks\ and \kl\ states as
$$\eqalign{
\sta{\ks}&=[(1+\epss)\sta{\ko}+(1-\epss)\sta{\kob}]/\sqrt 2\cr
\sta{\kl}&=[(1+\epsl)\sta{\ko}+(1-\epsl)\sta{\kob}]/\sqrt 2\cr
}$$
and define the parameters $\tilde\eps$ and $\delta$ through the 
identities
$$\epss\equiv\tilde\eps+\delta\qquad\epsl\equiv\tilde\eps-
\delta.$$
Following the usual convention, we introduce the ratios 
of the amplitudes for \K\ decay to a final state $f_i$, 
$\eta_i=A(\kl\to f_i)/A(\ks\to f_i)$, and define the parameters 
\eps\ and $\eps'$ with the identities
$$\eta_{+-}\equiv\eps+\eps'\qquad\eta_{00}\equiv\eps-
2\eps'\eqn\eqeta$$
From Eq. \eqeta, \eps\ is given, in terms of the measurable amplitude 
ratios $\eta$,  by:
$$\eqalign{
\eps&=(2\eta_{+-}+\eta_{00})/3\cr
\hbox{Arg}(\eps)&=\phi_{+-}+(\phi_{+-}-\phi_{00})/3.\cr
}$$

To treat consistently the possibility of $CPT$ violation we write 
the most general decay amplitudes for \K\to2$\pi$ as
$$\eqalign{
A(\ko\to2\pi,\ I)&\equiv\sqrt{3/2}(A_I+B_I)\cr
A(\kob\to2\pi,\ I)&\equiv\sqrt{3/2}(A^*_I-B^*_I),\cr
}\eqn\eqdec$$
where $I$ is the isospin of the 2 pion state and we use the Wu and 
Yang phase convention
\ie\ $A_0$ real and positive. The symmetry properties of the $A$ 
and $B$ amplitudes in Eq. \eqdec\ are given 
below:
$$\vbox{\halign{
\hfil#&\ \hfil#\hfil&\ \hfil#\hfil&\ \hfil#\hfil&\ \hfil#\hfil\cr
       &  $\Re A$    &  $\Im A$    &  $\Re B$    &  $\Im B$    \cr
 $CP$  &    $+$    &    $-$   &   $-$    &   $+$    \cr
  $T$   &    $+$    &    $-$   &    $+$   &   $-$    \cr
 $CPT$ &  $+$    &    $+$   &   $-$    &    $-$    \cr}
}.$$
If $CPT$ invariance is valid we have $\delta=0$, 
$\eps=\tilde\eps$, $B_I=0$, otherwise, including ``direct'' $CPT$ 
in the decay amplitude to two pions:
$$\eps=\tilde\eps-(\delta-{\Re B_0\over A_0})\qquad
\hbox{Arg}(\tilde\eps)=\phi_{SW}\equiv\tan^{-1}{2\Delta 
m\over\Delta\Gamma}.\eqn\eqthree$$

From unitarity\rlap,\refmark{\maia} with the most reasonable 
assumption that 
$\Gamma_{11}-\Gamma_{22}\ll\Gamma_S$ for all channels but 2 pions 
with $I$=0, it follows that $\delta-{\Re B_0/ A_0}$ is 
orthogonal to \eps, see fig. \epseps, from which:
$$\eqalign{\hbox{Arg}\Big(\delta-{\Re B_0\over 
A_0}\Big)&=\phi_{SW}\pm90\deg\cr
\Big|\delta-{\Re B_0\over A_0}\Big|&=|\eps|\x|\phi_{SW}-
\hbox{Arg}(\eps)|.\cr
}\eqn\eqdelboaa$$

\vbox{
\cl{\epsfig{file=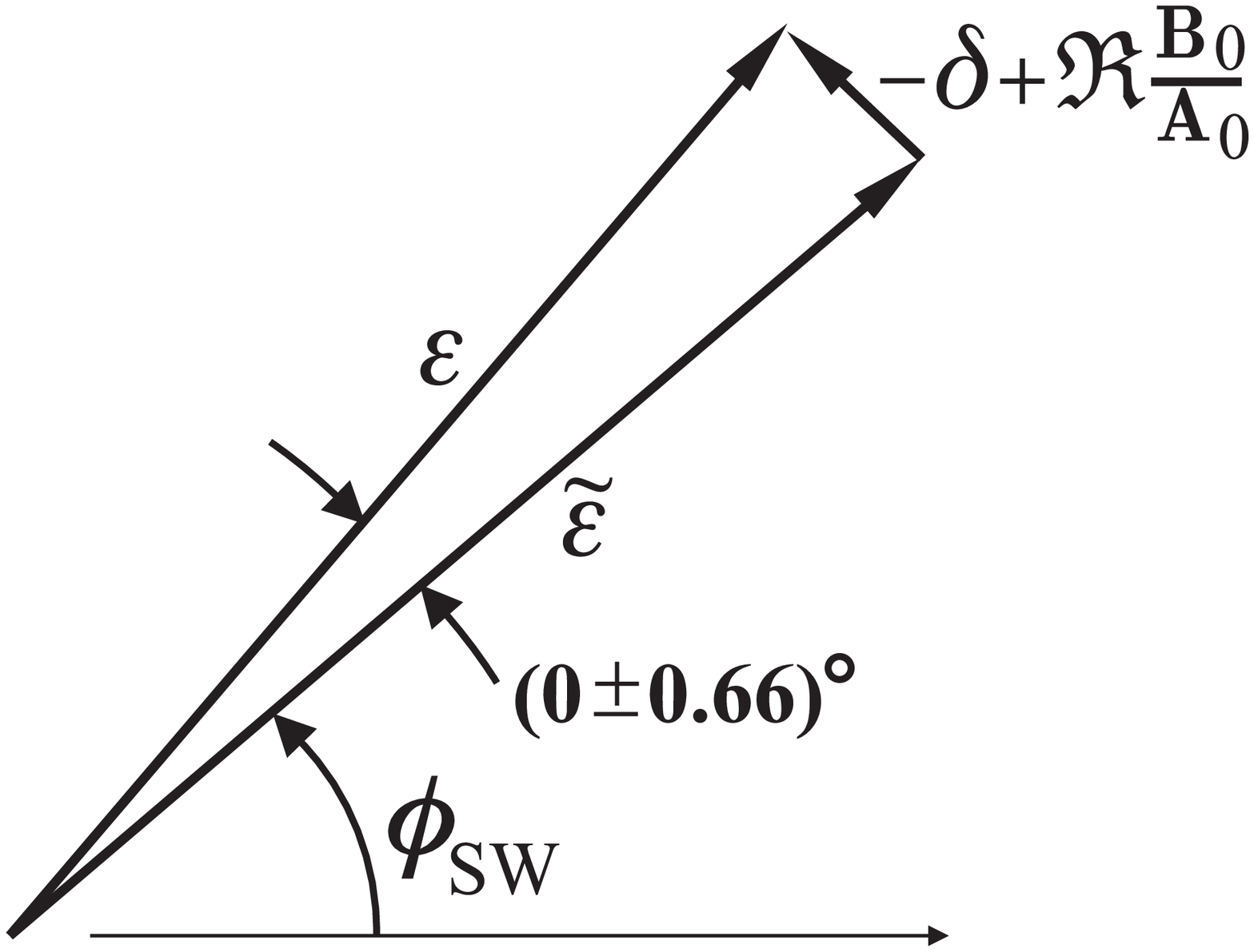,height=5truecm}}
\cl{\tenpoint {\bf Fig. \epseps}. Diagram of the complex quantities in Eq. 
\eqthree.}
}
\section{Experimental Data}

In table 1 we have collected the data relevant to \K\ decays, as 
known today, as well as the values of some derived quantities 
according to the definitions above.
From the values in the table
$$\eqalign{
\Big|\delta-{\Re B_0\over A_0}\Big|&=2.282\x10^{-
3}\x(0\pm0.66\deg)\cr
&=(0\pm2.6)\x10^{-5}.\cr
}\eqn\eqdelbobb$$
If there is no $CPT$ violation in the semileptonic decay 
amplitudes, the 
leptonic asymmetry for \kl\ decays, ${\cal 
A}^\ell_L=(\Gamma_{L,S}^{\ell^+}-\Gamma_{L,S}^{\ell^-
})/(\Gamma_{L,S}^{\ell^+}+\Gamma_{L,S}^{\ell^-})$, can be used 
together with this result for determining $\delta$ and ultimately 
put 
limits on the $CPT$ violating quantity $M(\ko)-M(\kob)$. Under 
this 
assumption one has 
$$\eqalign{
{\cal A}^\ell_L&=2\Re(\tilde\eps-\delta)\cr
\Re\delta&={2\Re\tilde\eps-{\cal A}^\ell_L\over 
2}={2|\tilde\eps|\cos\phi_{SW}- {\cal A}^\ell_L\over 2}\cr
&\hphantom{={2\Re\tilde\eps-{\cal A}^\ell_L\over 2}\ 
}=(1.6\pm6)\x10^{-
5}.\cr
}$$

\REF\pdg{Particle Data Group, ``Review of Particle Properties",
Phys. Rev. D  {\bf54}, 1 (1996).}
\def\dg{a }  \def\ddg{b }
\cl{\tenpoint  Table 1:  Parameters of the neutral \K\ mesons.}
\vglue2mm
{
\vbox{\begintable
 & Parameter & Value \cr
 \dg | $\Delta m$ & \pt$(0.534\pm0.0014)$,10, s\up{-1} \cr
   \dg | $\Gamma_S$ & \pt$(1.1202\pm0.0010)$,10, s\up{-1} \cr
  \ddg | $\phi_{SW}$ & $43.63\pm0.08\deg$ \cr
  \dg |  $\phi_{+-}$ & $43.7\pm0.6\deg$ \cr
  \dg | $\phi_{00}-\phi_{+-}$ & $-0.2\pm0.8\deg$ \cr
  \dg |  $\vert\eta_{+-}\vert$ & \pt$(2.285\pm0.019)$,-3, \cr
  \dg |  $\vert\eta_{00}\vert$ & \pt$(2.275\pm0.019)$,-3, \cr 
  \ddg | $\vert\eps\vert$ & \pt$(2.282\pm0.014)$,-3, \cr 
  \dg |  Arg(\eps) & $43.63\pm0.66\deg$ \cr
  \dg |  ${\cal A}^\ell_L $ & \pt$(3.27\pm0.12)$,-3,  \endtable}
{\baselineskip 12pt
\line{\tenpoint\hglue4cm \dg From reference \pdg.\hfill}
\line{\tenpoint\hglue4cm\ddg Derived using definitions in the 
text.\hfill}} }

Using the magnitude and phase of  $\delta-\Re B_0/A_0$ from eqs. 
\eqdelboaa\ and \eqdelbobb\ we find:
$$\eqalign{
\Im\delta&=(0\pm1.9)\x10^{-5}\cr
\Re B_0/A_0&=(-3\pm6)\x10^{-5}.\cr
}$$
From
$$|M(\ko)-M(\kob)|=|\Gamma_S-\Gamma_L|\,|\Re\delta\tan\phi_{SW}-
\Im\delta|$$
it follows that
$${|M(\ko)-M(\kob)|\over\langle M(K)\rangle}=(0.2\pm0.9)\x10^{-
18},$$
the uncertainty in this result being due mostly to the error on 
${\cal 
A}^\ell_L$.
Note that if there were no $CPT$ violation in the two pion decay 
amplitudes, the limit on the mass difference would be:
$${|M(\ko)-M(\kob)|\over\langle M(K)\rangle}=(0.0\pm0.3)\x10^{-
18}.$$
The complete relation between the various parameters in the text 
is illustrated in fig. \epseta.

\vbox{\vglue1mm
\cl{\epsfig{file=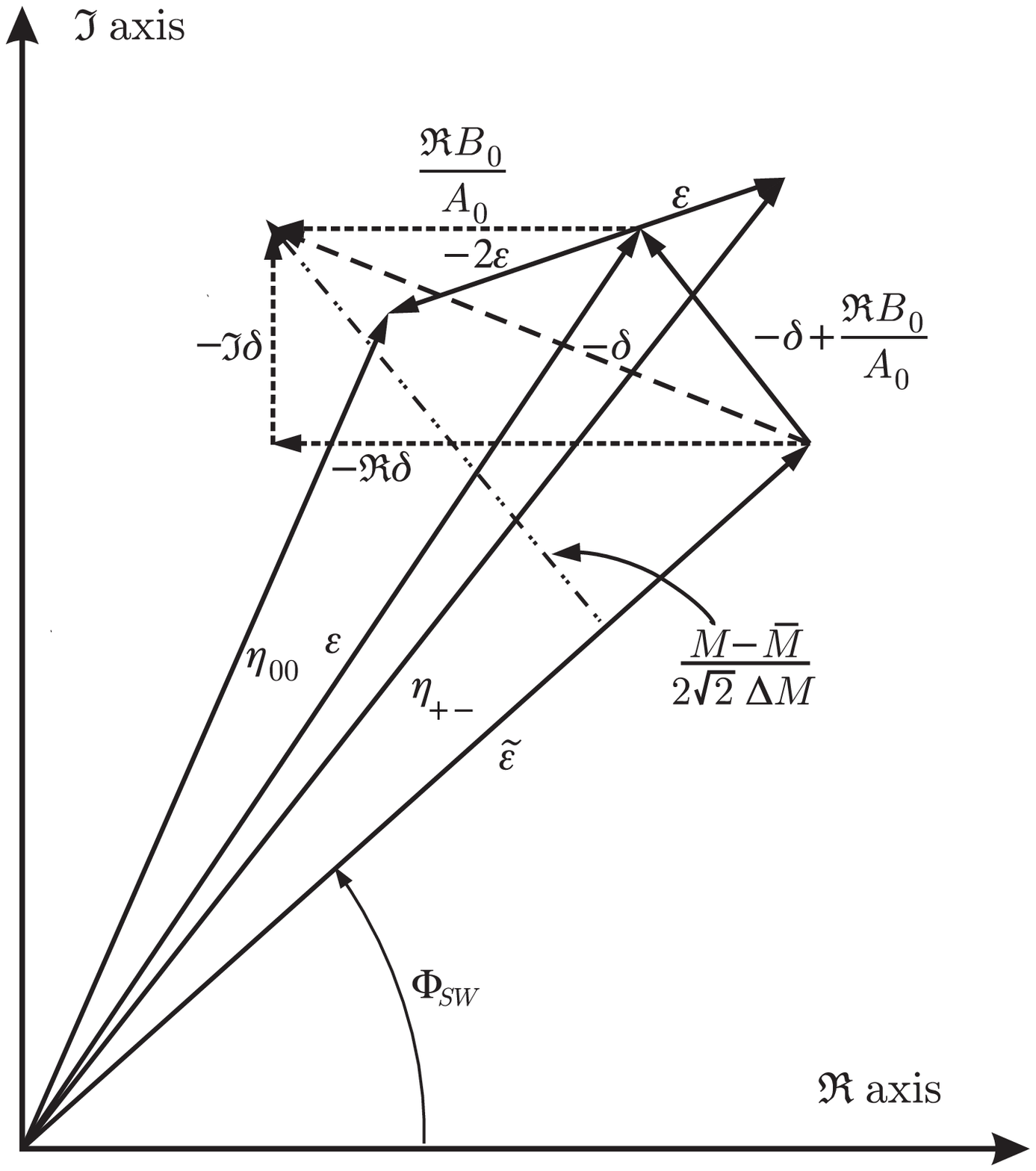,height=13truecm}}
\cl{\tenpoint {\bf Fig. \epseta}. Diagram of all the quantities 
discussed in 
the text.}\vglue2mm}

\section{\CPT\ Violation in $A(\K\to\ell^\pm\pi\nu)$}
To describe consistently \K\ decays without assuming \C\P\T, we 
must allow fo \nocpt\ in semileptonic decays. We therefore introduce 
four complex amplitudes $a$, $b$, $c$, and 
$d$, in terms of which we write
$$\eqalign{
A(\ko\to\ell^+)&=a+b\cr
A(\kob\to\ell^-)&=a^*-b^*\cr
A(\ko\to\ell^-)&=c+d\cr
A(\kob\to\ell^+)&=c^*-d^*,\cr
}$$
where the $c$ and $d$ amplitudes are for $\Delta S=-\Delta Q$ 
transitions. 
Their symmetry properties are displayed below:
$$\vbox{\halign{
\hfil#&\ \hfil#\hfil&\ \hfil#\hfil&\ \hfil#\hfil&\
\hfil#\hfil\cr
  &  $\Re(a,\,c) $ & $\Im(a,\,c) $ & $\Re(b,\,d) $ & $\Im(b,\,d) 
$\cr
 $CP$  &  $+$  &  $-$  &  $-$   &  $+$  \cr
 $T$     &  $+$  &  $-$  &  $+$  &  $-$    \cr
 $CPT$ & $+$  &  $+$ &   $-$  &  $-$    \cr}
}.$$

To first order in $\delta$, $\Im a$, $b$, $c$ and $d$ the leptonic 
asymmetries are
$$\eqalign{
{\cal A}^\ell_L&=2(\Re\tilde\eps-\Re\delta+{\Re b\over\Re a}+{\Re 
d\over\Re a})\cr
{\cal A}^\ell_S&=2(\Re\tilde\eps+\Re\delta+{\Re b\over\Re a}-{\Re 
d\over\Re a})\cr
}$$
which implies that \ks, \kl\ experiments cannot disentangle 
\nocpt\ from 
violation of the $\Delta S=\Delta Q$ rule. The validity or 
otherwise of the rule can be checked by studying the decays of 
strangeness tagged \ko\ and \kob\ states, if the tagging is done 
using strong interactions and not semileptonic decays.  This was 
successfully done by CPLEAR, unfortunately with relatively limited 
statistics.
In the standard model, it is very hard to imagine how $\Delta S=-
\Delta 
Q$ transitions can be induced at the level of 10\up{-5}. Assuming 
$c$=$d$=0, the remaining \nocpt\ term cancel in the difference of 
the leptonic asymmetries and we obtain
$${\cal A}^\ell_S-{\cal A}^\ell_L=4\Re\delta.$$

At \DAF, it should be possible to reach an accuracy of \ab\pt2.5,-
4, for ${\cal A}^\ell_S$ (we assume a tagging efficiency of order 
75\%, see section 9.3, which means a measurement of  $\Re\delta$ to 
an accuracy \ab\pt0.6,-4,, almost ten times better than 
CPLEAR, albeit with 
the assumption of the 
validity of the $\Delta S=\Delta Q$ rule. 
The rule itself can also be checked to good accuracy at \DAF\ by 
using strange\-ness tagged neutral \K\ mesons produced by charge 
exchange of  
$K^\pm$ which are produced even more copiously than  neutral \K's.
The ratio $\Gamma^\ell_L/\Gamma^\ell_S=1+4\Re(c/a)$ can put limits 
on 
the $CPT$ conserving part of the $\Delta S=-\Delta Q$ amplitude, 
to 
about the same sensitivity as above. Time ordered 
asymmetries\Ref\daip{G. D'Ambrosio, G. Isidori and A. Pugliese, 
``$CP$ and $CPT$ Measurements at DA$\Phi$\-NE", {\it The Second DA${\!\mit 
\Phi}$NE Physics Handbook},  Eds. L. Maiani \etal, (S.I.S., INFN, 
LNF, Frascati, 1995) 63.}
$(\ell^+\ell^--\ell^-\ell^+)/(\ell^+\ell^-+\ell^-\ell^+)$, where 
$+-$ means that the positive lepton appears earlier than the 
negative and viceversa for $-+$, are 
also sensitive to the various $CPT$ odd terms at small and large 
time 
differences.

The CPLEAR limit for the mass difference does
not assume $\Delta S=\Delta Q$ and uses their own new limits
on $\Im x$ and
$\Im \eta_{+-0}$. The limit on \nocpt\ is only slightly
weaker\rlap:\Ref\noulis{P. Pavlopoulus, 
``Measurements of $CP$ and $T$ Violation Parameters in the Neutral
Kaon System at CPLEAR", {\it Proc. of the 2\up{nd} 
Workshop on Physics and Detectors for DA${\mit\Phi}$NE}, eds. R. 
Baldini \etal\ (S.I.S., INFN, LNF, Frascati, 1995) 55.}
$${|m_{\ko}-m_{\bar K_0}|\over m_{\ko}}<2.2\x10^{-18}$$

\chapter{Rare \K\ Decays}

Rare \K\ decays offer several interesting possibilities, which 
could 
ultimately open a window beyond the standard model.
They allow the determination of the CKM matrix parameters, as for 
instance
from the \noc\nop\ decay \kl\to\po$\nu\bar\nu$, as well as from 
the \C\P\
conserving one \K\up+\to\pip$\nu\bar\nu$.
They also permit the verification of conservation laws which are 
not strictly
required in the standard model, for instance by searching for 
\ko\to$\mu e$
decays.

The connection between measurements of neutral \K\ properties and 
branching ratios and
the $\rho$ and $\eta$ parameters of the Wolfenstein 
parameterization of 
the CKM matrix, is shown {\it schematically} in fig. 
\etarho.

\vglue 2mm
\centerline{\epsfig{file=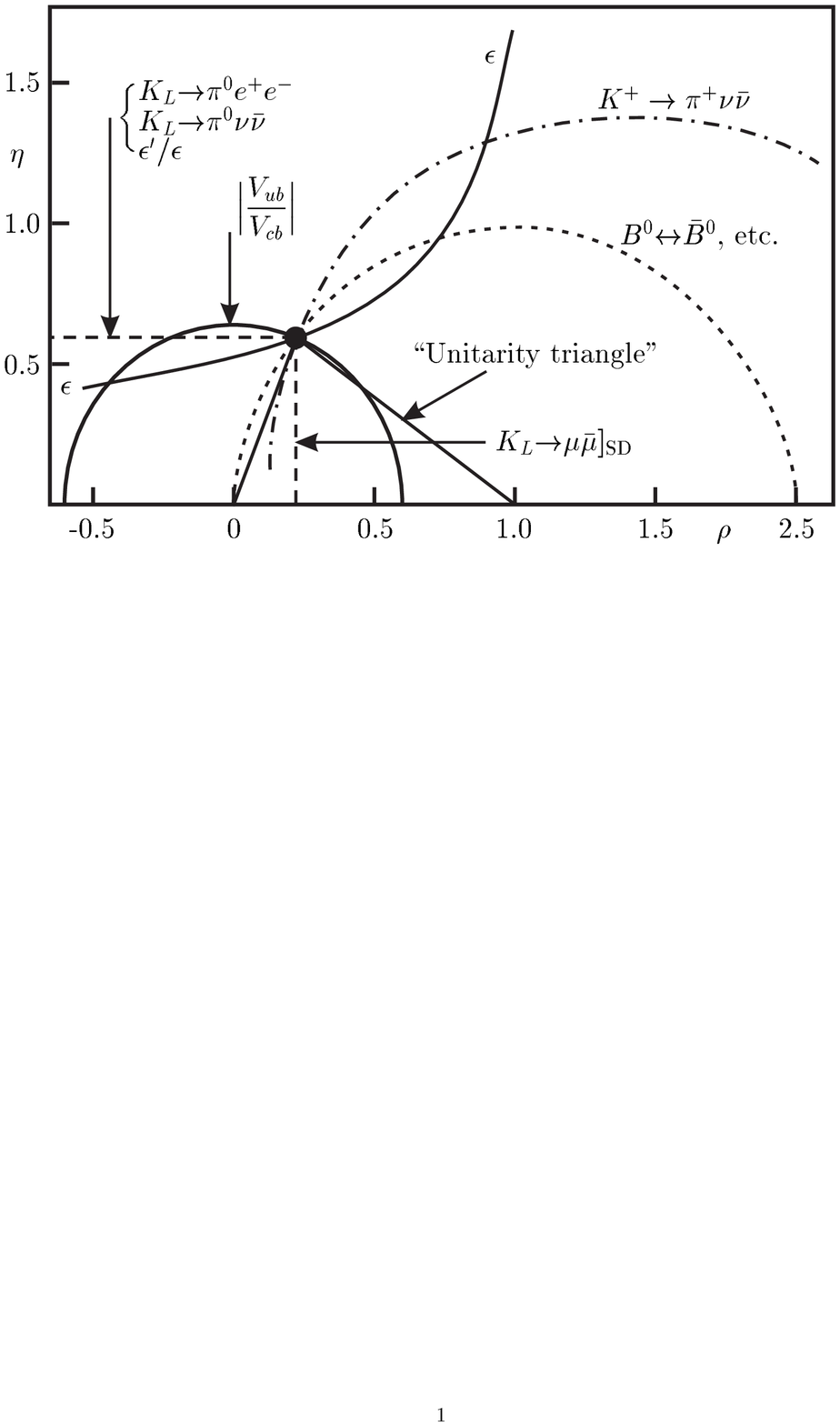,width=13truecm}}
\vglue 2mm
\cl{\tenpoint{\bf Fig. \etarho.} \ Constraints on 
$\eta$ 
and $\rho$ from measurements of \eps, $\eps'$, rare decays and $B$ meson 
properties.}
\vglue2mm

In general the situation valid for the more abundant \K\ decays, 
\ie\ that
the 
$\nocp\big|_{\rm direct}$ decays have much smaller rates then the 
$\nocp\big|_{\rm indirect}$ ones,
can be reversed for very rare decays. In addition, while the 
evaluation 
of $\eps'$ is particularly unsatisfactory because of the 
uncertainties 
in the calculation of the
hadronic matrix elements, this is not the case for some rare decays.
A classifications of measurable quantities according to increasing
uncertainties in the calculation of the hadronic matrix elements 
is
given by Buras\Ref\buras{A. J. Buras, 
``$CP$ Violation: Present and Future",in {\it Phenomenology of 
Unification from Present to Future}, eds. G. Diambri\-ni-Palazzi \etal, 
(World Scientific, 1994) 41.} as:
1. \B(\kl\to\po$\nu\bar\nu$),
2. \B($K^+\to\pi^+\nu\bar\nu$),
3. \B(\kl\to\po\epm), \ \eps\dn{K}, and
4. $\eps'_{K}$, \ \B(\kl\to$\mu\bar\mu]_{\rm SD}$), where SD 
stands for 
{\it short distance} contributions.
The
observation $\eps'\ne0$ remains a unique proof of direct
\nocp. Measurements of 1 through 3, plus present knowledge, over
determine the CKM matrix. 
Rare \K\ decay experiments are not easy however, just like 
measuring
\rep\ has turned out to be difficult. Typical expectations for some of 
the
interesting decays are:
$$\eqalign{
BR(\kl\to\po\epm,\ \noc\nop]_{dir})&\ab(5\pm2)\x10^{-12}\cr
BR(\kl\to\po\nu\bar\nu)&\ab(3\pm1.2)\x10^{-11}\cr
BR(K^+\to\pi^+\nu\bar\nu)&\ab(1\pm.4)\x10^{-10}\cr
}$$

The most extensive program in this field has been ongoing for a 
long time at
BNL and large statistics have been collected 
recently and are under analysis.
Sensitivities of the order of 10\up{-11} will be reached, although
10\up{-(12{\rm\ or\ }13)} are really necessary.
Experiments with high energy kaon beams have been making excellent 
progress
toward observing rare decays.
\REF\rarea{M. Weaver \etal, ``Limit on Branching Ratio \kl\to\po$\nu\bar\nu$",
\prl. {\bf72} (1994) 3758.}
\REF\rareb{P. Gu \etal, ``Measurement of Branching Ratio and Study of $CP$
of \kl\to\epm\epm",
\prl. {\bf72} (1994) 3000.}
\REF\rarec{D. Roberts \etal, ``Search for Decay \kl\to\ppo\gam",
Phys. Rev. {\bf D73} (1994) 1874.}
\REF\rared{T. Nakaya \etal, 
``Measurement of Branching Ratio of \kl\to\epm\gam\gam",
\prl. {\bf73} (1994) 2169.}
\REF\raree{M. B. Spencer \etal, 
``Measurement of the Branching Ratio anf Form Factor of \kl\to$\mu\mu$\gam",
\prl. {\bf74} (1995) 3323.}
\REF\raref{K. Arisaka \etal, 
``Search for the Lepton Family Number Violating Decay", EFI 95-08.}
\REF\rareg{P. Gu \etal, 
``First Evidence for the Decay \kl\to\epm$\mu^+\mu^-$",
\prl. {\bf 76} (1996) 4312.}
\REF\harea{G. D. Barr \etal, 
``Limit on the Decay \ks\to\po\epm",
Phys. Lett. {\bf B 304} (1993) 381.}
\REF\hareb{G. D. Barr \etal, 
``Search for the Decay \kl\to\ppo\gam",
Phys. Lett. {\bf B 328} (1994) 528.}
\REF\harec{G. D. Barr \etal, 
``Measurement of the Branching Ratio of the Double Dalitz Decay \kl\to\epm\epm",
Z. Phys. {\bf C65} (1995) 361.}
\REF\hared{A. Kreutz \etal, 
``Determination of the Branching Ratio of \kl\to\pppo/\kl\to$pi^+\pi^-
\break\pi^0$",
Z. Phys. {\bf C65} (1995) 67.}
\REF\haree{K. Kleinknecht, 
``Results on Rare Decays of Neutral Kaons from NA31 Experiment",
MZ-ETAP/95-4.}
\REF\harez{G. D. Barr \etal, 
``Measurement of \kl\to3\gam",
Phys. Lett. {\bf B 358} (1995) 399.}
\REF\haref{G. D. Barr \etal, 
``A Test of CHPT from measurement of the decay \ks\to\gam\gam",
Phys. Lett. {\bf B 351} (1995) 579.}
We will discuss new results from E799-I\refmark{\rarea-\rareg}
(E731 without regenerators) and NA31\rlap.\refmark{\harea-\haref}
The results obtained by the two experiments are 
summarized in the tables below.
\def\less{\ifm{<}}  
\def\pen{\ifm{\pi e\nu}}   

\def\nunu{\ifm{\nu\bar\nu}}  \def\mumu{\ifm{\mu^+\mu^-}}
\def\mepm{\ifm{\mu^\pm e^\mp}}
\vglue4mm
\cl{\tenpoint{\bf Table 3.} E799-I Rare \K-decays Results.}
\vglue3mm
\begintable
  Reaction | Events | \B\ or limit |  Ref.  \cr
 \kl\to\po\nunu |   | \less\pt5.8,-5, | \rarea\   \cr
 \kl\to\epm\epm | 27 | (4.0\plm0.8\plm0.3)\x10\up{-8} | \rareb\   
\cr
 \kl\to\pio\gam) |    | \less\pt2.3,-4, |  \rarec\  \cr
 \kl\to\epm\gam\gam, $E_\gamma>$5 MeV | 
|(6.5\plm1.2\plm0.6)\x10\up{-7} | 
\rared\ \cr
 \kl\to\mumu\gam | 207 | (3.23\plm0.23\plm0.19)\x10\up{-7} | 
\raree\   \cr
 \kl\to\po\mepm | | \less\pt6.4,-9, | \raref\   \cr
 \kl\to\epm\mumu | 1 | $(2.9^{+6.7}_{-2.4})$\x10\up{-9} | \rareg\ 
\endtable

\vglue 4mm
\cl{\tenpoint{\bf Table 4.} NA31 Rare \K-decays Results.}
\vglue3mm
\begintable
  Reaction | Events |  \B\ or limit | Ref.   \cr
 \ks\to\po\epm | 0  |  \less\pt1.1,-6, | \harea\   \cr
 \kl\to\pio\gam | 3 |  \less\pt5.6,-6, | \hareb\   \cr
 \kl\to\epm\epm\ | 8 | (10.4\plm3.7\plm1.1)\x10\up{-8} | \harec\   
\cr
 \kl\to\pio\po  |   |  0.211\plm0.003 |  \hared\ \cr 
   |\multispan{2} 
\Gam(\kl\to3\po)/\Gam(\kl\to\pic\po)=1.611\plm0.037 | 
\hared\ \cr
   |\multispan{2} \Gam(\kl\to3\po)/\Gam(\kl\to\pen)=0.545\plm0.01 
| \hared\ \cr
 \kl\to\po\gam\gam | 57 | (1.7\plm0.3)\x10\up{-6} | \haree\   \cr
 \kl\to\epm\gam | 2000 | (9.1\plm0.3\plm0.5)\x10\up{-6} | \haree\ 
\cr
 \kl\to3\gam    |      |  \less\pt2.4,-7, | \harez\ \cr
 \ks\to\gam\gam | 16   | \pt(2.4\plm0.9),-6, | \haref\  \endtable

\noindent
These new results do not yet determine $\rho$ and $\eta$. 
They do however confirm the feasibility of such a program.

\section{Search for \K\up+\to\pip\nunu} This decay, $CP$ allowed, 
is
best for determining $V_{td}$. At present there is no information, 
other
than E787-BNL's limit \B\less\pt2.4,-9,\rlap.\Ref\old{
S. Adler \etal, ``Search for the Decay $K^+$\to$\pi^+\nu\bar\nu$",
\prl. {\bf 76} (1996) 1421.}
The new E787\Ref\new{S. Kettell, ``E787: A Search for the Rare Decay 
$K^+$\to$\pi^+\nu\bar\nu$", 
in {\it Heavy Quarks at Fixed Target}, 
Ed. L. K\"opke, (S.I.S., INFN, LNF, Frascati, 1996) 397.}
detector, which has found 12 events of
\K\to$\pi$\mumu, \B\ab10\up{-8}, has collected data for a total of 
\pt2.55,12, stopped kaons, \ab 7 times the previous statistics.
This corresponds to about two
\K\up+\to\pip\nunu\ event. At least 100 are necessary for a first
$V_{td}$ measurements. 

\section{\K\to\gam\gam}
Direct \nocp\ is possible in this channel.
Defining the two photon states, where $L$ and $R$ refer to the 
photon 
polarizations,
$$\eqalign{
\sta+ &=(\sta{LL}+\sta{RR})/\sqrt2\cr
\sta- &=(\sta{LL}-\sta{RR})/\sqrt2\cr}$$
we have four possibilities for \kl,\ks\to\gam\gam, given below, 
with the 
expected \B's:
$$\vbox{\halign{\ #\ &\quad#\quad\hfil   &\quad#\hfil\cr
     &\sta+         &\sta-      \cr
\noalign{\vglue 2mm}
\kl  &\pt7,9,, \nocp&\pt6,-4,   \cr
\ks  &\pt2,-6,      &\pt5,-12,, \nocp \cr} }$$

The \nocp\ channels can be isolated by measuring the \gam\ 
polarization, using Dalitz conversion. The present results confirm 
expectations on the \C\P\ conserving channels. Both E799-I and 
NA31 
have detected \kl\to\epm\epm\ decays, 27 and 8 events 
respectively, finding \B=(3.9\plm0.8, 10\plm4)\x10\up{-8} to be 
compared 
with the expectation (3.4\plm0.2)\x10\up{-8}. They also have 
determined 
that $CP\sta{K_2}=-\sta{K_2}$.
NA31 has also observed 69 \K\to\gam\gam\ events, of which 52 are 
from 
\kl\ and one is background. From this they derive 
\B(\ks\to\gam\gam)=\nextline
(2.4\plm0.9)\x10\up{-6}. 
These results are in agreement with expectations, still one needs 
sensitivities of 10\up{-12}.

\section{\K\to\mumu}

Second order weak amplitudes give contributions which depend on
$\rho$, with\break
 $BR|_{\rm SD}\ab10^{-9}$. Measurements of the muon polarization 
are necessary. One however 
needs to confirm the calculations for \K\to\gam\gam\to\mumu, which 
can 
confuse the signal. The following results are relevant
\pointbegin NA31 with 2000 \kl\to\epm\gam\ events finds
\B=(9.1\plm0.3\plm0.5)\x10\up{-6}.
The \B\ depends on the \K\gam\up*\gam\ form factor, with 
contributions 
from vector meson dominance and the $KK^*\gam$ coupling,
$f(q^2)=f_{VMD}+\alpha_{K^*}f_{KK^*\gam}$. The measured \B\ 
corresponds
to $\alpha_{K^*} =- 0.27\pm0.1.$ 
\point E799-I observes 207 \kl\to\mumu\gam\ events, giving 
\B=(3.23\plm0.23\plm0.19)\x10\up{-7} and 
$\alpha_{K^*}=0.13^{+0.21}_{-0.35}$
\point E799-I has found one \kl\to\epm\mumu\ 
event\rlap,\refmark{\rareg}
on the basis of 
which they estimate the branching ratio as
\B=$(2.9^{+6.7}_{-2.4})$\x10\up{-9}. Expectations are \pt2.3,-9,, 
from VMD 
and \pt8,-10, for $f(q^2)$=const. Previous limits were 
\B\less\pt4.9,-6,.

\noindent
At BNL the experiment E871\Ref\twomu{W. Molzon,
``Leptonic Decays of Neutral Kaons",
in {\it Heavy Quarks at Fixed Target}, 
Ed. L. K\"opke, (S.I.S., INFN, LNF, Frascati, 1996) 415.}
has completed collecting data which should allow the observation of one 
event for a \B\ of 10\up{-12}.

\section{\kl\to\po\epm}

The direct \nocp\ \B\ is expected to be \ab\pt5,-12,.
There are however three contributions to the rate plus a 
potentially
dangerous background.
\pointbegin \K\dn2\to\po\gam\gam\to\po\epm, a \C\P\ allowed 
transition.
\point \kl\to\po\epm, from the \kl\ \C\P\ impurity 
($\eps\sta{K_1}$). 

\point Direct \nocp\ from short distance, second order 
weak contributions, via $s\to d+Z$, the signal of interest.
\point Background from 
\kl\to\gam\gam\up*\to\epm\gam\to\epm\gam\gam, 
with a photon from final state radiation.

The relevant experimental results are:
\pointbegin NA31 finds 57 \kl\to\po\gam\gam\ events corresponding to 
\B=(1.6\plm0.3)\x10\up{-6}, equivalent to 
\B(\kl\to\po\epm) = \pt5,-13,
\point NA31 finds no \ks\to\po\epm\ events or \B\less\pt1.1,-6,, 
from 
which \B(\kl\to\po\epm)\nextline
\ab$|\eps|^2(\Gamma_S/\Gamma_L)\B(\ks)<3.2\x10^{-9}$,
which is not quite good enough yet.
\point 799-I finds 58 \kl\to\epm\gam\gam\ events, 
\B=(6.5\plm1.2\plm0.6)\x10\up{-7}. 

The background from  point 3 above will not be dangerous for the 
new 
proposed experiments (KTEV and NA48), because of the superior 
resolution 
of their new electromagnetic
calorimeter.
The observation of direct \nocp\ contributions to \kl\to\po\epm\ 
should be convincing when the necessary sensitivity is reached.

\section{\kl\to\po\nunu}
This process is a pure direct \nocp\ signal. The present limits 
are far 
from the goal. The 
sensitivities claimed for E799-II and at KEK are around 10\up{-9}. 
Another factor of 100 improvement is necessary.

\chapter{Future}
Three new experiments: NA48\Ref\epsi{M. Calvetti, 
``Status Report on NA48 at CERN",
in {\it Proc. of the 2\up{nd} 
Workshop on Physics and Detectors for DA${\mit\Phi}$NE}, eds. R. 
Baldini \etal\ (S.I.S., INFN, LNF, Frascati, 1995) 81.}
in CERN, KTEV\Ref\nktev{B. Winstein, 
``Status of KTEV",
in {\it Proc. of 
the 2\up{nd} Workshop on Physics and Detectors for DA${\mit\Phi}$NE}, 
eds. R. Baldini \etal\ (S.I.S., INFN, LNF, Frascati, 1995) 73.}
at FNAL and KLOE\Ref\kloe{J. Lee-Franzini, 
``Status Report on KLOE",
in {\it Proc. 
of the 2\up{nd} Workshop on Physics and Detectors for 
DA${\mit\Phi}$NE}, 
eds. R. Baldini \etal\ (S.I.S., INFN, LNF, Frascati, 1995) 31.} at 
LNF, are
under construction and will begin taking data in '97 -- '98, with 
the
primary aim to reach an ultimate error in \rep\ of ${\cal 
O}$(10\up{-4}).
The sophistication of these experiments takes advantage of our 
experience of two decades of
fixed target and \epm\ collider physics. Fundamental in KLOE
is the possibility of continuous self-calibration while running, 
via 
processes like Bhabha scattering and charged \K\ decays.

\section{NA48}
The layout of the NA48 experiment, with its main components is shown in 
fig. \nafe.

A new feature of NA48, with respect to its predecessor NA31, is 
that 
\kl\ and \ks\ 
beams simultaneously illuminate the detector, by the very clever 
use of 
a bent crystal to deflect a portion of the incident proton beam. 
This
deflected beam
is brought to a \ks\ production target located close to the 
detector, reducing systematic errors due to different dead times 
when 
detecting \pic\ or \pio\ \K\ decays.

\vbox{\cl{\epsfig{file=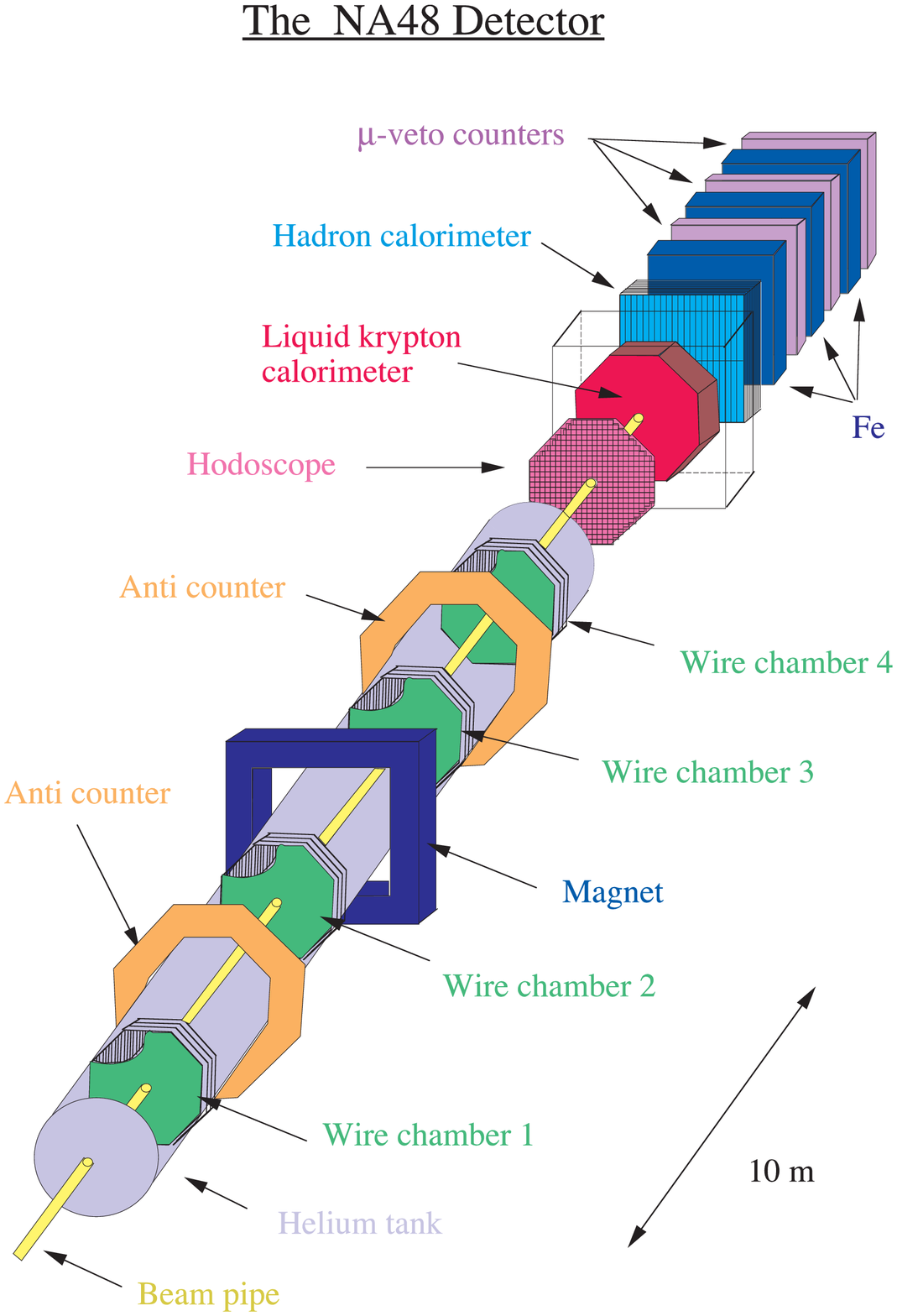,width=13cm}}
\cl{\tenpoint{\bf Fig. \nafe.} The NA48 experiment at 
CERN.}}

The superior resolution of 
the 
liquid krypton calorimeter further improves the definition of the 
fiducial regions and improves rejection of 3\po's 
decays. A magnetic spectrometer has also been 
added in order to improve resolution and background rejection for the
\ko\to\pic\ decays.

\section{KTEV}

Fig. \ktev\ gives a plan view of the KTEV experiment at FNAL; note the 
different longitudinal and transverse scales.

\vbox{\cl{\epsfig{file=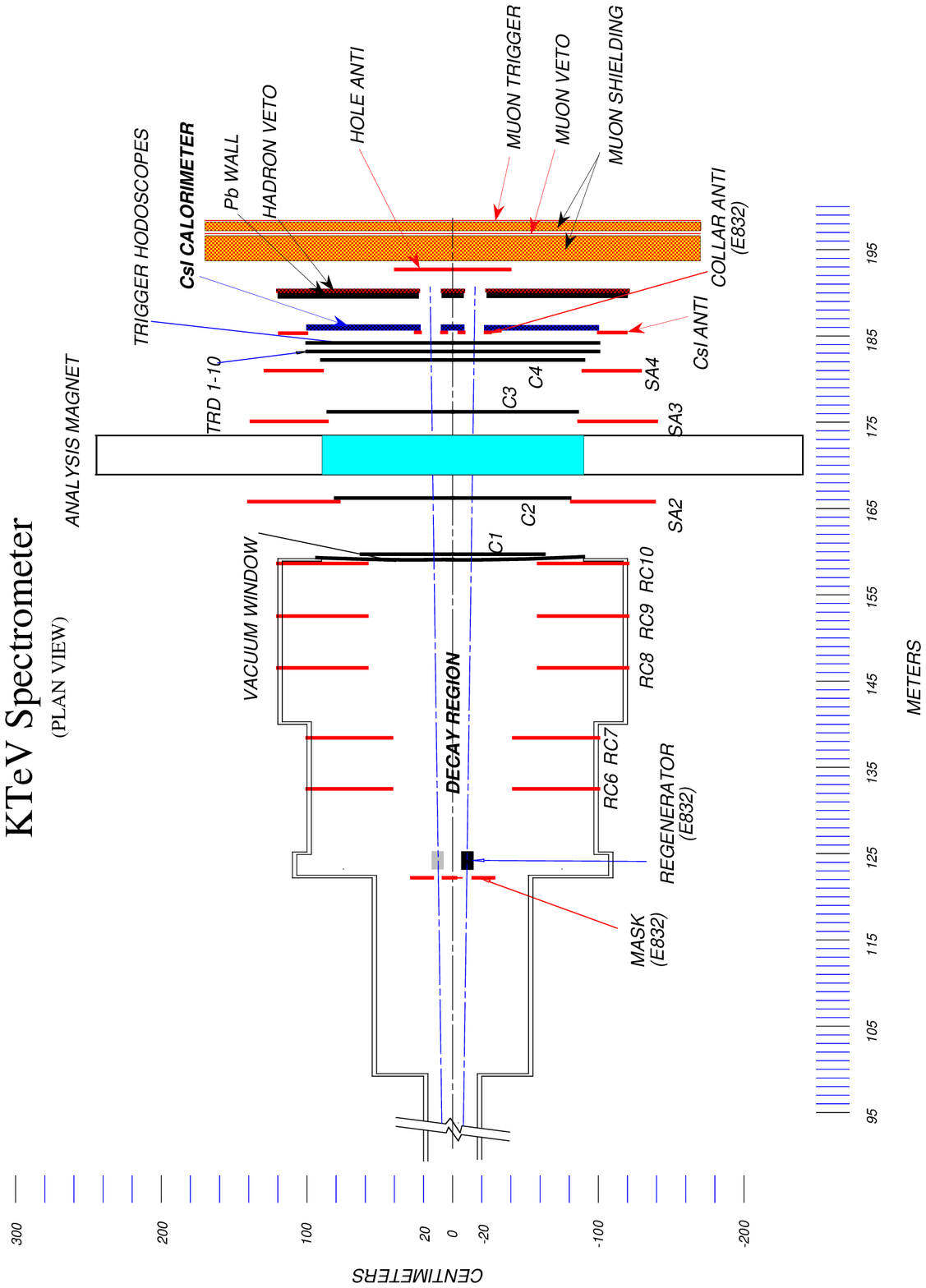,height=17cm}}
\cl{\tenpoint{\bf Fig. \ktev.} Plan view of the KTEV 
experiment at FNAL.}}
\vglue2mm

The KTEV experiment retains the basic principle of E731, with several
significant improvements, the most important being the use of CsI
crystals for the electromagnetic calorimeter. This results in better
energy resolution which is important for background rejection in the
\pio\ channel as well as in the search for rare \K\ decays. 

\section{KLOE}

The KLOE detector\rlap,\Ref\tecprop{KLOE Collaboration,
``KLOE, A General Purpose Detector for \DAF",   Internal Report LNF-
019, April 1, 1992.} designed by the KLOE collaboration and under 
construction by the collaboration at the Laboratori Nazionali di 
Frascati, is shown in cross section in fig. \klosec.
The KLOE detector looks very much like a collider detector and 
will be 
in fact operated at the \DAF\ collider under construction at the 
Laboratori Nazionali di Frascati, LNF. 
At \DAF\ \K-meson are produced in 
pairs at rest in the laboratory, via the reaction \epm\to\f\to2\K.
\ab5000 \f-mesons are produced per second 
at a total 
energy of W=1020 MeV and full \DAF\ luminosity. 

The main motivation behind the whole KLOE venture is the 
observation of direct $CP$ violation from a measurement of \rep\ 
to a sensitivity of 10\up{-4}. The first requirement for achieving 
such accuracy is to be able to collect enough statistics, which in 
turn requires studying of the order of few\x10\up{10} \kl\ decays. 
The dimensions of the detector are dictated by one parameter, the 
mean free path for decay of \kl's which is about 3.4 m.

\vbox{\vglue3mm
\cl{\epsfig{file=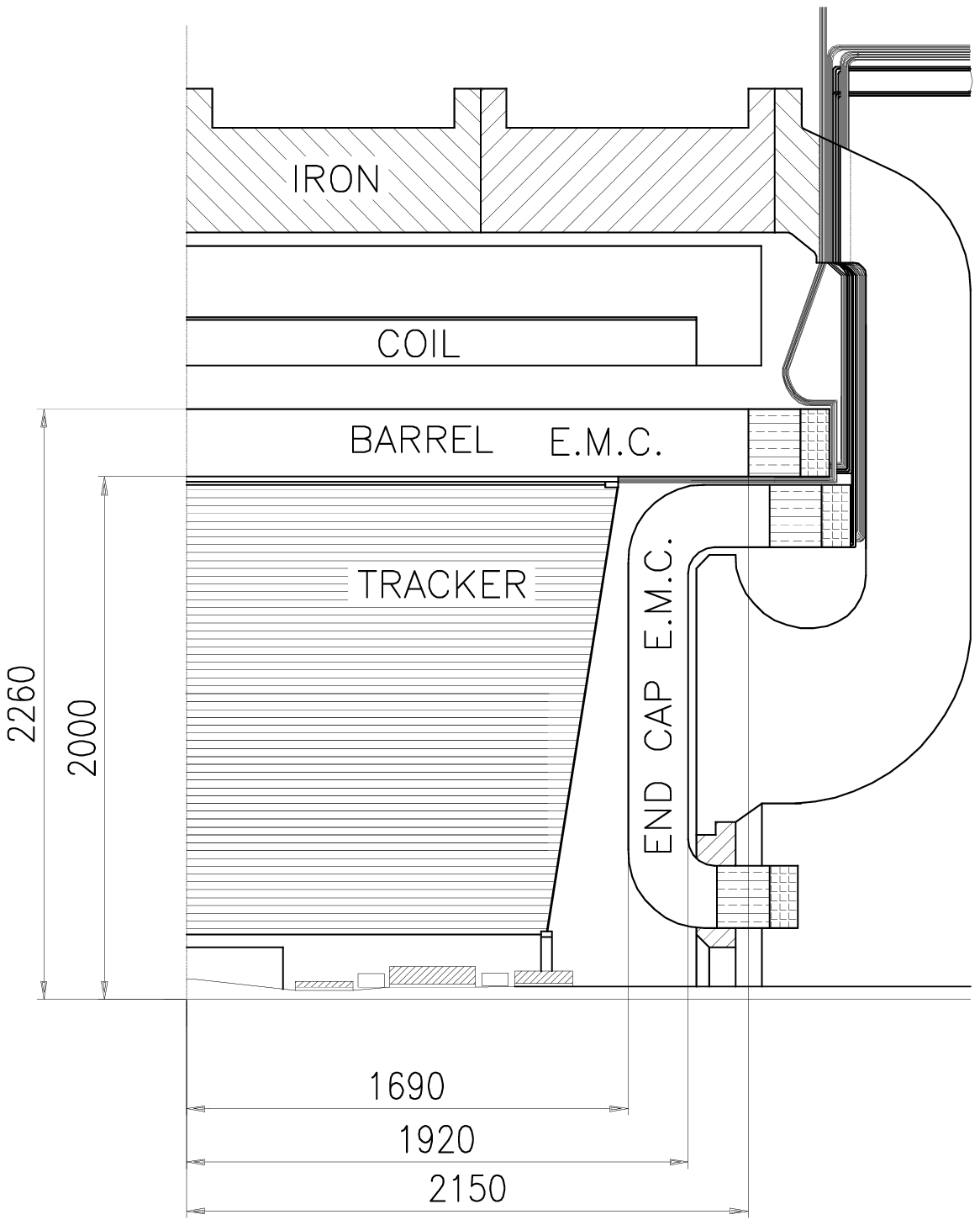,height=13truecm}}
\cl{\tenpoint {\bf Fig. \klosec}. Cross section of the KLOE detector at 
\DAF.}\vglue3mm}

The detectors consists of a 2 m radius drift chamber, employing 
helium rather than argon, to control multiple scattering at  
energies below 500 MeV and to minimize regeneration. The chambers 
has 13,000 W sense wires plus 39,000 Al field wires. The chamber 
is surrounded by a sampling electromagnetic calorimeter consisting 
of 0.5 mm Pb foils and 1 mm diameter scintillating fibers. The 
calorimeter resolution in energy is $\sigma(E)/E$=4.7\%
at 1 GeV and timing resolution is $\sigma(t)$=55 ps, also at 1 
GeV.

At full \DAF\ luminosity, \L=10\up{33} cm\up{-2} s\up{-1}, KLOE 
will collect almost 2000 \ksl\ decays per second. Measurements of 
the leptonic decays mentioned in section 3.1 is possible with KLOE 
because of the large statistics and the tagged \ks\ beam unique to 
a \ff.
The two neutral \K\ mesons are produced in a pure 
$C$-odd quantum state. This implies that, to a very 
high level of accuracy, the final state is always \ks\kl$-$\kl\ks\ 
or 
\ko\kob$-$\kob\ko. Tagging of \ks, \kl, \ko, \kob\ is therefore 
possible.
The produced \K\ mesons are monochromatic, with $\beta$\ab0.2. This 
allows 
measurement of the flight path of neutral \K's by time of flight. 
\REFS\rosner{I. Dunietz, J. Hauser and J. Rosner, 
``Proposed Experiment Addressing $CP$ and $CPT$ Violations in the \ko\kob\
System", Phys. Rev. D {\bf35}, 2166 (1987).}
\REFSCON\paolo{P. Franzini, 
``$CP$ Physics at \DAF", in {\it Proc. of the 
Workshop on Physics and Detectors for DA${\mit\Phi}$NE}, ed. G. 
Pancheri (S.I.S., INFN, LNF, Frascati, 1991) 733.}
A pure \ks\ beam of about 10\up{10} per year is a unique 
possibility at 
\DAF\ at full luminosity. 
A very high \ks\ tagging efficiency, \ab75\%, can be 
achieved in KLOE by detecting  \kl\ interactions in the 
calorimeter, in addition to \kl\ decays in the tracking volume. 

Finally 
because of the well defined quantum state, spectacular 
interference 
effects are observable\rlap,\refsend allowing a totally different 
way of measuring 
\rep, in addition to the classical method of the double ratio 
${\cal R}$.

\chapter{Conclusions}

Ultimately three independent measurements
performed with very different techniques should
be able to determine whether \rep$\ne$0, as long as \rep\ab 
few\x10\up{-4}.
Each experiment has additional by-products of interest in kaon
physics.
From KTEV and NA48, more precise values of \f\dn{+-} and $\Delta 
\f$ will
be obtained. KTEV expects to reach an error of 0.5\deg\ in the 
experimental determination
of \f\dn{f} or \f\dn{\rho} using semileptonic decays. NA48 can 
measure 
\f\dn{+-} by oscillations of the decay rate behind their 
production targets, 
if $n(\ko)\ne n(\kob)$. The strong correlation between $\Delta m$ 
and 
\f\dn{+-} does not change. However all errors will be smaller.
Likewise other parameters relevant to testing $CPT$ invariance 
will be
measured to higher accuracy, {\it e.g.} the charge asymmetry $A_L$ 
in 
semileptonic decays. In this respect the 
uniqueness of 
\DAF\ is that of providing a tagged, pure \ks\ beam which allows 
KLOE to 
measure
the charge asymmetry ${\cal A}^\ell_S$ in leptonic decays of \ks-mesons to an 
accuracy of a few\x10\up{-4}. The value of $\Gamma_L$ is 
becoming
relevant in the analysis of the \ko--\kob, \ks--\kl\ system. This 
is
a measurement which KLOE can perform, improving the accuracy by 
\ab\x15.

Concerning rare decays the number of events collected by KTEV and 
NA48 
should increase by a
factor of 100, corresponding to putting limits of few\x10\up{-11} 
on unobserved
decays and an improvement of a factor ten in the measurable rates.
The statistics available at \DAF\ for \kl\ decays cannot compete 
with that
of KTEV and NA48. However the tagged \ks\ beam will allow us to 
improve
the measurements of rare \ks\ decays by three orders of magnitude.

One last open question is a better test of the $\Delta S=\Delta Q$ 
rule. 
This is
not possible with the \ko-\kob\ state produced at \DAF\ (without 
invoking $CPT$) nor with high energy \K\ beams. \K's tagged via 
strong interactions are required to test the rule.
The copious $K^+K^-$ \ production at \DAF\ provides tagged
\K\up{+}(\K\up{-}) beams which, via charge exchange, results in 
strangeness
tagged \ko(\kob)'s, much in the same way it is done in CPLEAR. 
CPLEAR has
collected tens of million events, KLOE can do at least a factor of 
ten 
better.

\chapter{Acknowledgements}
We wish to thank P. Pavlopoulus for the CPLEAR results,
M. Calvetti for the NA48 setup, B. Weinstein for the KTEV
set up and L. Maiani for discussions on CPT tests.
One of us, JLF, wishes to thank Misha Danilov and the organizers of the 
Winter school for a most enjoyable stay.

\immediate\closeout\referencewrite
\referenceopenfalse
\chapter{References}

\input referenc.texauxil

\bye